\documentclass[a4paper,11pt]{article}
\pdfoutput=1 % if your are submitting a pdflatex (i.e. if you have images in pdf, png or jpg format)

\usepackage{jheparxiv}

\usepackage[T1]{fontenc} % if needed

\usepackage{tensor}
\usepackage{amsmath}
\usepackage{amssymb}
\usepackage{mathrsfs}
\usepackage{mathtools}
\usepackage{bbm}
\usepackage{graphicx}
\usepackage{stmaryrd}
\usepackage{subcaption}
\usepackage{twistor}
\usepackage[compat=1.1.0]{tikz-feynman}
\tikzfeynmanset{warn luatex=false}
\usepackage{xcolor}

\title{Quantizing the non-linear graviton}

\author[a]{Roland Bittleston,}
\author[b]{David Skinner}
\author[c]{and Atul Sharma}

\affiliation[a]{Perimeter Institute for Theoretical Physics,\\ 51 Caroline Street, Waterloo, Ontario, Canada\vspace{0.1cm}}
\emailAdd{rbittleston@perimeterinstitute.ca}

\affiliation[b]{Dept. Applied Maths \& Theoretical Physics,\\ University of Cambridge, Wilberforce Road, CB3 0WA, United Kingdom\vspace{0.1cm}}
\emailAdd{d.b.skinner@damtp.cam.ac.uk}

\affiliation[c]{The Mathematical Institute,\\ University of Oxford, Woodstock Road, OX2 6GG, United Kingdom\vspace{0.2cm}}
\emailAdd{atul.sharma@maths.ox.ac.uk}

\begin{document}

\abstract{
    We consider holomorphic Poisson-BF theory on twistor space. Classically, this describes self-dual Einstein gravity on space-time, but at the quantum level it is plagued by an anomaly. The anomaly corresponds to the fact that integrability of the self-dual vacuum Einstein equations does not survive in self-dual quantum gravity. We compute the anomaly polynomials in the Poisson-BF theory, as well as in this theory coupled to a holomorphic BF theory on twistor space describing self-dual Yang-Mills. We show that all anomalies may be cancelled by further coupling to a twistor field representing a type of axion on space-time. When the twistor anomalies are cancelled, all $n\geq4$-pt amplitudes vanish and integrability is restored.
}

\maketitle
\flushbottom

%%%%%%%%%%%%%%%%%%%%%%%%%%%%%%%%%%
%%%%%%%%%%%%%%%%%%%%%%%%%%%%%%%%%%
%%%%%%%%%%%%%%%%%%%%%%%%%%%%%%%%%%

\section{Introduction}
\label{sec:introduction}

The self-dual vacuum Einstein equations, \begin{equation}
\label{sdVEeqs}
    {\rm Ric} = 0 = {\rm Weyl}^-\,,
\end{equation} 
describe hyperk{\"a}hler metrics, including many gravitational instantons. They have long been recognised as a rare example of an integrable system in four dimensions. For example, if $\nabla_{\dal\al}$ is the connection on the tangent bundle, self-duality implies the integrability of the Lax pair $L_{\dal}(\lambda)=\lambda^{\al}\nabla_{\dal\al}$ for every 2-spinor $\lambda_\al$. The vacuum Einstein condition can be reformulated as either the first or second Plebanski heavenly equations~\cite{Plebanski:1975wn}, and the interplay between these leads to a recursion operator for higher flows in the associated hyperk{\"a}hler hierarchy~\cite{Dunajski:2000iq}. If the four-manifold admits a Killing vector $K$, symmetry reduction along $K$ reduces~\eqref{sdVEeqs} to known integrable systems such as the  SU($\infty$) Toda system~\cite{Ward:1990qt} and the dispersionless KP system~\cite{Dunajski:2000rf}. The self-dual vacuum Einstein equations themselves have also been interpreted as the equations of motion of an integrable 2d non-linear sigma model in the large $N$ limit~\cite{Park:1989vq}, and also play a prominent role in the target space interpretation of the $\cN=2$ string~\cite{Ooguri:1991fp}.\\

This integrability is closely associated to twistor theory. Penrose's non-linear graviton construction~\cite{Penrose:1976js} shows that solutions of~\eqref{sdVEeqs} are in 1:1 correspondence with twistor spaces. These are complex 3-folds $\CPT$ that admit a fibration $\CPT\to\CP^1$ and that contain at least one rational curve whose normal bundle $N\cong\cO(1)\oplus\cO(1)$, and that possess an $\cO(2)$-valued symplectic form on each fibre. In the Euclidean case~\cite{Atiyah:1978wi}, twistor space can be viewed as the bundle of complex structures on the hyperk{\"a}hler manifold. The twistor construction `explains' the integrability of~\eqref{sdVEeqs}, in the sense that $\CPT$ may be built without solving any pdes. See {\it e.g.}~\cite{Mason:1991rf} for a review of twistor theory and its relation to integrable systems.\\

Self-dual gravity is also interesting as a finite, though non-unitary, theory of quantum gravity in four dimensions. However, integrability does not survive at the quantum level. Perturbatively around flat space, the obstruction can be seen in the existence of $n$-particle scattering amplitudes\footnote{In Euclidean signature, these amplitudes require that the external momenta are complex.}, where all external particles have positive helicity. The amplitudes appear at 1-loop and were first calculated by Bern {\it et al.}~\cite{Bern:1998xc,Bern:1998sv}, following similar calculations in self-dual Yang-Mills~\cite{Bern:1993qk,Mahlon:1993fe}. The non-trivial S-matrix implies, as the contrapositive to the Coleman-Mandula theorem, that quantum self-dual gravity cannot have the higher conserved charges we would expect in an integrable theory. The amplitudes themselves have many remarkable properties. In versions of self-dual gravity  where the action involves a Lagrange multiplier imposing the field equations~\eqref{sdVEeqs} (or some equivalent version), the amplitudes are 1-loop exact. Furthermore, despite being loop amplitudes, they are rational functions of the external momenta. For these reasons, as in self-dual Yang-Mills, it has long been suspected that the amplitudes should be viewed as an anomaly in integrability~\cite{Bardeen:1995gk}.\\

The philosophy of this paper, following Costello~\cite{Costello:2021bah}, is that quantum field theories which exist as local theories on twistor space are {\it necessarily} equivalent to integrable quantum theories on space-time. Thus, the failure of quantum integrability of self-dual gravity indicates an obstruction to defining the twistor theory at the quantum level. More specifically, non-linear gravitons corresponding to solutions of~\eqref{sdVEeqs} can be obtained from the field equations of a certain chiral theory -- holomorphic Poisson-BF theory -- on twistor space, first investigated in~\cite{Mason:2007ct}. As we review in section~\ref{sec:twistoractions}, the action for this theory makes use of the $\cO(-2)$-valued Poisson structure on $\CPT$, and its field equations ensure the integrability of an almost complex structure $\bar\p + \{h, \ \}$ on $\CPT$ obtained from Hamiltonian deformations of a background $\bar\p$-operator. The classical action enjoys a gauge redundancy under diffeomorphisms that preserve this Poisson structure. However, at the quantum level, these Hamiltonian diffeomorphisms become anomalous, rendering the quantum theory ill-defined. We remark that anomalies in holomorphic Poisson-BF theory on $\C^{2n}$ have been studied previously in~\cite{Williams:2018ows}, but the case of 3 complex dimensions (and we expect odd dimensions more generally) turns out to be significantly different.\\

The situation is parallel to that of self-dual Yang-Mills, whose twistor space anomalies were investigated in~\cite{Costello:2021bah}. In fact, self-dual Yang-Mills can be coupled to self-dual gravity, and we show that the corresponding twistor theory suffers from gauge, gravitational and mixed anomalies. Since, as a smooth manifold, twistor space is six-dimensional, these anomalies appear as the failure of certain box diagrams to be diffeomorphism and/or gauge invariant. It is important to emphasise that the anomalies do not signal any breakdown of the self-dual theories on space-time. Indeed, in the gauge theory case Costello demonstrates that the anomaly can be cancelled using a non-local term on twistor space which does not contribute to the space-time action. Rather, they simply mean that these self-dual space-time theories do not have twistor space avatars at the quantum level.\\

One of the main results of this paper is that all three anomalies can be cancelled by adjusting only two couplings to a field $\eta\in\Omega^{2,1}_{\rm cl}$ that plays the role of a type of axion on space-time.  In more detail, as we explain below, the action for $\eta$ is
\be
S_\eta =\frac{1}{4\pi}\int_\PT \p^{-1}\eta\,\overline{\nabla}\eta +  \lambda_\fg\,\eta\,\tr(a\p a) + \mu\,\eta\, \tr(s\p s)\,, 
\ee
where $a$ and $s$ represent the gauge and gravitational fields on twistor space, and where $\lambda_\fg$ and $\mu$ are coupling constants. We find that all three anomalies in 1-loop box diagrams may be cancelled by tree-level $\eta$ exchange provided we tune
\be
\lambda_\fg^2 = \frac{C_\fg}{3!}\left(\frac{\im}{2\pi}\right)^2\qquad\text{and}\qquad
\mu^2 = \frac{2+\dim\fg}{2\cdot 5!}\left(\frac{\im}{2\pi}\right)^2
\ee
whenever the gauge algebra $\fg =\fsl_2$, $\fsl_3$, $\fso_8$ or any of the exceptional Lie algebras. (These choices of $\fg $ admit an identity $\tr_{\rm adj}(X^4) = C_\fg\,\tr(X^2)^2$ with $C_\fg = 10({\bf h}^\vee)^2 / (2+\dim\fg)$ and ${\bf h}^\vee$ the dual Coxeter number.) These same $\fg$ were found in~\cite{Costello:2021bah} by cancelling just the gauge anomaly, so in particular no new constraints on $\fg$ arise from also cancelling the gravitational and mixed anomalies. 

The twistor space theory remains 1-loop exact even when coupled to $\eta$, so no higher loop anomalies are possible. Thus, for the above $\fg$ and couplings $(\lambda_\fg,\mu)$, the twistor theory exists and the corresponding space-time theory is expected to be quantum integrable. Indeed, we show explicitly that all 4-pt amplitudes with external gravitons or gluons vanish once the contribution of the axion is included. We also argue that this implies there can be no $n>4$-pt amplitudes.\\

In fact, our calculations make it clear that there are many ways to cancel these three anomalies, perhaps the most obvious being to make the theory (even minimally) supersymmetric.  The reason for our focus on axion exchange as the mechanism for anomaly cancellation is twofold. Firstly, and most importantly, it is closely related to the Green-Schwarz mechanism for the type I topological string~\cite{Costello:2007ei}. While a version of the B-model can be formulated on twistor space~\cite{Costello:2021bah}, its closed string sector corresponds to self-dual space-time geometries that are scalar-flat K{\"a}hler, rather than Ricci flat. These are more closely related to conformal gravity than to Einstein gravity. It is hoped that the Green-Schwarz mechanism here encourages the belief that some topological string exists in twistor space which describes (non-supersymmetric) self-dual Einstein gravity. It seems likely that such a theory would be a twistor space version of the $\cN=2$ string~\cite{Ooguri:1990ww,Ooguri:1991fp}, whose string field theory would involve a Poisson-Chern-Simons theory on $\CPT$, towards which the Poisson-BF theory considered here may be viewed as a warm-up example. The anomaly cancellation mechanism explored here may help to understand, without invoking space-time supersymmetry, the fact that worldsheet calculations in~\cite{Berkovits:1994vy,Berkovits:1994ym,Ooguri:1995cp} show that $n\geq4$-pt amplitudes in this theory vanish to all orders. At a more mundane level, the fact that loop diagrams in self-dual gravity may be cancelled by a tree-level axion exchange diagram allows for a simpler method of obtaining loop level results by computing trees.\\

The relation between four dimensional integrable systems and theories in twistor space is strikingly similar to the relation between two dimensional integrable systems and a certain 4d mixed topological-holomorphic Chern-Simons theory rediscovered by Costello in~\cite{Costello:2013sla} (it originally appears in \cite{nekrassov1996four}) and studied in~\cite{Costello:2017dso,Costello:2018gyb,Costello:2019tri}. In both cases, the spectral parameter $\lambda$ of the integrable system becomes part of the geometry of the higher dimensional theory. Indeed, at least in the gauge theory case, these interpretations of four- and two-dimensional integrable systems are themselves closely related. It is well-known that many classical 2d integrable systems arise from symmetry reduction of the self-dual Yang-Mills equations at the classical level~\cite{Mason:1991rf}, and it was shown in~\cite{Bittleston:2020hfv} that performing this symmetry reduction at the level of the twistor space action reduces it to exactly the 4d Chern-Simons theories considered in~\cite{Costello:2019tri}. The  4d Chern-Simons theory involves a choice of meromorphic measure, which arises in part from the similar choice of meromorphic measure on twistor space and in part from the vector fields used in the reduction. In this regard, it would be interesting to investigate whether the gravitational theories considered in the present paper can descend to the 4d Beltrami-Chern-Simons theory constructed in~\cite{Costello:2020lpi}, as well as to understand how to perform the reduction at the quantum level.\\

Our paper is structured as follows. In section~\ref{sec:twistoractions} we review the basic construction of twistor space and the non-linear graviton, together with the twistor actions that correspond to self-dual gravity and self-dual Yang-Mills at the classical level. These holomorphic BF and holomorphic Poisson-BF theories are the main objects of our study. Next, we consider the anomalies in these theories. We construct the anomalies in two complementary ways: in section~\ref{sec:anomaly} via family index theorems associated to the determinant line bundle and adapted to our case of Hamiltonian deformations of the complex structure, and again in section~\ref{sec:anomaly-calc} via a direct evaluation of the failure of 1-loop box diagrams to be invariant under Hamiltonian diffeomorphisms. We give the form of the anomaly polynomials for the pure gauge, pure gravitational and mixed anomalies. Along the way, we give a more general discussion of how anomalies on twistor space relate to anomalies on space-time.  Then, in section~\ref{sec:anomaly-calc} we show that all three anomalies may be cancelled by coupling the twistor theories to a field $\eta\in\Omega^{2,1}_{\rm cl}$, corresponding on space-time to a scalar field with a 4$^{\rm th}$-order kinetic term. In the pure gauge theory case, this result was first found in~\cite{Costello:2021bah}; our work generalises this to the gravitational and mixed anomalies. We proceed in section~\ref{sec:space-time-theory} to determine the space-time theory to which our twistor theory corresponds, where $\eta$ appears with both gauge and gravitational axion-like couplings. Finally, in section~\ref{sec:amplitudes} we consider  amplitudes in this axion coupled theory. We explicitly compute all 4-pt amplitudes with external gravitons and/or gluons, finding that they vanish. We also argue that the  vanishing of these 4-pt amplitudes implies the vanishing of all higher $n$-pt amplitudes.\\

Let us summarize our conventions. We let $\alpha,\beta,\ldots\in\{0,1\}$ denote left Weyl spinor indices, whilst $\dot\alpha,\dot\beta,\ldots\in\{\dot0,\dot1\}$ denote right Weyl spinor indices. Spinor indices can be raised or lowered using the Levi-Civita symbols $\eps^{\dal\dot\beta}$ and $\eps_{\dal\db}$ with the conventions $u_\dal = \eps_{\dal\db}u^\db$, along with $\eps^{\dal\dg}\eps_{\dg\db} = \delta^\dal_{~\,\db}$. $\SL(2,\C)$-invariant spinor contractions will be denoted  $[uv] = u^\dal v_\dal$ and $\langle ab\rangle = a^\alpha b_\alpha$. Greek letters $\mu,\nu,\dots$ from the middle of the alphabet denote space-time indices. Letters in sans serif typeface $\sfa,\sfb,\dots$ refer to Lie algebra indices. 

%%%%%%%%%%%%%%%%%%%%%%%%%%%%%%%%%%
%%%%%%%%%%%%%%%%%%%%%%%%%%%%%%%%%%

\section{Twistor actions for self-dual theories} 
\label{sec:twistoractions}

The self-dual Yang-Mills equations and self-dual Einstein equations are rare examples of non-linear pdes in four dimensions that are nonetheless integrable. From the twistor perspective, this integrability arises because solutions of these equations are in one-to-one correspondence with certain holomorphic data on twistor space that may be chosen essentially freely, despite the non-linearity of the equations. 

In this section, we provide a brief review of some of the twistor constructions -- the Penrose-Ward correspondence and non-linear graviton constructions -- that we use in this work. Much of the classical work on twistor space uses the language of sheaf cohomology in the \v{C}ech framework. To aid the passage to the quantum theory, it will be more convenient to work in the Dolbeault picture. We review the construction of actions on twistor space whose classical field equations describe the holomorphic structures that arise in twistor formulations of self-dual theories.

%%%%%%%%%%%%%%%%%%%%%%%%%%%%%%%%%%

\subsection{The twistor space of Euclidean space-time}
\label{sec:basicPT}

The basic twistor space of flat, four-dimensional space-time is called $\PT$.\footnote{In this work space-time refers to a (possibly curved) 4-dimensional Euclidean manifold. It's an important result of \cite{Costello:2021bah} that a consistent quantum theory on twistor space retains the complex symmetries of the classical theory, and in particular all correlators extend analytically. In this way our results also apply in both Lorentzian and ultrahyperbolic signatures.} As a complex manifold, $\PT$ is the total space of the rank two bundle
\be\label{eq:PTdefn}
	\PT \ \cong  \ \cO(1)\oplus\cO(1)\to\CP^1\,.
\ee
Equivalently, it can be thought of as $\CP^3$ with a $\CP^1$ removed. Note that $\PT$ comes with a natural action of the (complexified) Lorentz group $\SL(2,\C)\times\SL(2,\C)$, where one copy of $\SL(2,\C)$ acts on the $\CP^1$ base of~\eqref{eq:PTdefn} and the other mixes the two $\cO(1)$ fibres. We will often describe twistor space using homogeneous co-ordinates $\lambda_\alpha$ (for $\alpha=0,1$) on $\CP^1$ and $\mu^{\dot\alpha}$ ($\dot\alpha=\dot0,\dot1$) on the fibres. Thus $(\mu^{\dot\alpha},\lambda_\alpha)\sim (r \mu^{\dot\alpha},r\lambda_\alpha)$ for any $r\in\C^*$, and the two components of $\lambda_\alpha$ cannot vanish simultaneously.   Under Lorentz transformations, $\lambda_\alpha$ and $\mu^{\dot\alpha}$ transform as left- and right-handed Weyl spinors respectively.

Flat, four-dimensional (complexified) space-time can be reconstructed from $\PT$ as the space of degree one holomorphic sections $\CP^1\hookrightarrow\PT$. Explicitly, any such section can be described by the \emph{incidence relations}
\be\label{eq:incidence}
	\mu^{\dot\alpha} = x^{\dot\alpha\alpha}\lambda_\alpha
\ee
for some $x^{\dal\alpha}\in\C^4$. We will sometimes call this section a \emph{twistor line}, and denote it by $L_x$. Note that the space $H^0(\CP^1,\cO(1)\oplus\cO(1))\cong\C^4$ of sections inherits the action of $\SL(2,\C)\times \SL(2,\C)$ from twistor space.

We can obtain a real space-time, rather than $\C^4$, by introducing a reality condition on our twistor lines. In this paper, we will mostly consider Riemannian signature, where the appropriate condition is an antiholomorphic involution $\sigma:\PT\to\PT$, acting as
\be\label{eq:involdefn}
	\sigma: (\mu^{\dot\alpha},\lambda_\alpha)\mapsto (\hat{\mu}^{\dot\alpha} ,\hat{\lambda}_\alpha)\,,
\ee
where $(\hat{\mu}^{\dot0},\hat{\mu}^{\dot1}) = (-\overline{\mu^{\dot1}},\overline{\mu^{\dot0}})$ and $(\hat{\lambda}_0,\hat{\lambda}_1)= (-\overline{\lambda_1},\overline{\lambda_0})$. This involution has no fixed points on $\PT$, acting in particular as the antipodal map on the $\CP^1$ base. There is a unique twistor line through any point $Z\in\PT$ and its image $\sigma(Z)$, and such twistor lines correspond via~\eqref{eq:incidence} to points $x$ in a real slice $\R^4\subset\C^4$.  We will refer to these as \emph{real} twistor lines, though they remain $\CP^1$s. Explicitly, in terms of the co-ordinates above we have
\be
\left.	
\begin{aligned}
	\mu^{\dot\alpha} &= x^{\dot\alpha\alpha}\lambda_\alpha\\
	\hat{\mu}^{\dot\alpha} & =x^{\dot\alpha\alpha}\hat{\lambda}_\alpha
\end{aligned}
\right\}
\ \Rightarrow\ 
x^{\dot\alpha\alpha} = \frac{\hat{\mu}^{\dot\alpha}\lambda^\alpha-\mu^{\dot\alpha}\hat{\lambda}^{\alpha}}{\langle\lambda\hat\lambda\rangle}\,,
\ee
and it is easy to see that this $x$ obeys $\hat{x}^{\dot\alpha\alpha}=x^{\dot\alpha\alpha}$ so that the corresponding twistor line is fixed by $\sigma$. Note that, since $\hat\lambda$ is antipodal to $\lambda$, $\langle\lambda\hat\lambda\rangle$ is nowhere vanishing. 

This construction shows that $\PT$ fibres over $\R^4$ with the twistor lines $L_x\cong S^2$ as fibres, so that as a smooth six-manifold, $\PT\cong S^2\times\R^4$. Furthermore, requiring compatibility with the antiholomorphic involution breaks $\SL(2,\C)\times\SL(2,\C)$ to $\SU(2)\times\SU(2)$. Thus our $\R^4$ carries a natural action of the (double cover of) $\SO(4)$ and can be identified as flat Euclidean space with metric $\d s^2 = \eps_{\dot\alpha\dot\beta}\,\eps_{\alpha\beta}\,\d x^{\dot\alpha\alpha}\odot\d x^{\dot\beta\beta}$.

In the other direction, if we treat $\R^4$ with its flat Euclidean metric as a hyperk{\"a}hler manifold, its twistor space is the total space of the $S^2$-bundle of complex structures compatible with the metric and standard orientation of $\R^4$. The incidence relations~\eqref{eq:incidence} then provide an identification $\R^4\simeq\C^2$ in the complex structure determined by $[\lambda_\alpha]\in\CP^1$, with $\mu^{\dot\alpha}$ being holomorphic co-ordinates on this $\C^2$.

%%%%%%%%%%%%%%%%%%%%%%%%%%%%%%%%%%

\subsection{The Penrose-Ward correspondence and holomorphic BF theory}
\label{sec:sdYM}

The first example of a 4d integrable system we will consider are the self-dual Yang-Mills equations $F^-=0$. The Penrose-Ward correspondence \cite{Ward:1977ta} states that there is a one-to-one correspondence between
\begin{itemize}
\item[-] solutions of the self-dual Yang-Mills (sdYM) equations with gauge group $\GL(n)$, considered up to gauge equivalence, and 
\item[-] holomorphic rank $n$ vector bundles $E\to\PT$ that are topologically trivial on each twistor line $L_x$.\footnote{Generically, a topologically trivial holomorphic vector bundle over $\CP^1$ will also be holomorphically trivial.}
\end{itemize}
Other gauge groups are obtained by imposing additional structure on the fibres of $E$. In particular, for any semi-simple gauge group the twistor bundle has vanishing first Chern class ${\rm c}_1(E)=0$, whereupon $E|_{L_x}$ is necessarily topologically trivial. In this way, the Penrose-Ward transform `trivialises' the sdYM equations, explaining their integrability.\\

Holomorphic bundles arise as solutions to the field equations of a holomorphic BF theory. On twistor space, holomorphic BF theory was first considered in~\cite{Witten:2003nn}, where it appeared as the open string sector of the B-model on $\mathcal{N}=4$ supersymmetric twistor space, truncated to $\mathcal{N}=0$. The present paper will mostly be interested in this non-supersymmetric truncation. Note that, unlike holomorphic Chern-Simons theory, holomorphic BF theory does not need the space to be Calabi-Yau. This is the holomorphic analogue of the fact that, unlike real Chern-Simons theory, BF theory on a real three-manifold does not need the manifold to be oriented.

The action of holomorphic BF theory is
\be\label{gaugeBFaction}
	S[b,a] = \frac{1}{2\pi \im}\int_\PT b\wedge F^{0,2}(a)\,,
\ee
where 
\be
F^{0,2}(a) = \bar{D}^2=\bar\p a +\frac{1}{2}[a,a]\,,
\ee
is the $(0,2)$-form part of the curvature of $\bar D=\bar\p+a$. We take the partial connection $(0,1)$-form  $$a\in\Omega^{0,1}(\PT,\fg)$$ 
for some choice of complex semi-simple Lie algebra $\fg$. The remaining field is
\be 
b\in \Omega^{3,1}(\PT,\fg^\vee)
\ee
where $\fg^\vee$ is the dual of $\fg$. The action involves the canonical pairing between $\fg$ and $\fg^\vee$. Equivalently, we can treat $b$ as living in $\Omega^{0,1}(\PT,K(\fg^\vee))$ where $K(\fg^\vee) =K_\PT\otimes\fg^\vee$ with $K_\PT\cong\cO(-4)$ the canonical bundle of $\PT$. 
 
The action~\eqref{gaugeBFaction} is invariant under the gauge transformations 
\begin{subequations}
\be\label{gaugeBFgaugetrans1}
\delta a = \bar\p c + [a,c] \,,\qquad\qquad
\delta b = [b,c]
\ee
where $c\in\Omega^0(\PT,\fg)$ is smooth gauge parameter. As usual, $F^{0,2}(a)$ transforms in the adjoint. Holomorphic BF theory is also invariant under further transformations\footnote{To avoid cluttering our notation we use the symbols $c,d$ for both the gauge parameters and their corresponding fermionic ghosts, which are introduced in section \ref{sec:anomaly}. Which is meant should be clear from the context.} 
\be\label{gaugeBFgaugetrans2}
\delta a = 0 \,,\qquad\qquad \delta b =  \bar\p d + [a,d]
\ee
\end{subequations}
parametrized by $d\in\Omega^{3,0}(\PT,\fg^\vee)$.  These transformations act non-trivially only on $b$ and leave the action invariant (up to possible boundary terms) by virtue of the Bianchi identity for $F$.

The equations of motion of~\eqref{gaugeBFaction} state that
\be\label{gaugeBFeom}
F^{0,2}(a)=0\,,\qquad\qquad \bar{D}b=0\,.
\ee
Thus, on-shell $\bar{D}^2=0$ and we have a holomorphic bundle on $\PT$, together with a field $b \in H^{0,1}(\PT,K(\fg^\vee))$. By the Penrose-Ward correspondence, this holomorphic bundle is equivalent to a gauge field $A$ on $\R^4$ obeying the sdYM equations. To interpret $b$ on $\R^4$, in the Abelian case we simply integrate $b$ over a twistor line:
\be \label{eq:B-transform}
	B^-(x) = \int_{L_x} b\qquad\qquad\text{(Abelian)}
\ee
The integral over $L_x\cong S^2$ uses up a $(1,1)$-form from $b$, leaving us with a 2-form $B^-$ that depends only on $x$. This 2-form is anti-self dual on account of the remaining form indices on $b$ being holomorphic on $\PT$. $B^-$ depends only on the cohomology class of $b$, and obeys $\d^*B^-=0$ on account of $b$ being $\bar\p$-closed. In the non-Abelian case, where $b$ transforms in the coadjoint, we must first fix a holomorphic frame for the twistor bundle $E$ over $L_x$ before applying equation \eqref{eq:B-transform}. Doing so modifies the equation obeyed by $B^-$ to $D^*B^-=0$, where $D$ is the covariant derivative with respect to the space-time gauge field. We can interpret $B^-$ as a field of helicity $-1$ propagating freely in the background of the self-dual YM field.

In fact, at the classical level it is easy to reduce the action~\eqref{gaugeBFaction} to the action
\be\label{gaugeBFR4}
\int_{\R^4} B^-\wedge F(A)
\ee
describing sdYM on $\R^4$. In brief, one uses the gauge transformations~\eqref{gaugeBFgaugetrans1}-\eqref{gaugeBFgaugetrans2} to ensure that the restrictions of both $a$ and $b$ to the $S^2$ fibres of $\PT\to\R^4$ are harmonic with respect to the Fubini-Study metric on these fibres. This sets $a|_{L_x}=0$ on any twistor line $L_x$, whilst 
\be
	b|_{L_x} = \d x^{\dal\alpha}\wedge\d x^{\ \beta}_{\dal} \,B_{\gamma\delta}(x)\, \lambda_\alpha\lambda_\beta  \hat{\lambda}^\gamma\hat{\lambda}^\delta\,\frac{\la\lambda \d\lambda\ra\la\hat\lambda \d \hat\lambda\ra}{\la \lambda\hat\lambda\ra^4}
\ee
in terms of the co-ordinates introduced above. Integrating out the other components of the $(0,1)$-form $b$, one learns that $a$ takes the form
\be
 a = A_{\alpha\dot\alpha}(x) \lambda^\alpha \frac{\d x^{\beta\dot\beta} \hat{\lambda}_\beta}{\la\lambda\hat\lambda\ra}\,.
\ee
At this point, all the dependence of $a$ and $b$ on the fibre directions is explicit, so the $S^2$ fibres can be integrated out, whereupon~\eqref{gaugeBFaction} reduces to~\eqref{gaugeBFR4}.
We refer the reader to {\it e.g.}~\cite{Mason:2005zm, Boels:2006ir} for further details.

%%%%%%%%%%%%%%%%%%%%%%%%%%%%%%%%%%

\subsection{The non-linear graviton and Poisson-BF theory}
\label{sec:sdG}

Twistor theory can be extended to self-dual curved space-times by means of Penrose's non-linear graviton construction~\cite{Penrose:1976js,Penrose:1976jq,Atiyah:1978wi}. While the Penrose-Ward correspondence moves between different solutions to the sdYM equations on a fixed space-time by deforming the complex structure of a holomorphic bundle $E\to\PT$, to describe self-dual gravity we must change the complex structure of twistor space itself. More precisely, the non-linear graviton construction states that there is a one-to-one correspondence between
\begin{itemize}
\item[-] four-dimensional manifolds $M$ with a conformal class $[g]$ of Riemannian metrics whose Weyl curvature is self-dual, and
\item[-] complex 3-folds $\cPT$ that possess at least one rational curve $\cL_x$ with normal bundle $N=\cO(1)\oplus\cO(1)$, together with a free antiholomorphic involution $\sigma:\cPT\to\cPT$ acting as the antipodal map on $\cL_x$.
\end{itemize}
Ignoring the involution, once there is a single such rational curve, Kodaira theory ensures there is a four-parameter family of them, and their moduli space $M_\C$ is the complexification of the curved space-time $M$.  The space $H^0(\cL_x,N)\cong\C^4$ of degree one sections of the normal bundle now becomes the holomorphic tangent space $T_xM_\C$.  As in $\PT$, the real Riemannian space is the moduli space of twistor lines that are fixed by $\sigma$.
In perturbation theory, we can  construct a curved $\mathcal{PT}$ by deforming the flat twistor space $\PT$. As usual, deformations of the almost complex structure can be encoded in the deformation 
\be\label{Dolbeaultdef}
	\dbar\ \mapsto\ \nbar = \dbar + \cL_V\,,
\ee
of the Dolbeault operator, where $\cL_V$ is the Lie derivative along a Beltrami differential $V\in\Omega^{0,1}(\PT,T^{1,0}_\PT)$. This almost complex structure is integrable iff the Nijenhuis tensor $N^{0,2} = [\nbar,\nbar] = \bar\p V + \frac{1}{2}[V,V]$ vanishes. The correspondence is conformally invariant and so yields self-dual solutions of \emph{conformal} gravity; in particular, the condition $N^{0,2}=0$ becomes the condition that the space-time Weyl tensor is self-dual.\\

We obtain a unique Ricci-flat metric $g$ in the conformal class if the curved twistor space $\mathcal{PT}$ admits a holomorphic fibration $\mathcal{PT}\to\CP^1$ with an $\cO(2)$-valued holomorphic symplectic form on the fibres. (It follows that such twistor spaces again have $K_{\mathcal{PT}}\cong\cO(-4)$.) In the case of $\PT$ for flat space, the $\cO(2)$-valued symplectic form on the fibres of $\PT\to\CP^1$ is just $\d\mu^\dal\wedge\d\mu_\dal/2$. The deformations~\eqref{Dolbeaultdef} preserve this form, and so generate a self-dual solution of Einstein's equations, if $V$ is Hamiltonian in the sense
\be\label{eqn:HamVdef}
V = \{h,\,\cdot\,\} = \eps^{\db\dal}(\cL_{\p_\dal}h)\,\p_\db
\ee
for some $h\in\Omega^{0,1}(\PT,\cO(2))$ playing the role of a Hamiltonian. Here, $\{\ ,\ \}$ is the Poisson bracket of weight $-2$ on the fibres, defined by
\be
\{f,g\} = \eps^{\db\dal}\,\cL_{\p_{\dal}}f\wedge\cL_{\p_{\dot\beta}}g
\ee
for any $(p,q)$-forms $f,g$ on $\PT$ and $\p_{\dot\alpha}=\p/\p\mu^{\dot\alpha}$. Often when $(1,0)$ vector fields act on forms we suppress the Lie derivatives and just write the vector field, {\it e.g.}, $V = (\p^\dal h)\,\p_\dal$. Since the Poisson structure has weight $-2$, the weight $+2$ of $h$ ensures that the associated vector $V$ is weightless. Such Hamiltonian deformations alter the complex structure purely in directions normal to the $\CP^1$ base of $\mathcal{PT}$. In particular, for  Hamiltonian deformations, the Nijenhuis tensor takes the form
$$
N^{0,2} = (\bar\p + \{h,\ \})^2 = \{T^{0,2}(h),\ \}
$$
where $T^{0,2}(h) = \bar\p h + \frac{1}{2}\{h,h\}\in\Omega^{0,2}(\PT,\cO(2))$.\\

A twistor action that describes self-dual Einstein gravity was first identified in~\cite{Mason:2007ct}. This action can be viewed as the $\mathcal{N}=0$ truncation of the constant map sector of the $\mathcal{N}=8$ twistor string constructed in~\cite{Skinner:2013xp}. This action involves a choice of background complex twistor space and so, unlike the holomorphic BF theory of section~\ref{sec:sdYM}, it is not fully covariant. Taking the background twistor space to be the basic $\PT$, the action is
\be\label{PoissonBFaction}
S[g,h] = \frac{1}{2\pi \im}\int_\PT g \wedge T^{0,2}(h) = \frac{1}{2\pi \im}\int_\PT g \wedge \left(\bar\p h + \frac{1}{2}\{h,h\}\right)
\ee
where $h \in \Omega^{0,1}(\PT,\cO(2))$ is our Hamiltonian as above. The remaining field is
\be
	g\in\Omega^{3,1}(\PT,\cO(-2))
\ee
and we can equivalently view $g\in\Omega^{0,1}(\PT,K(-2))$ with $K(-2)= K_\PT\otimes\cO(-2)\cong\cO(-6)$. We will refer to this action as \emph{holomorphic Poisson-BF theory}, or often just Poisson-BF theory. Holomorphic Poisson-BF theories on $\C^{2n}$ with a non-degenerate Poisson bracket have been studied in~\cite{Williams:2018ows}. We will see that there are notable differences in the twistor case.

Similar to gauge theory, \eqref{PoissonBFaction} is invariant under transformations acting as
\be\label{PoissonBFgaugetrans}
\begin{aligned}
\delta h &=  \bar\p \chi + \{h,\chi\}\\
\delta g & = \{g,\chi\} + \bar\p \phi + \{h,\phi\}
\end{aligned}
\ee
where the parameters $\chi\in\Omega^{0}(\PT,\cO(2))$ and $\phi\in\Omega^{0}(\PT,K(-2))$. The transformations parametrized by $\chi$ generate Hamiltonian diffeomorphisms of $\PT$, for which the corresponding deformations $\bar\p\mapsto\nbar = \bar\p + \{h,\ \}$ of the $\bar\p$-operator are trivial. The transformations generated by $\phi$ are analogous to the $d$-transformations~\eqref{gaugeBFgaugetrans2} in the gauge theory case.

The equations of motion following from~\eqref{PoissonBFaction} state that
\be\label{PoissonBFeom}
	T^{0,2}(h) =0\,, \qquad\qquad \nbar g=0\,,
\ee
where $\nbar = \bar\p + \{h,\ \}$ is the covariant derivative in the deformed almost complex structure. The first equation in~\eqref{PoissonBFeom} ensures that the deformed almost complex structure is integrable, placing us in the realm of the non-linear graviton construction. In section~\ref{sec:space-time-theory} (see also~\cite{Sharma:2021pkl}), we will see that the twistor action~\eqref{PoissonBFaction} is classically equivalent to the action~\cite{Capovilla:1991qb,Krasnov:2021cva}
\be\label{sdGaction}
 	S[\Psi,\Sigma,\Gamma] = \int_{\R^4} \Sigma^{\alpha\beta} \wedge \d\Gamma_{\alpha\beta} + \frac{1}{2}\,\Psi_{\alpha\beta\gamma\delta}\,\Sigma^{\alpha\beta}\wedge\Sigma^{\gamma\delta}
\ee
which describes self-dual (sd) Einstein gravity on $\R^4$.\footnote{In this work we will always assume vanishing cosmological constant.} For example, integrating out $\Psi$ imposes the condition that $\Sigma^{\alpha\beta} = e^{\dal\al}\wedge e_{\dal}^{\ \ \beta}$ for some vierbein 1-form $e^{\dal\al}$. Then~\eqref{sdGaction} reduces to the usual action
\be
\label{Plebanski_action}
S[e,\Gamma] = \int_{\R^4}e^{\al\dal}\wedge e^\beta{}_{\dal}\wedge\d\Gamma_{\al\beta}
\ee
for sd gravity, often obtained as the $\kappa\to0$ limit of the Plebanski action for full gravity \cite{Ashtekar:1987qx,Smolin:1992wj}.\\

It will be useful to note that any deformation $\bar\p \mapsto \bar \p + V$ of the almost complex structure of twistor space induces a deformation of the holomorphic tangent bundle $T_\PT$. Specifically, let $U$ be a smooth vector field pointing in the holomorphic directions and expand this as $U = U^a\p_a$ in terms of a holomorphic frame $\{\p_a\}$ for $T_\PT$. Then
\be
\begin{aligned}	
	\nbar U &= \bar\p U + \cL_V U = (\bar\p U^a)\p_a + \cL_V(U^a\p_a) \\
	&= \nbar(U^a)\p_a + U^a[V,\p_a] = (\nbar U^a - (\p_b V^a) U^b)\p_a\,,
\end{aligned}
\ee
which shows that $A^a_{\ b} = - \p_bV^a$ is the induced connection $(0,1)$-form. In the Einstein context where $V = \{h,\ \}$ is Hamiltonian, this reduces to an upper triangular partial connection $A^{\dot\alpha}_{~\,b} = -\p_b(\p^\dal h)$. In fact, the only vector fields we will meet point only along the fibres of $\mathcal{PT}\to\CP^1$, whereupon our connection restricts to an $\fsl_2(\C)$ connection
\be\label{eq:s-Nconn-def}
s^{\dot\alpha}_{~\,\dot\beta} = - \p^{\dot\alpha}\p_{\dot\beta} h = - \epsilon^{\dot\alpha\dot\gamma}\p_{\dot\gamma}\p_{\dot\beta}h
\ee
on the normal bundle to any twistor line. (More precisely $s$ is a connection on the vertical tangent bundle $\cN\subset T_\PT$ with respect to the holomorphic fibration $\PT\to\CP^1$. The pullback of $\cN$ to a twistor line can be identified with the normal bundle $N$.) Indeed, under the transformations~\eqref{PoissonBFgaugetrans} of $h$ with parameter $\chi$, one finds that $s$ transforms as  
\be 
\label{eq:gaugetransNormal}
\delta s =\nbar\psi + [s,\psi] \,,
\ee
where $\psi^{\dot\alpha}_{\ \dot\beta} = - \p^{\dot\alpha}\p_{\dot\beta}\chi$, $[\ ,\ ]$ denotes the commutator in $\fsl_2(\C)$ and $\nbar$ is Dolbeault operator on the curved twistor space $\cPT$. The $(0,2)$-curvature of this connection is
\be\label{eq:Normalcurvature}
	 F^{0,2}(s)^{\dot\alpha}_{~\,\dot\beta}
	 = \Big(\nbar s +\frac{1}{2}[s,s]\Big)^{\dot\alpha}_{\ \dot\beta} 
	 = - \p^{\dot\alpha}\p_{\dot\beta}T^{0,2} 
\ee
in terms of $T^{0,2}=\bar\p h+\frac{1}{2}\{h,h\}$. Hence the equations of motion of Poisson-BF theory also imply that this normal bundle is holomorphic.

%%%%%%%%%%%%%%%%%%%%%%%%%%%%%%%%%%

\subsection{Self-dual Einstein-Yang-Mills theory}
\label{sec:sdEYM}

Self-dual gauge theory can be coupled to self-dual gravity and, at the classical level, the corresponding coupled field equations remain integrable \cite{Atiyah:1978wi}. This coupling is straightforward to achieve on twistor space, where we minimally couple the two theories as
\be\label{EYMtwistoraction}
\begin{aligned}
S[b,a;g,h] &= \frac{1}{2\pi i}\int_\PT g \wedge T^{0,2}(h) + b\wedge\left(\nbar a + \frac{1}{2}[a,a]\right)\\
& = \frac{1}{2\pi i}\int_\PT g\wedge T^{0,2}(h) + b\wedge \cF^{0,2}(a)\,,
\end{aligned}
\ee
where in the second line we have introduced
\be
\cF^{0,2}(a) = \nbar a + \frac{1}{2}[a,a] = \bar\p a + \{h,a\} + \frac{1}{2}[a,a]  \,.
\ee
We shall sometimes refer to this as Poisson-gauge-BF theory. Accordingly, the gauge transformations are modified by replacing $\bar\p \mapsto\nbar$, so that
\begin{subequations}\label{twistorEYMgaugetrans}
\bea\label{twistorEYMgaugetrans1}
\delta a  &= \nbar c + [a,c] + \{a,\chi\}\\
\delta b &= [b,c] + \nbar d + [a,d] + \{b,\chi\}
\eea
for the fields in the gauge theory, and
\bea\label{twistorEYMgaugetrans2} 
\delta h &= \nbar\chi\\
\delta g &= \{g,\chi\} + \nbar\phi + \{b,c\}+\{a,d\}
\eea
\end{subequations}
for the gravitational fields. Note that on a twistor space with only an almost complex structure, by itself, holomorphic BF theory is not invariant under the transformations~\eqref{twistorEYMgaugetrans1} because the deformed Dolbeault operator $\nbar$ fails to be nilpotent off-shell.  This is compensated by the terms $\{b,c\}$ and $\{a,d\}$ in the transformation of $g$ in~\eqref{twistorEYMgaugetrans2}. (These terms are constructed using both the weight $-2$ Poisson bracket and the canonical pairing between $\fg$ and $\fg^\vee$.) The commutation relations among successive gauge transformations of various types is somewhat complicated, and is best handled using the BV formalism introduced below.

At the classical level, the theory~\eqref{EYMtwistoraction} is equivalent to the action
\be\label{EYMaction}
 S[B,A;\Sigma,\Gamma,\Psi] = \int_{\R^4} \Sigma^{\alpha\beta}\wedge\d\Gamma_{\alpha\beta} +\frac{1}{2}\Psi_{\alpha\beta\gamma\delta}\Sigma^{\alpha\beta}\wedge\Sigma^{\gamma\delta}+ B^-\wedge F
\ee
for self-dual Einstein-Yang-Mills (sdEYM), where now $B^-= B^-_{\al\beta}\Sigma^{\al\beta}$ couples to $\Sigma$.

%%%%%%%%%%%%%%%%%%%%%%%%%%%%%%%%%%
%%%%%%%%%%%%%%%%%%%%%%%%%%%%%%%%%%

\section{Anomalies on twistor space} 
\label{sec:anomaly}

The twistor actions above are inherently chiral and so are susceptible to possible anomalies. In this section, we will show that the twistor space theory defined by the action~\eqref{EYMtwistoraction} indeed suffers from gauge, gravitational and mixed local\footnote{In this paper we shall consider only local anomalies.} anomalies, any one of which is enough to render the theory meaningless at the quantum level. We will establish the presence of these anomalies both by constructing the anomaly polynomial as an obstruction to gauge invariance of the partition function, and also by explicitly computing the gauge variation of 1-loop box diagrams. In section~\ref{sec:GS-mechanism}, we will see that, for certain choices of gauge group, all three anomalies can be cancelled by the introduction of a single field on twistor space.

%%%%%%%%%%%%%%%%%%%%%%%%%%%%%%%%%%

\subsection{The partition function and holomorphic Ray-Singer torsion} 
\label{subsec:PBF-anomaly}

In this section we compute the partition function of the theory~\eqref{EYMtwistoraction}. Formally, this can be viewed as $1/\det{\bar{\mathcal{D}}}$ where $\bar{\mathcal{D}}$ is a background complex structure on our bundles over twistor space. We will see that the absolute value of the partition function can be given precise meaning in terms of the Ray-Singer analytic torsion for complex manifolds~\cite{Ray:1973sb}. However, the phase of the partition function is ill-defined, indicating the presence of anomalies.\\

Let us first consider the gauge theory. In writing our twistor space action~\eqref{EYMtwistoraction}, we implicitly chose a background complex structure $\bar\p$. More generically, we can write the background complex structure on the gauge bundle as 
\be
\bar{\mathcal{D}} = \bar\p + \{h^{(0)},\ \}+[a^{(0)},\ ]
\ee
in terms of background fields $h^{(0)}$ and $a^{(0)}$. It is important that this background complex structure is integrable, $\mathcal{\bar{D}}^2=0$, so that $h^{(0)}$ and $a^{(0)}$ are both on-shell. In particular, expanding around $\bar{\mathcal{D}}$ gives the action
\be
S = \frac{1}{2\pi\im}\int_{\PT}b\wedge\bar{\mathcal{D}}a\,
\ee
correct to quadratic order in the fluctuations $a, b$.

To quantize the theory on this background, as in any gauge theory we must first fix the redundancies~\eqref{twistorEYMgaugetrans}. To do this, first pick a Hermitian metric on the twistor space, such as the one induced by the Fubini-Study metric on $\CP^3$. This gives us a Hermitian inner product $(\ ,\ )$ on forms defined by
\be\label{eqn:innerprod}
	(\xi , \omega) = \int_\PT \xi \wedge \star\,\omega
\ee
with the help of the antilinear\footnote{Here we define $\star :\Omega^{p,q}\to\Omega^{3-p,3-q}$ by $\star\omega = \overline{*\omega}$ where $*:\Omega^{p,q}\to\Omega^{3-q,3-p}$ is the usual Hodge star associated to the Hermitian metric.} Hodge star. 

We now use the standard BRST formalism to impose the gauge conditions 
\be\label{eqn:gauge-fix}
	\bar{\mathcal{D}}^\dagger a = 0\,,\qquad\qquad \bar{\mathcal{D}}^\dagger b=0\,,
\ee
where $\bar{\mathcal{D}}^\dagger = - \star\bar{\mathcal{D}}\,\star\, :\, \Omega^{p,q}(\fg)\to\Omega^{p,q-1}(\fg)$ is the adjoint of $\bar{\mathcal{D}}$ with respect to the inner product. That is, we promote the gauge parameters to fermionic ghosts $c\in\Pi\Omega^{0,0}(\PT,\fg)$ and $d\in\Pi\Omega^{3,0}(\PT,\fg^\vee)$ respectively, with corresponding antighosts $\bar{c}$ and $\bar{d}$. The BRST variations of the antighosts are bosonic Nakanishi-Lautrup fields $n\in\Omega^{3,3}(\fg^\vee)$ and $m\in\Omega^{0,3}(\fg)$. The gauge-fixed action is then
\be
S_{\rm gf} = \frac{1}{2\pi \im} \int_\PT b\wedge\bar{\mathcal{D}}a + n\, \bar{\mathcal{D}}^\dagger a+ b\,\bar{\mathcal{D}}^\dagger m + \bar{c}\,\bar{\mathcal{D}}^\dagger\bar{\mathcal{D}}c + \bar{d}\,\bar{\mathcal{D}}^\dagger\bar{\mathcal{D}}d
\ee
to quadratic order in fluctuations around the background.

The two sets of ghosts can be integrated out straightforwardly, giving a factor of 
\be
	{\det}'(\Delta_\fg^{0,0})\,{\det}'(\Delta_\fg^{3,0})\,,
\ee
where $\Delta_{\fg}^{p,q}=\bar{\mathcal{D}}^\dagger\bar{\mathcal{D}} + \bar{\mathcal{D}} \bar{\mathcal{D}}^\dagger$ is the background Laplacian acting on $\fg$-valued $(p,q)$-forms and $\det'$ is the determinant with zero-modes removed. To perform the path integral over the remaining fields, we first define
\bea
	\cA &= a+m \in \Omega^{0,1}(\PT,\fg)\oplus\Omega^{0,3}(\PT,\fg)\\
	\cB &= b+ n\in \Omega^{3,1}(\PT,\fg^\vee)\oplus\Omega^{3,3}(\PT,\fg^\vee)
\eea
and note that the remaining terms in the action can be succinctly written using the inner product~\eqref{eqn:innerprod} as
\be
	\frac{1}{2\pi \im}(\cB,L\cA) = \frac{1}{2\pi \im}\int_\PT b\wedge\bar{\mathcal{D}}a + n\,\bar{\mathcal{D}}^\dagger a + b\,\bar{\mathcal{D}}^\dagger m
\ee
where the operator 
\be
L= \bar{\mathcal{D}}^\dagger+\bar{\mathcal{D}}\,:\,\Omega^{0,1}(\PT,\fg)\oplus\Omega^{0,3}(\PT,\fg)\to \Omega^{0,0}(\PT,\fg)\oplus\Omega^{0,2}(\PT,\fg)
\ee
and $(\ ,\ )$ denotes the pairing of $\Omega^{0,0}(\PT,\fg)\oplus\Omega^{0,2}(\PT,\fg)$ with $\Omega^{3,3}(\PT,\fg^\vee)\oplus\Omega^{3,1}(\PT,\fg^\vee)$.  Furthermore, with the help of an inner product on $\fg$ that identifies $\fg \cong \fg^\vee$, we can define the adjoint $L^\dagger$. We then have
\be
L^\dagger L = \Delta_\fg^{0,0} +\Delta_\fg^{0,1}+\Delta_\fg^{0,2}+\Delta_\fg^{0,3}
\ee
where $\Delta_\fg^{p,q}$ is the Laplacian acting on a $\fg$-valued $(p,q)$-form.  The absolute value of the path integral over the remaining fields is then
\be	
 \left({\det}'(L^\dagger L)\right)^{-1/4} = \left(\prod_{q=0}^3{\det}'(\Delta_\fg^{0,q})\right)^{-1/4}\,,
\ee
where the power $1/4$ appears because we are viewing the operator $L^\dagger L$ as acting on the combined field $(\cA+\star \cB) \in\bigoplus_{q=0}^3\Omega^{0,q}(\PT,\fg)$. \\

Thus, including the ghosts, in the presence of a holomorphic background (integrable) complex structure on the gauge bundle over twistor space the absolute value of the 1-loop partition function of holomorphic BF theory is
\be
	|\cZ_{\rm BF}| = \frac{(\det'{\Delta_\fg^{0,0}})(\det'{\Delta_\fg^{3,0}})}{\left(\prod_{q=0}^3(\det'{\Delta_\fg^{0,q}})\right)^{1/4}}
=\left[\frac{(\det'{\Delta_\fg^{0,0}})^3(\det'{\Delta_\fg^{3,0}})^4}{(\det'{\Delta_\fg^{0,1}})(\det'{\Delta_\fg^{0,2}})(\det'{\Delta_\fg^{0,3}})}\right]^{\frac{1}{4}}\,.
\ee
Hodge duality shows that $\det'{\Delta_\fg^{3,q}}=\det'{\Delta_\fg^{0,3-q}}$, so we can write this as
\be \label{eqn:BFpartition}
|\cZ_{\rm BF}|= \left(\tau_{\fg,0}(\bar{\mathcal{D}})\,\tau_{\fg,3}(\bar{\mathcal{D}})\right)^{\frac{1}{2}}
\ee
in terms of the combinations
\be\label{eqn:RaySingerdef}
	\tau_{\fg,p}(\bar{\mathcal{D}})= \left[\frac{(\det'{\Delta_\fg^{p,1}})(\det'{\Delta_\fg^{p,3}})^3}{(\det'{
	\Delta_\fg^{p,2}})^2}\right]^{\frac{1}{2}}\,.
\ee
$\tau_{\fg,p}(\bar{\mathcal{D}})$ is the complex manifold version of the Ray-Singer analytic torsion~\cite{Ray:1973sb} for $\bar{\mathcal{D}}$ acting on a complex of $\fg$-valued $(p,0)$-forms. \cite{Ray:1973sb} proved that this torsion is independent of the choice of Hermitian metric made in fixing the gauge~\eqref{eqn:gauge-fix}. Notice that~\eqref{eqn:BFpartition} isn't sensitive to whether we view the $\bar{\mathcal{D}}$-operator in the original action as acting on $a$ or $b$. We also remark that, for holomorphic BF theory on a Calabi-Yau 3-fold, we have $\tau_{\fg,3}(\bar{\mathcal{D}})=\tau_{\fg,0}(\bar{\mathcal{D}})$ so that the partition function $|\cZ_{\rm BF}^{\rm CY}| = \tau_{\fg,0}(\bar{\mathcal{D}})$, which is the square of the partition function of holomorphic Chern-Simons theory~\cite{Bershadsky:1993cx}.\\

In just the same way, the 1-loop partition function coming from integrating out fluctuations in the fields $h$ and $g$ of Poisson-BF theory has absolute value
\be\label{eqn:PoissonBFpartition}
	|\cZ_{\rm PBF}|  = \left[\frac{(\det'{\Delta_{2}^{0,0}})^3(\det'{\Delta_{-2}^{3,0}})^4}{(\det'{\Delta_{2}^{0,1}})(\det'{\Delta_{2}^{0,2}})(\det'{\Delta_{2}^{0,3}})}\right]^{\frac{1}{4}}=\left(\tau_{2,0}(\bar{\mathcal{D}})\,\tau_{-2,3}(\bar{\mathcal{D}})\right)^{\frac{1}{2}}
\ee
where $\Delta^{0,q}_{n}$ is the Laplacian acting on $\Omega^{0,p}(\PT,\cO(n))$ and $\tau_{n,p}(\bar{\mathcal{D}})$ the associated Ray-Singer holomorphic torsion.\footnote{If the vacuum is isolated then we obtain an invariant, but more generally this defines a top form on the moduli space of on-shell backgrounds. Note that the only compact hyperk{\"a}hler 4-manifolds are K3 surfaces and 4-tori, which do have moduli.}\\

Since the theory~\eqref{EYMtwistoraction} is 1-loop exact, the full partition function of self-dual Einstein-Yang-Mills theory in twistor space is just the product $\cZ=\cZ_{\rm BF}\cZ_{\rm PBF}$. It should be interpreted as a section of a holomorphic determinant line bundle Det over the space of complex structures (both on the bundle and on the twistor space itself). Since the absolute value $|\cZ|$ is gauge invariant, this determinant line bundle has a natural Hermitian metric and hence a preferred Chern connection, first constructed in classic work by Bismut-Gillet-Soul{\'e}~\cite{bismut1988analytic1,bismut1988analytic2,bismut1988analytic3} following work of Quillen~\cite{quillen1985determinants} in the case of bundles on a Riemann surface. Any consistent definition of the phase of $\cZ$ is obstructed if this connection has non-zero curvature $\mathscr{F}_{\rm Det}$, for then the phase changes when we parallel transport $\cZ$ around a closed loop in the space of complex structures. This indicates a failure of gauge invariance of the quantum theory.

The curvature of Det was computed in~\cite{bismut1988analytic1,bismut1988analytic2,bismut1988analytic3}, as the holomorphic version of the family index theorem. There, the authors considered varying over all complex structures on a Dolbeault complex for an arbitrary holomorphic bundle $\cV\to X$ over a compact complex manifold $X$. Adapting their results to our case gives
\begin{multline}
\label{eqn:Anomaly-family-index}
	\mathscr{F}_{\rm Det}= \frac{1}{2}\left[\int_\cPT {\rm Td}(T_\cPT)\wedge{\rm Ch}(\fg)\wedge\left(1+{\rm Ch}(\cO(-4))\right) \right]_{(1,1)}\\
	+\frac{1}{2}\left[\int_\cPT {\rm Td}(T_\cPT)\wedge\left({\rm Ch}(\cO(2))+{\rm Ch}(\cO(-6))\right)\right]_{(1,1)}\,,
\end{multline}
where the terms in the first line come from the holomorphic BF theory, while those in the second come from Poisson-BF theory.\footnote{More formally, the anomaly polynomial is defined on the total space of the fibre bundle $\mathscr{X}\to\mathscr{M}$ where $\mathscr{M}$ is the moduli space of on-shell backgrounds, and the fibre over a point is the corresponding background. The curvature of the determinant line bundle (over $\mathscr{M}$) is obtained by integrating the anomaly polynomial over the fibres. As is common in the physics literature, we will abuse notation by using a particular on-shell background as the argument in our characteristic classes instead of $\mathscr{X}$.} The two terms in each expression correspond to the two different Ray-Singer torsions appearing in each of $\cZ_{\rm BF}$ and $\cZ_{\rm PBF}$. In~\eqref{eqn:Anomaly-family-index}, ${\rm Td}(T_\cPT) = \sum_n {\rm td}_n(T_\cPT)$ denotes the total Todd class of the curvature of the background connection on the holomorphic tangent bundle, while ${\rm Ch}(\fg) = \sum_n {\rm ch}_n(\fg)$ likewise denotes the total Chern character of the curvature of the background gauge bundle. The subscript $(1,1)$ indicates that we are to extract the $(1,1)$-form part of the resulting expression, viewed as the curvature of Det over the space of complex structures.

To simplify this expression, first notice that since the adjoint representation is self-conjugate, only the terms in ${\rm Ch}(\fg)$ that involve even powers of the curvature can be non-zero. Hence we have ${\rm Ch}(\fg) = \mathrm{rk}(\fg) + {\rm ch}_2(\fg) + {\rm ch}_4(\fg)$. Second, recall that the Todd class arises essentially because we are considering the determinant line bundle associated to a Dolbeault complex. In our context, since we are only varying among background \emph{Hamiltonian} complex structures on twistor space, all our curved twistor spaces admit a holomorphic fibration $\cPT\to\CP^1$ with fibre $\C^2$. The line bundles $\cO(n)\to\cPT$ are defined as the pullbacks of $\cO(n)\to\CP^1$ by this map. Therefore $\chern_1(\cO(n)) = - (n/2)x$ for $x$ the pullback to $\cPT$ of the symplectic form associated to the Fubini-Study metric on $\CP^1$. Similarly $\chern_1(T_\cPT) = - \chern_1(K_\cPT) = - \chern_1(\cO(-4)) = - 2x$.\footnote{As noted above, these bundles are really over $\mathscr{X}$, and we're using the fibration $\mathscr{X}\to\CP^1$ defined fibrewise by $\cPT\to\CP^1$. In extending these maps to $\mathscr{X}$ we're assuming that they vary holomorphically as we deform the Hamiltonian complex structure on $\cPT$.}

Since $\CP^1$ has complex dimension one, $x^2=0$. In particular, terms in~\eqref{eqn:Anomaly-family-index} that are of order $c_1(T_\cPT)^2$ must vanish. Thus, for our Hamiltonian deformations of $\PT$, we find the simplification
\be
	{\rm Td}(T_\cPT) = 1 + \frac{1}{2}{\rm c}_1(T_\cPT) - \frac{1}{12}{\rm ch}_2(T_\cPT) - \frac{1}{4!}{\rm c}_1(T_\cPT)\wedge{\rm ch}_2(T_\cPT) + {\rm td}_4(T_\cPT)
\ee
in terms of the Chern characters.

For the same reason, ${\rm Ch}(\cO(n))$ simplifies to $1 - (n/2)x = 1 + (n/4)\chern_1(T_\cPT)$. The gauge theory contribution in equation \eqref{eqn:Anomaly-family-index} is therefore
\be
\frac{1}{2}{\rm Td}(T_\cPT)\wedge\left(1+{\rm Ch}(\cO(-4))\right)= 
1 - \frac{1}{12}\chern_2(T_\cPT) + \todd_4(T_\cPT)\,,
\ee
and the Poisson-BF contribution is
\be
\frac{1}{2}{\rm Td}(T_\cPT)\wedge\left({\rm Ch}(\cO(2))+{\rm Ch}(\cO(-6))\right) = 1 - \frac{1}{12}\chern_2(T_\cPT) + \todd_4(T_\cPT)\,.
\ee
That these two factors are the same in both the holomorphic and Poisson-BF theories is just the usual fact that pure gravitational anomalies are proportional to a count of the chiral degrees of freedom in the theory weighted by Grassmann parity.

The 4\textsuperscript{th} Todd class can be expressed in terms of Chern classes as
\be \label{eq:todd4} \todd_4(T_\cPT) = \frac{\chern_1(T_\cPT)\chern_3(T_\cPT) + 3 \chern_2(T_\cPT)^2}{720}\,, \ee
where we've discarded terms of order $\chern_1(T_\cPT)^2$ or higher for the reasons given above, and $\chern_4(T_\cPT)=0$ since the tangent bundle has rank three. Now, the holomorphic map $\cPT\to\CP^1$ induces a short exact sequence of vector bundles
\be 0\to \cN\hookrightarrow T_\cPT\to\cPT \times_{\CP^1}T_{\CP^1}\to0\,, \ee
where here $\cN$ is the vertical tangent bundle and $\cPT\times_{\CP^1}T_{\CP^1}$ is the pullback of the tangent bundle of $\CP^1$ to $\cPT$. (Recall that the pullback of $\cN$ to a twistor line can be identified with the normal bundle.) By additivity we therefore have
\be \chern_3(\cPT) = \chern_3(\cN) + \chern_2(\cN)\wedge\chern_1(\cPT\times_{\CP^1}T_{\CP^1}) = - x\wedge\chern_2(\cN)\,, \ee
where in the second equality we've used the fact that $\chern_3(\cN) = 0$ since $\cN$ has rank two.
%Finally, since the standard complex structure of $\PT$ has vanishing $\chern_2$, $\chern_3$ and $\chern_4$, the $\todd_4$ term can only come from our deformation $h^{(0)}$. This appears in the Todd class only via the partial connection $s^{\dot\alpha}_{\ b} = - \p^{\dot\alpha}\p_{b}h^{(0)}$ defined above. This connection is upper triangular, and only its $\fsl_2$ part $s^{\dal}_{\ \db}$ contributes to the curvature polynomials. 
Hence the term involving $\chern_3(T_\cPT)$ drops out of equation \eqref{eq:todd4} and the 4\textsuperscript{th} Todd class simplifies to
\be
	\todd_4(T_\cPT) = \frac{\chern_2(T_\cPT)^2}{240} = \frac{1}{20}\chern_4(T_\cPT)\,.
\ee
We emphasise again that these simplifications arise because our twistor space $\mathcal{PT}$ fibres over $\CP^1$, obtained via Hamiltonian deformations of $\PT$. \\

Combining these results shows that the curvature of the determinant line bundle over the gauge redundant configuration space of Hamiltonian complex structures on $\PT$ and complex structures on the gauge bundle is given by
\be\label{eqn:Anomaly-poly}
	\mathscr{F}_{\rm Det}= \left[\int_\cPT {\rm ch}_4(\fg) + \frac{1+{\rm dim}(\fg)}{20}\int_\cPT{\rm ch}_4(T_\cPT) - \frac{1}{12}\int_\cPT{\rm ch}_2(\fg)\wedge{\rm ch}_2(T_\cPT)\right]_{(1,1)}\,.
\ee
These three terms are, respectively, the pure gauge, pure gravitational, and mixed gauge-gravitational anomalies of Poisson-BF theory coupled to holomorphic BF theory on twistor space. These anomalies indicate that the theory is not well-defined at the quantum level. We will explore how these anomalies may be cancelled in section~\ref{sec:GS-mechanism}.

%%%%%%%%%%%%%%%%%%%%%%%%%%%%%%%%%%

\subsection{Anomalies from 1-loop box diagrams} 
\label{subsec:anom-box}

\begin{figure}[!ht]
\centering
  \includegraphics[scale=0.3]{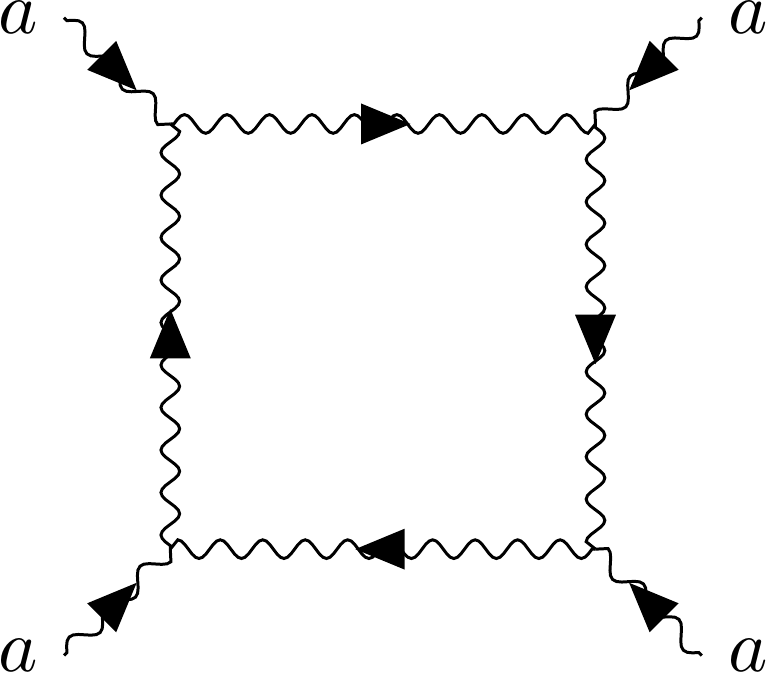}
\caption{\emph{The pure gauge anomaly in holomorphic BF theory can be viewed as the failure of this box diagram to be gauge invariant.}} \label{fig:HBF-anomaly}
\end{figure}

In six real dimensions, anomalies can also be seen as the failure of 1-loop box diagrams to be gauge invariant. In particular, the gauge variation of the 1-loop diagram of holomorphic BF theory, shown in figure~\ref{fig:HBF-anomaly}, was explicitly computed in~\cite{Costello:2015xsa,Gwilliam:2018lpo} and found to be\footnote{These works used BV to quantize the theory, and so in fact the BRST variation of the 1-loop diagram was expressed as
\[
\frac{1}{4!}\bigg(\frac{\im}{2\pi}\bigg)^3\int_{\PT}\tr_{\rm{ad}}(\ba(\p \ba)^3)\,,
\]
in terms of a polyform field $\ba$ that includes $a$ as its physical field. The more familiar expression~\eqref{eq:HBF-anomaly} is obtained by restricting the to its ghost and physical part, $\ba = c + a$. Note that the BV anomaly also receives contributions from the antifields in $\ba$. See subsection \ref{subsec:BV-classical} for further details.}
\be\label{eq:HBF-anomaly}
	\frac{1}{3!}\bigg(\frac{\im}{2\pi}\bigg)^3\int_\PT\tr_\ad(c(\p a)^3)\,. 
\ee
As usual, this is related to the pure gauge term $\int_\PT\chern_4(\fg)$ in our anomaly polynomial by descent.

%To see this recall that here $\fg$ refers to the adjoint bundle over the fibre bundle $\mathscr{X}_\mathrm{gauge}\to\mathrscr{M}\mathrm{gauge}\times\cPT$ where $\mathrscr{M}_\mathrm{gauge}$ is the configuration space of on-shell configurations $\nbar + a^{(0)}$ without quotienting by gauge transformations. The Chern class of this bundle can be computed by specifying a connection...

To see this note that, since our background field $a^{(0)}$ is a $(0,1)$-form, only the $(1,1)$-form part $F^{1,1}=\p a^{(0)}$ of its curvature can appear in~\eqref{eqn:Anomaly-poly}. Thus, dropping the superscript $\!^{(0)}$ on the background field, the pure gauge anomaly polynomial is 
\be
	\int_\PT {\rm ch}_4(\fg) = \frac{1}{4!}\left(\frac{\im}{2\pi}\right)^4\int_\PT \tr_\ad((\p a)^4)\,.
\ee
That is, the gauge anomaly is an invariant polynomial that can be written as the exterior derivative of a Chern-Simons form. The BRST variation of this Chern-Simons form is $\bar\p$-exact for the $(3,3)$-form on twistor space given in~\eqref{eq:HBF-anomaly}.

\begin{figure}[!ht]
\centering
  \begin{subfigure}[t]{0.49\textwidth}
  \centering
    \includegraphics[scale=0.3]{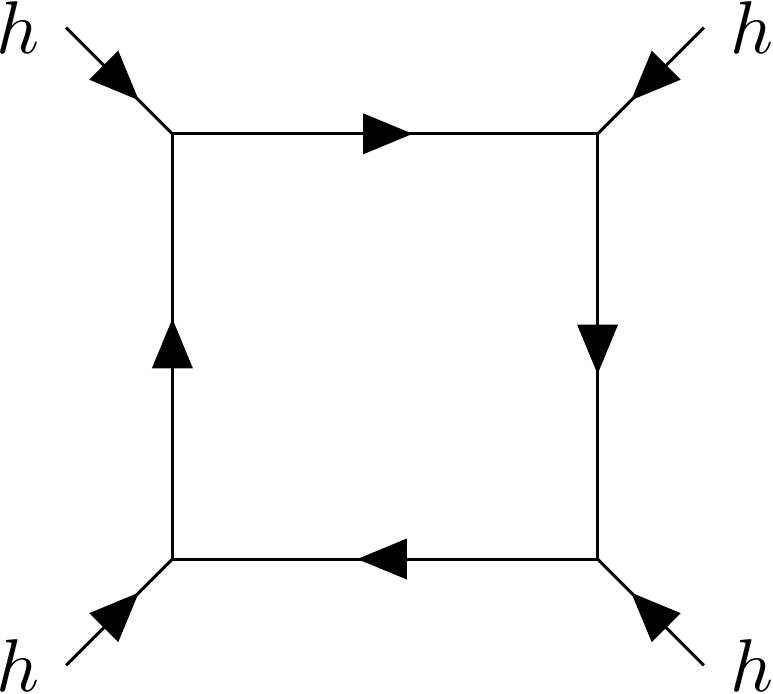}
  \end{subfigure}
  \begin{subfigure}[t]{0.49\textwidth}
  \centering
    \includegraphics[scale=0.3]{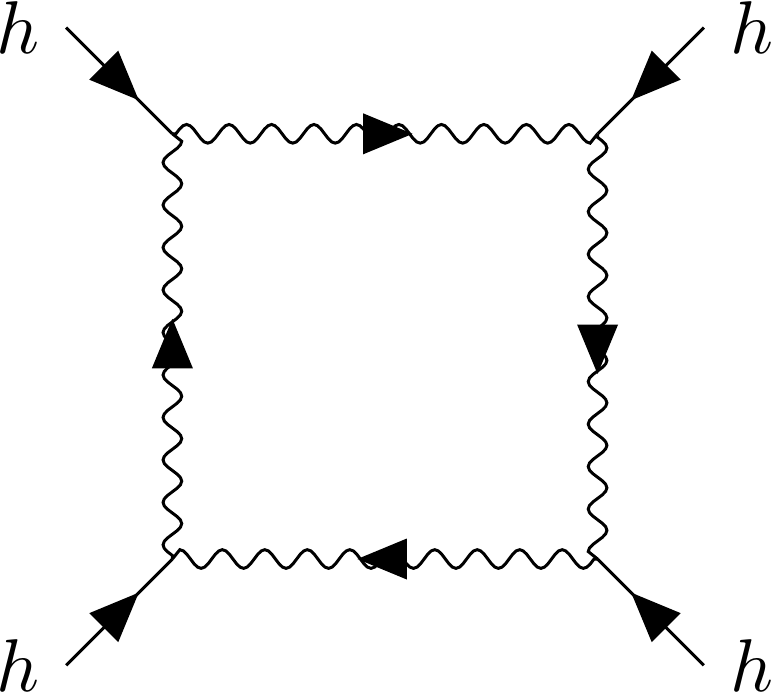}
  \end{subfigure}
\caption{\emph{Diagrams contributing to the pure gravitational anomaly. The first diagram involves quantum fluctuations of the gravitational fields $(g,h)$ (and associated ghosts) running around the loop, whilst the second it is the gauge fields $(b,a)$ (and their ghosts) that circulate. The two diagrams yield the same result, up to a factor counting the number of degrees of freedom in the loop. In each diagram, the gauge and gravitational fields can run around the loop in either direction.}}
\label{fig:PBF-anomaly}
\end{figure}

The remaining two terms in the anomaly polynomial involve the gravitational fields of Poisson-BF theory and are new to this paper. The second term in~\eqref{eqn:Anomaly-poly} is the pure gravitational anomaly, coming from the terms in~\eqref{eqn:Anomaly-family-index} proportional to ${\rm td}_4(T_\PT)$.  This anomaly receives contributions from quantum fluctuations of both the gauge and gravitational fields running around the loop, as we see from the coefficient $1+{\rm dim}(\fg)$.  It thus measures the failure of the two box diagrams in figure~\ref{fig:PBF-anomaly} to be invariant under diffeomorphisms of $\PT$. The combined variation of the two loop diagrams is evaluated in section~\ref{sec:anomaly-calc}, where it is found to be
\be \label{eq:PBF-anomaly} 
\frac{1+\dim(\fg)}{5!}\bigg(\frac{\im}{2\pi}\bigg)^3\int_\PT\tr(\psi(\p s)^3)
\ee
in terms of the $(1,1)$-curvature $\p s$ of the connection $(0,1)$-form $s^{\dot\alpha}_{\ \dot\beta} = - \p^{\dot\alpha}\p_{\dot\beta}h$ and the corresponding ghost $\psi^\dal_{~\,\db} = - \p^\dal\p_\db\chi$. This is in perfect agreement with descent of the pure gravitational anomaly
\be
\frac{1+{\rm dim}(\fg)}{20}\int_\PT{\rm ch}_4(T_\PT) = \frac{1+{\rm dim}(\fg)}{4\cdot 5!}\left(\frac{\im}{2\pi}\right)^4 \int_\PT {\rm tr}((\p s)^4)\,,
\ee
using $s$ as the connection to be used in computing the Todd class. We remark that the evaluation of the box diagram in Poisson-BF theory is substantially more involved than that of holomorphic BF theory.

\begin{figure}[!ht]
\centering
  \begin{subfigure}[t]{0.49\textwidth}
  \centering
    \includegraphics[scale=0.3]{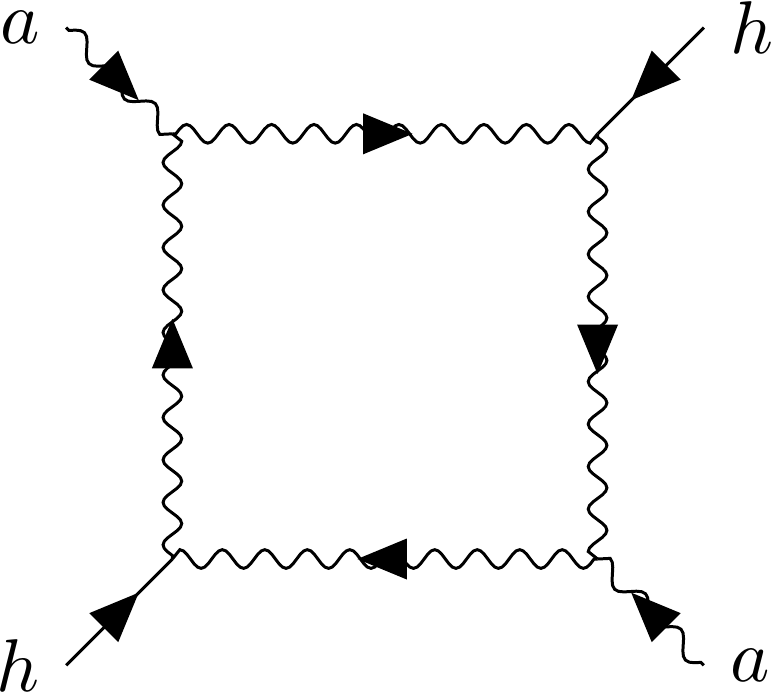}
  \caption*{(O)}
  \end{subfigure}
  \begin{subfigure}[t]{0.49\textwidth}
  \centering
    \includegraphics[scale=0.3]{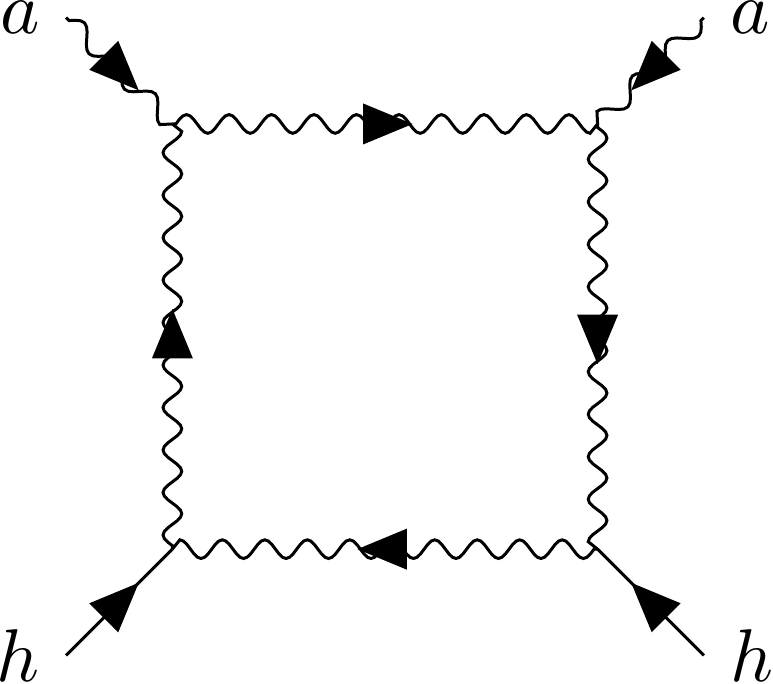}
  \caption*{(A)}
  \end{subfigure}
\caption{\emph{The mixed gauge-gravitational anomaly in twistor space arises from both these two diagrams, each involving two external gauge fields and two external gravitational fields. In each diagram, only the quantum fields of holomorphic BF theory run around the loop, while the gauge fields may be radiated from opposite (O) or adjacent (A) vertices of the box.}} 
\label{fig:PG-mixed-anomaly}
\end{figure}

The final term in~\eqref{eqn:Anomaly-poly} is the mixed gauge-gravitational anomaly associated with the two diagrams in figure~\ref{fig:PG-mixed-anomaly}. These diagrams are invariant neither under gauge transformations nor diffeomorphisms. We evaluate their combined variation in appendix~\ref{app:mixed-anomaly-calc}, finding 
\begin{subequations} \label{eq:PGBF-mixed-cocycle}
\be
	-\frac{1}{24}\bigg(\frac{\im}{2\pi}\bigg)^3\int_\PT\tr_{\rm ad}(c\p a)\wedge\tr((\p s)^2) 
\ee
for the gauge variation, and
\be
	-\frac{1}{24}\bigg(\frac{\im}{2\pi}\bigg)^3\int_\PT\tr_{\rm ad}((\p a)^2)\wedge\tr(\psi\p s) 
\ee
\end{subequations}
for the variation under Hamiltonian diffeomorphisms. Once again, these explicit loop calculations are in perfect agreement with the descent of the mixed term in the anomaly polynomial~\eqref{eqn:Anomaly-poly}. 

To the best of our knowledge, the gravitational and mixed anomalies involving the holomorphic Poisson structure on twistor space have not been computed before. By contrast, in~\cite{elliott2020holomorphic} it was shown that holomorphic Poisson-BF theory on $\C^{2n}$ is anomaly-free for any $n$. (The authors considered a non-degenerate Poisson bi-vector. They also studied a slightly more general mixed topological/holomorphic variant.) While it is perhaps surprising that no anomaly occurs on $\C^{2n}$, this is consistent with the observation that Poisson-BF theory is gravitational in nature. Anomaly polynomials in gravitational theories are non-vanishing only in real dimension $4k+2$.

%%%%%%%%%%%%%%%%%%%%%%%%%%%%%%%%%%

\subsection{Space-time anomalies from twistor space} \label{subsec:space-time-anomalies}

In the anomaly polynomial~\eqref{eqn:Anomaly-poly} for holomorphic BF and Poisson-BF theory, all terms involving $c_1(T_\cPT) = -2x$ cancelled. It is interesting to understand this result from another perspective. Thus, suppose we had obtained a contribution to the anomaly polynomial that was proportional to $x$. Recall that $x$ is the pullback of $\chern_1(K_{\CP^1})$ to $\cPT$, so has components only along the $\CP^1$ fibres of twistor space over space-time. Conversely the coefficient of $x$ will be a 6-form that we may assume has no components along these fibres. Integrating over the $\CP^1$ fibres, this contribution to the twistor anomaly polynomial will thus reduce\footnote{For example, we could compute $\chern_1(K_{\CP^1})$ using a representative whose curvature is concentrated at a single point on the $\CP^1$.} to a 6-form representing the anomaly polynomial of the space-time theory coming from triangle diagrams. These would be potential gauge and gravitational anomalies of the space-time theory. However, there is no candidate for these anomalies: gravitational anomaly polynomials vanish in $4k$ dimensions \cite{Alvarez-Gaume:1983ihn}, as do gauge anomaly polynomials for all self-dual representations of the gauge group \cite{Frampton:1983ez,Townsend:1983ana,Zumino:1983rz}.

\begin{figure}[!ht]
\centering
  \begin{subfigure}[t]{0.49\textwidth}
  \centering
    \includegraphics[scale=0.3]{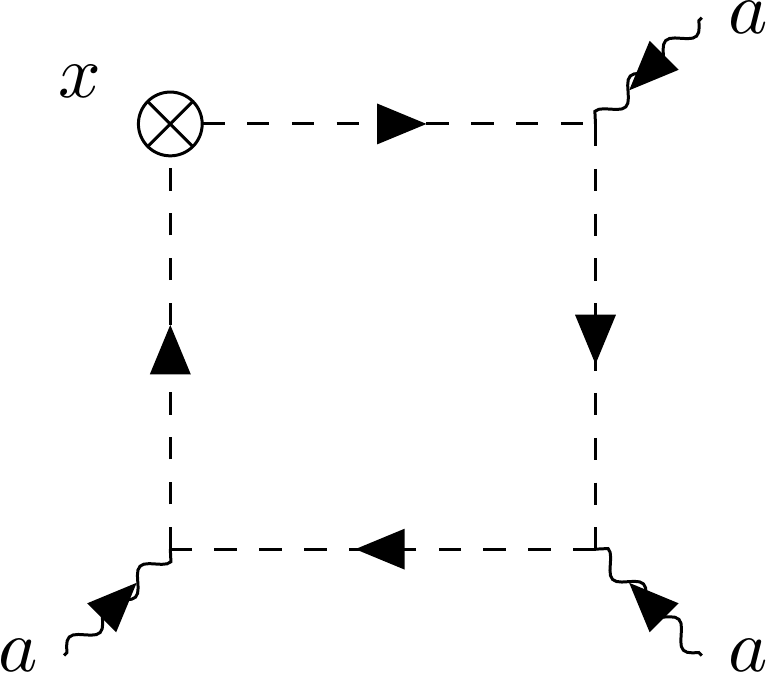}
  \end{subfigure}
  \begin{subfigure}[t]{0.49\textwidth}
  \centering
    \includegraphics[scale=0.33]{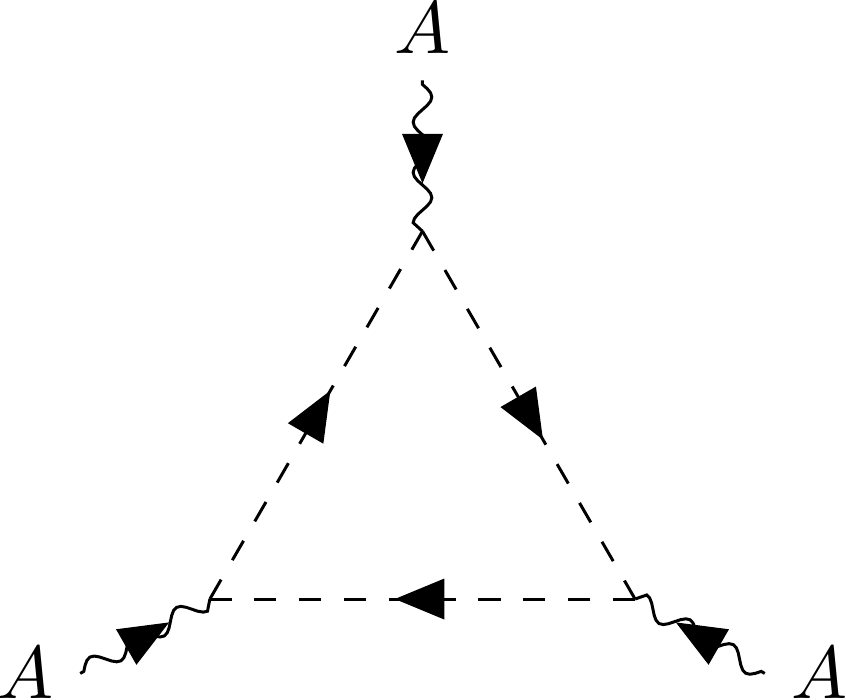}
  \end{subfigure}
\caption{\emph{A term in the twistor anomaly polynomial that is proportional to $\chern_1(T_\cPT) = -2x$ corresponds to a triangle anomaly upon reduction to $\R^4$. No such potential anomalies exist if all the fields in the theory transform in self-dual representations.}}
\label{fig:chiral-anomaly}
\end{figure}

To see the relation to gauge anomalies on $\R^4$ more clearly, let us couple holomorphic BF theory to fermionic matter
\bea
\psi&\in\Pi\Omega^{0,1}(\PT,V\otimes\cO(2s-2))\\
\tilde\psi&\in\Pi\Omega^{3,1}(\PT, V^\vee\otimes\cO(2-2s))\cong\Pi\Omega^{0,1}(\PT,V^\vee\otimes\cO(-2-2s))\,,
\eea
where $V$ is a non-self-dual representation of $\fg$.  We take the matter action to be
\be 
	\frac{1}{2\pi\im}\int_\PT \la\tilde\psi,\bar{D}\psi\ra_V\,, 
\ee
where $\bar{D}$ acts in the representation $V$ and $\la\cdot,\cdot\ra_V$ is the natural pairing between $V^\vee$ and $V$. Note that the coupling to the background complex structure on $\PT$ is implicit in $\bar\p$. On $\R^4$ this action describes a massless Weyl fermion with helicity $s$ in the representation $V$ and helicity $-s$ in the representation $V^\vee$, coupled to the self-dual Yang-Mills field.

The chiral fermions $\psi,\tilde\psi$ contribute to both to the pure gauge and mixed gauge-gravitational anomalies. Their partition function contributes a term
\be
	-\frac{1}{2}\left[\int_\cPT {\rm Td}(T_\cPT)\wedge \left({\rm Ch}(V)\wedge{\rm Ch}(\cO(2s-2))+{\rm Ch}(V^\vee)\wedge{\rm Ch}(\cO(-2s-2))\right)\right]_{(1,1)}
\ee
to the anomaly polynomial in twistor space. In particular, the space-time gauge anomaly comes from the term proportional to ${\rm ch}_3(V)$. Using the fact that ${\rm ch}_3(V^\vee)=-{\rm ch}_3(V)$, this is
\be
	s\int_\cPT \chern_3(V)\wedge x = 2s\int_{\R^4} \chern_3(V)\,.
\ee
The familiar contribution to the space-time gauge anomaly from a Weyl fermion of helicity $s$ transforming in a chiral representation of the gauge group. This identification between twistorial and space-time anomalies can be understood in terms of Feynman diagrams as illustrated in figure \ref{fig:chiral-anomaly}.

To get a non-vanishing mixed gauge-gravitational anomaly on space-time we need to relax the condition that $\fg$ be semi-simple, and allow an abelian summand. The mixed gauge-gravitational space-time anomaly then comes from the term proportional to $\chern_1(V)=-\chern_1(V^\vee)$ in the twistor anomaly polynomial, given by
\be
	s\int_\cPT x\wedge\chern_1(V)\wedge\todd_2(T_\cPT) = -\frac{s}{6}\int_{\R^4}\chern_1(V)\wedge\mathrm{p}_1(T)\,.
\ee
Here $\mathrm{p}_1(T) = -\tr(R\wedge R)/8\pi^2$ is the first Pontryagin class of the Levi-Civita connection on the tangent bundle to space-time. Again, the r.h.s. is familiar from {\it e.g.} the Standard Model.

%%%%%%%%%%%%%%%%%%%%%%%%%%%%%%%%%%
%%%%%%%%%%%%%%%%%%%%%%%%%%%%%%%%%%

\section{Direct evaluation of the gravitational anomaly in Poisson-BF theory}
\label{sec:anomaly-calc}

In section \ref{sec:anomaly} we analysed the anomalies present in Poisson-gauge-BF theory by invoking the family index theorem. In particular, we identified a 1-loop anomaly in pure Poisson-BF theory which could be attributed to the failure of the box diagram illustrated in figure \ref{fig:PBF-anomaly} to be BRST invariant.

Given the perhaps unfamiliar context in which we are applying the index theorem, in this section we provide an independent derivation of the anomaly by performing an explicit diagram computation. In order to so, we first rewrite the theory in the BV formalism. For us, the BV formalism provides an elegant means of performing loop computations without having to treat the ghost and physical fields separately. We will see, however, that the classical theory can also be more straightforwardly understood after recasting in BV. In appendix~\ref{app:mixed-anomaly-calc} we similarly directly evaluate the mixed gauge-gravitational anomaly in Poisson-gauge-BF theory.

%In appendix \ref{app:eta-box} we employ similar techniques to compute the contribution of $\eta$ to the gravity anomaly on $\PT$ when it runs through the loop of the box diagram.

%%%%%%%%%%%%%%%%%%%%%%%%%%%%%%%%%%

\subsection{A brief introduction to the classical BV formalism}
\label{subsec:BV-classical}

For many purposes, it will be convenient to describe twistor space theories using the BV formalism. For BF-type theories, this entails organising the physical fields, ghosts and antifields into a pair of superfields~\cite{Wallet:1989wr,Costello:2007ei,Williams:2018ows}. Doing so allows us to evaluate Feynman diagrams (especially at loop level) using the corresponding super-Feynman rules, instead of treating the ghosts and physical fields separately. (See {\it e.g.} \cite{Budzik:2022mpd} for a recent application of BV to the computation of loop diagrams in holomorphic theories.)

We first consider the holomorphic BF theory~\eqref{gaugeBFaction}. The BV version of this theory extends the fields $a$ and $b$ to polyform fields
\be\label{eq:BV-HBF-fields}
\ba\in\Omega^{0,\bullet}(\PT,\fg)[1]\,,\qquad\qquad \bb\in\Omega^{3,\bullet}(\PT,\fg^\vee)[1]\,.
\ee 
The $\bullet$ here indicates that $\ba$ and $\bb$ are not simply $(0,1)$-forms, but rather a sum of forms of degree $(0,p)$ for $0\leq p\leq 3$.  For example, we can expand
\be
\ba = c + a + b^\vee +d^\vee
\ee
where $c$ is a $\fg$-valued 0-form that is the ghost for standard gauge transformations, $a$ is the original $(0,1)$-form gauge field, $b^\vee$ is a $\fg$-valued $(0,2)$-form of ghost number $-1$ that is the antifield for $b$; it is closely related to the antighost $\tilde d$ in the BRST formalism. The $(0,3)$-form $d^\vee$ is the antifield of the ghost $d$ for gauge transformations~\eqref{gaugeBFgaugetrans2} of the $b$-field.

The $(0,p)$-form fields with $p$ even have fermionic components, while those with $p$ odd have bosonic components so that the whole BV multiplet is anticommuting. The symbol $[1]$ in~\eqref{eq:BV-HBF-fields} indicates this shift of cohomological degree, and reminds us that fields of form degree $(0,p)$ have ghost number $1-p$. Polyforms can naturally be interpreted as superfields: in particular~\cite{Axelrod:1991vq,Alexandrov:1995kv} they can be thought of as functions on the odd tangent bundle $\Pi T^{0,1}_\PT$. 

One reason the BV formalism is convenient is that the full action, including the ghost and gauge-fixing terms, is written simply as
\be \label{eq:BV-HBF-action}
	S[{\bf b},{\bf a}] = \frac{1}{2\pi\im}\int_\PT \bb\,F(\ba)
\ee
in terms of BV superfields. Here $F({\bf a})= \bar\p{\bf a} + \frac{1}{2}[{\bf a},{\bf a}] \in \Omega^{0,\bullet}(\PT,\fg)[2]$ is the polyform curvature, and the integral is understood to extract the top form part of the integrand. When working with polyform fields we suppress the wedge product of forms. Expanding $\ba$ as above and $\bb = d + b  + a^\vee + c^\vee$, this action becomes
\be 
\frac{1}{2\pi\im}\int_\PT\bigg(b\,F^{0,2}(a) + a^\vee\bar{D}c + b^\vee\bar{D}d + b^\vee[c,b] + \frac{1}{2}c^\vee[c,c]  + d^\vee[c,d]\bigg)
\ee
where again $\bar{D} = \bar\p + [a,\ ]$ is the usual gauge covariant $\bar\p$-operator.

Note that the $(0,3)$-form fields $(c^\vee,d^\vee)$ are non-dynamical. The terms in the action involving these fields encode the structure constants of the gauge algebra, as we will see below when we compute the BRST transformations.\\

In order to make the connection with the ordinary BRST construction we reverse the procedure outlined in \cite{Axelrod:1991vq}. The first step is to gauge fix, which in the BV formalism involves specifying a Lagrangian subspace in the extended space of fields. See below for the definition of the symplectic structure on this space. A natural choice is
\be 
\bar\p^\dag\ba = 0\,,\qquad\bar\p^\dag\bb = 0\,,
\ee
where $\bar\p^\dag$ is defined with respect to an arbitrary Hermitian metric on $\PT$, together with a choice of background $\bar\p$-operator. These conditions leave the ghosts $(c,d)$ unconstrained, but kill their antifields $(c^\vee,d^\vee)$. We can solve the constraints on the antifields of the physical fields by writing
\be 
a^\vee = -\bar\p^\dag\tilde{c}\,,\qquad 
b^\vee = -\bar\p^\dag\tilde{d} 
\ee
for new fields $(\tilde{c},\tilde{d})$ that we interpret as BRST antighosts. Finally, the gauge conditions $\bar\p^\dag a = \bar\p^\dag b = 0$ on the physical fields are imposed using Lagrange multipliers which can be identified with Nakanishi-Lautrup fields. Note that the determinants which appear when we integrate out these Lagrange multipliers cancel those generated by the field redefinition $(\tilde c,\tilde d)\to(a^\vee,b^\vee)$. The action \eqref{eq:BV-HBF-action} together with the new Lagrange multiplier terms then coincides with the standard BRST gauge fixed action.\\

The BV formalism for Poisson-BF theory works similarly. (See also~\cite{elliott2020holomorphic} for a description of the BV formalism for mixed holomorphic-topological Poisson-BF theory on $\C^{2n}\times\R^d$.) We combine the ghosts, physical fields and their antifields into a pair of polyform fields 
\bea \label{eq:BV-PBF-fields}
	\bh &= \chi + h + g^\vee + \phi^\vee\ \in\ \Omega^{0,\bullet}(\PT,\cO(2))[1]\\
	\bg &= \phi + g + h^\vee + \chi^\vee\ \in\ \Omega^{3,\bullet}(\PT,\cO(-2))[1]\,.
\eea
Again, $(h,g)$ are the physical fields, while $\chi,\phi$ are the corresponding form ghosts. $\chi$ has a straightforward interpretation as the Hamiltonian for an odd vector field. The antifields $(h^\vee,g^\vee)$ of ghost number $-1$ are closely related to antighosts in the BRST formalism. Finally the antifields $(\phi^\vee,\chi^\vee)$ are required to make the extended space of fields symplectic. In the BV action, they encode the local Lie algebra of Poisson diffeomorphisms and its module $\Omega^{3,0}(\PT,\cO(-2))$.

The classical BV action corresponding to holomorphic Poisson-BF theory is
\be 
\label{eq:BV-PBF-action} 
S[{\bf g},{\bf h}] = \frac{1}{2\pi\im}\int_\PT\bg\,T(\bh) = \frac{1}{2\pi\im}\int_\PT\bg\,\bigg(\bar\p\bh + \frac{1}{2}\{\bh,\bh\}\bigg)\,. 
\ee
This should be quantized after imposing the conditions $\bar\p^\dagger\bh=0$, $\bar
\p^\dagger\bg=0$ which kill the ghost antifields $(\phi^\vee,\chi^\vee)$, impose a gauge condition on the physical fields $(h,g)$ and leave the ghosts $(\chi,\phi)$ unaffected.\\

The BV action for Poisson-gauge-BF theory is similarly
\be
\label{eq:BV-PGBF-action}
	S[\bb,\ba;\bg,\bh] = \frac{1}{2\pi\im}\int_\PT \bg\,T(\bh) + \bb\,\cF(\ba)\,,
\ee
where again $\cF({\bf a}) = \bar\p{\bf a}+ \{{\bf h},{\bf a}\}+ \frac{1}{2}[{\bf a},{\bf a}]$ is the polyform curvature, now coupled to the BV gravitational field ${\bf h}$. \\

The BV formalism equips a theory with an \emph{antibracket}. This is an odd (fermionic) Poisson bracket $\{\ ,\ \}_{\rm BV}$ (not to be confused with the Poisson bracket on $\PT$) defined as the inverse of the odd symplectic form. In the given case of Poisson-gauge-BF theory, this symplectic form is defined by
\be \label{eq:BV-symplectic}
\omega = \frac{1}{2\pi\im}\int_\PT \delta{\bf g}\,\delta{\bf h} + \delta{\bf b}\,\delta{\bf a}\,,
\ee
where $\delta$ is the exterior derivative on the space of fields and the integral is understood to extract the $(3,3)$-form component on $\PT$.

One of the main uses of the antibracket is to determine the BRST transformations. That is, given the action $S$ and antibracket $\{\ ,\ \}_{\rm BV}$, we define the BRST variation of any combination of fields by 
\be \label{eq:BV-BRST-operator}
\delta_{\rm BRST} (\ \cdot \ ) = \{ S,\,\cdot\,\}_{\rm BV}
\ee
Thus, in the BV version of Poisson-gauge-BF theory, we have  BRST transformations
\bea \label{eq:BV-PGBF-transformation}
\delta\ba &= \{S,\ba\}_{\rm BV} = \nbar\ba + \frac{1}{2}[\ba,\ba]\\
\delta\bb &= \{S,\bb\}_{\rm BV} = \nbar\bb + [\ba,\bb]\\
\delta\bh &= \{S,\bh\}_{\rm BV} = \bar\p\bh + \frac{1}{2}\{\bh,\bh\}\\
\delta\bg &= \{S,\bg\}_{\rm BV} = \bar\p\bg + \{\bh,\bg\} + \{\ba,\bb\}\,.
\eea
Restricted to their $(0,1)$-form (or $(3,1)$-form) parts, these transformations coincide with the BRST transformations~\eqref{twistorEYMgaugetrans1}-\eqref{twistorEYMgaugetrans2} given in the text. However, they contain far more. For example, by restricting to their $(0,0)$-form (or $(3,0)$-form) parts, we obtain the BRST transformations of the ghosts
\bea
	\delta c &= \{\chi,c\}+\frac{1}{2}[c,c]\qquad \qquad &\delta d &= \{\chi,d\} + [c,d] \\
	\delta\chi & = \frac{1}{2}\{\chi,\chi\}\qquad\qquad &\delta \phi &= \{\chi,\phi\} + \{c,d\}\,.
\eea
More succintly, BRST invariance of the action is the condition
\be\label{eq:classical-master}
	\delta S=\{S,S\}_{\rm BV}=0
\ee
known as the \emph{classical master equation}. This condition encodes the fact that $\Omega^{0,\bullet}(\PT,\fg)$ and $\Omega^{0,\bullet}(\PT,\cO(2))$ are local dg Lie algebras, with $\bar\p$ the differential and Lie brackets induced from those on $\fg$ and the twistor Poisson bracket, respectively. In particular, the classical master equation ensures the BRST operator obeys $\delta^2=0$.

\subsection{BV quantization} \label{subsec:BV-quantum}

Thus far our discussion of BV has been entirely classical, in this subsection we sketch how to perform BV quantization of Poisson-BF theory.\\

Before doing so, we first restrict to the patch $\C^3\cong\{\lambda\neq\hat\alpha\}\subset\PT$ for some left-handed spinor $\alpha$, and use the inhomogeneous co-ordinates $\{z^a\}_{a=0,1,2}$ where
\be z^0 = z = \frac{\la\lambda\,\alpha\ra}{\la\lambda\,\hat\alpha\ra}\,,\qquad z^{\dot\alpha} = v^\dal = \frac{\mu^\dal}{\la\lambda\,\alpha\ra}\,. \ee
%for $\alpha=1,2$. 
In this patch we can trivialize the line bundles $\cO(n)$: explicitly $\Gamma(\cO(n))\ni\Phi\mapsto\Phi/\la\lambda\,\hat\alpha\ra^n$. For example, $\D^3Z$ is represented by
\be \d^3z = \d z\d v^{\dot 1}\d v^{\dot 2} = \frac{1}{3!}\varepsilon_{abc}\d z^a\d z^b\d z^c = \frac{\D^3Z}{\la\lambda\,\hat\alpha\ra^4}\,, \ee
where $\varepsilon_{012}=1$. We abuse notation by continuing to write $\bh,\bg$ for the trivializations of the corresponding fields on $\PT$. The Poisson-BF action \eqref{eq:BV-PBF-action} takes the same form in terms of the trivialized fields.

Since the gravitational anomaly in Poisson-BF theory is local, it is enough to evaluate it in this patch.\footnote{Note, however, that this kind of calculation is not sensitive to the potential mixed anomaly involving the background complex structure on $\PT$ discussed in subsection \ref{subsec:space-time-anomalies}.}\\

In order to quantize we follow the systematic approach developed by Costello in \cite{Costello:2007ei} and specialised to the case of holomorphic theories by Williams in \cite{Williams:2018ows}. This procedure broadly speaking involves two steps, which we outline below. For a more complete introduction see the aforementioned references.

The first, termed \emph{prequantization} by the above authors, broadly corresponds to the physicist's notion of renormalization. It entails constructing a family of (length) scale $L$ effective interactions $I[L]$ for $L\in\R_+$ all related by renormalization group flow. These effective interactions are formal power series in $\hbar$, and the restriction to $\hbar=0$ should match the interacting part of the classical action. Given a classical action, we can construct a prequantization as follows. First we gauge fix and regularise by introducing a UV cutoff $0<\epsilon<L$. Ideally one wishes to define the vertices in the scale $L$ effective interaction by taking the $\epsilon\to0$ limit of regularised Feynman diagrams, which guarantees that the effective interactions are related by RG flow. Unfortunately this limit is usually singular, and counterterms must be introduced to compensate for this. However, it is proven by Williams in \cite{Williams:2018ows} that for a holomorphic theory on $\C^d$ this limit exists for all 1-loop Feynman diagrams, immediately leading to a prequantization of holomorphic BF-type theories on account of the fact they are 1-loop exact.\footnote{Since the theories we consider are 1-loop exact, we can consistently set $\hbar=1$. Note that the number of external Lagrange multiplier fields $b,g$ in a Feynman diagram determine its loop order, so we lose no information in doing this.}

The second step is verify that the scale $L$ effective actions obey the regularised \emph{quantum master equation}
\be \hbar\Delta_LI[L] + \frac{1}{2}\{I[L],I[L]\}_L = 0\,. \ee
The object on the left hand side is termed the anomaly, since if it's non-vanishing then the regularised theory fails to be BRST invariant. Here $\{\ ,\ \}_L$ is the regularised BV bracket, and $\Delta_L$ is the regularised BV Laplacian. The precise definition of these objects will not be important for us, because Williams \cite{Williams:2018ows} has proven that the order $\hbar$ contribution to the anomaly can be obtained as a sum over 1-loop wheel diagrams with $d+1$ external legs and with one internal propagator replaced by a regularised delta function. These wheel diagrams are evaluated with a finite UV cutoff $0<\epsilon<L$ and at the end of the calculation the limit $\epsilon\to0$ is taken. Since the quantum master equation is compatible with RG flow, it is enough to compute the anomaly in the $L\to0$ limit also. Finally, since any holomorphic BF-type theory is 1-loop exact, the 1-loop anomaly represents the only potential failure of BRST invariance.

\begin{figure}[!ht]
\centering
  \includegraphics[scale=0.3]{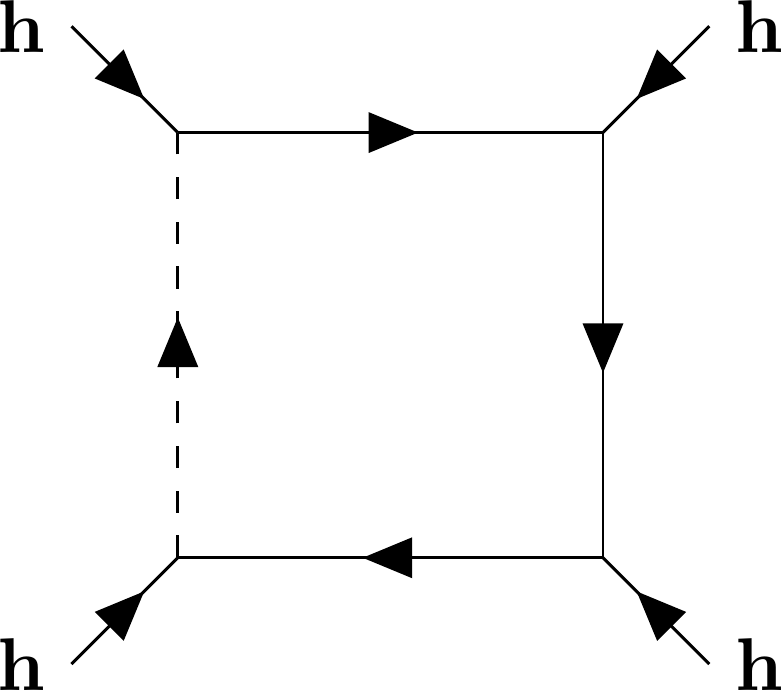}
\caption{\emph{BRST variation of the box diagram in Poisson-BF theory.}} \label{fig:PBF-anomaly-BRST}
\end{figure}

In the case of Poisson-BF theory on $\C^3$ the relevant wheel diagram has 4 external legs and is therefore a box. Up to isomorphism there is a unique way of choosing an internal propagator, illustrated in figure \ref{fig:PBF-anomaly-BRST}. This diagram can naturally be interpreted as the BRST variation of that appearing in figure \ref{fig:PBF-anomaly}, where the $\Z_4$ symmetry factor is compensated by the 4 choices of internal edge. In order to evaluate it we must first fix the gauge and regularise.\\

We gauge fix using the operator
\be \bar\p^\dag = -\ast\p\,\ast \ee
where $\ast$ is defined using the Hermitian structure $\delta = \d z^a\d\bar z_a = \delta_{a\bar b}\d z^a\d\bar z^{\bar b}$ on $\C^3$. Here, and from now on, we use the Hermitian form to lower the index on the conjugate co-ordinates: $\bar z_a = \delta_{a\bar b}\bar z^{\bar b}$. The restriction of this form to the $v^\dal$ directions is preserved by the subgroup $\SU(2)\subset\SL_2(\C)$, allowing us to adopt the same conventions for spinors. In particular complex conjugation lowers the index on a spinor: $\xi^\dal\mapsto{\bar\xi}_\dal$ 

This gauge fixing has the property
\be \{\bar\p,\bar\p^\dag\} = \Delta_{\bar\p} = \frac{1}{2}\Delta_\d \ee
for $\Delta_{\bar\p}$ and $\Delta_\d$ the Dolbeault and de Rham Laplacians respectively. The propagator is the pullback by the difference map 
\be \D_{12}:(\C^3)_1\times (\C^3)_2\to\C^3\,,\qquad(z_1,z_2)\mapsto z_{12} = z_1-z_2 \ee
of a $(0,2)$-form propagator $P$ obeying $\bar\p^\dag P=0$. Before regularisation it obeys
\be \bar\p P = -2\pi\bar\delta^3 \ee
for $\bar\delta^3$ the $(0,3)$-form $\delta$-function with support at the origin and normalized so that
\be \int_{\C^3}\d^3z\,\bar\delta^3 = \bigg(\frac{\im}{2}\bigg)^3\int_{\C^3}\d^3z\,\d^3\bar z\,\delta^{(6)}(z) = - 1\,. \ee
(Here $\delta^{(6)}(z)$ is the real $\delta$-function normalized against the standard measure on $\R^6$.) Explicitly we have
\be P = -4\pi\bar\p^\dag\Delta_\d^{-1}\bar\delta^3 = \frac{\im}{8\pi^2}\bar\p^\dag\bigg(\frac{\d^3\bar z}{\|z\|^4}\bigg)\,. \ee
Note that pulling this back by $\D_{12}$ will give a $(0,2)$ form on $(\C^3)_1\times(\C^3)_2$ which is not of definite degree on either of the two factors. This occurs precisely because we are working with polyform fields. The part of the propagator which is a $(0,0)$-form on $(\C^3)_1$ and a $(0,2)$-form on $(\C^3)_2$ encodes the ghost-antifield propagator.\\

To regularise we introduce length scales $\epsilon$ and $L$, which play the role of UV and IR cutoffs. The regulated $\delta$-function $\bar\delta^3$ is $\overline K^T = k^T \d^3\bar z$ for
\be \label{eq:heat-kernel} k^T(z) = \bigg(\frac{\im}{8\pi T}\bigg)^3\exp\bigg(- \frac{\|z\|^2}{4T}\bigg)\,, \ee
the heat kernel. We recover the $(3,0)$-form $\delta$-function as a distribution in the limit $\lim_{L\to\infty}\lim_{\epsilon\to0}$. The mollified propagator $P^{[\epsilon,L]}_{12} = \D^*_{12}P^{[\epsilon,L]}$, obeying $\bar\p^\dag P^{[\epsilon,L]}=0$ and
\be \label{eq:mollified-eom} \bar\p P^{[\epsilon,L]} = 2\pi(\overline K^L - \overline K^\epsilon)\,, \ee
is
\be \label{eq:mollified-prop}
P^{[\epsilon,L]} = -4\pi\bar\p^\dag\int_\epsilon^L\d T\,\overline K^T = 4\pi\ast\p\ast\int_\epsilon^L\d T\,\overline K^T\,.
\ee
$P^{[\eps,L]}$ obeys \eqref{eq:mollified-eom} by the heat equation for \eqref{eq:heat-kernel}. Using the identities
\be \ast\d^3\bar z = -\im\d^3\bar z\,,\qquad\ast(\d z^a\d^3\bar z) = \im\varepsilon^{abc}\d\bar z_b\d\bar z_c\,, \ee
we can rewrite it as
\be \begin{split} \label{eq:reg-prop}
&P^{[\epsilon,L]} = -4\pi\im\ast(\d z^a\d^3\bar z)\int_\varepsilon^L\d T\p_ak^T \\
&= 4\pi\varepsilon^{abc}\d\bar z_b\d\bar z_c\int_\varepsilon^L\d T\p_ak^T = -2\pi\varepsilon^{abc}\bar z_a\d\bar z_b\d\bar z_c\int_\epsilon^L\frac{\d T}{2T}k^T\,.
\end{split} \ee
This is the form of the propagator we shall use in the next section.

%%%%%%%%%%%%%%%%%%%%%%%%%%%%%%%%%%

\subsection{Evaluating the anomaly} \label{subsec:loopsec}

We can now compute the diagram in figure \ref{fig:PBF-anomaly-BRST} directly. Writing $z_i$ for the position of the $i$\textsuperscript{th} vertex with $i=1,\dots 4$ it evaluates to
\be \label{eq:anomaly-calc-1} -\bigg(\frac{1}{2\pi}\bigg)^4\int_{(\C^3)^4}\d^3z_1\{\bh_1,P^{[\epsilon,L]}_{12}\}_1\d^3z_2\{\bh_2,P^{[\epsilon,L]}_{23}\}_2\d^3z_3\{\bh_3,P^{[\epsilon,L]}_{34}\}_3\d^3z_4\{\bh_4,\overline K^\epsilon_{41}\}_4\,. \ee
Here $\bh_i$ denotes an insertion of the background field at $z_i$, and the subscript on the Poisson brackets indicate on which set of co-ordinates they act. Substituting in the explicit form of the propagator and employing the identity
\be \label{eq:epsilon-identity} \prod_{i=1}^3(\varepsilon^{a_ib_ic_i}\bar z_{i(i+1)a_i}\d\bar z_{i(i+1)b_i}\d\bar z_{i(i+1)c_i}) \d^3\bar z_{41} = -8\varepsilon^{a_1a_2a_3}{\bar z}_{12a_1}{\bar z}_{23a_2}{\bar z}_{34a_3}\d^3\bar z_{12}\d^3\bar z_{23}\d^3\bar z_{34} \ee
gives
\bea \label{eq:anomaly-calc-2}
&\frac{1}{4^4}\int_{(\C^3)^4}\d^3z_1\p^{\dal_1}\bh_1\d^3z_2\dots\d^3z_4\p^{\dal_4}\bh_4\d^3\bar z_{12}\d^3\bar z_{23}\d^3\bar z_{34}\,\varepsilon^{a_1a_2a_3}{\bar z}_{12a_1}{\bar z}_{23a_2}{\bar z}_{34a_3} \\
&\qquad\int_{[\epsilon,L]^3}\frac{\d^3\vec T}{\epsilon T_1^2T_2^2T_3^2}\p_{\dal_1}k^{T_1}(z_{12})\p_{\dal_2}k^{T_2}(z_{23})\p_{\dal_3}k^{T_3}(z_{34})\p_{\dal_4}k^\epsilon(z_{41})\,.
\eea
It's convenient to introduce co-ordinates $w_i = z_{i(i+1)} = z_i - z_{i+1}$ for $i\in\Z_4$ obeying the constraint $\sum_{i=1}^4w_i=0$. Using the fact
\be \p_ak^T = -\frac{\bar z_a}{4T}k^T\,. \ee
we can then rewrite equation \eqref{eq:anomaly-calc-2} as
\be \begin{split}
&\frac{1}{4^4}\int_{(\C^3)^4}\d^3w_1\p^{\dal_1}\bh_1\d^3w_2\dots\d^3z_4\p^{\dal_4}\bh_4\d^3\bar w_1\d^3\bar w_2\d^3\bar w_3\,\varepsilon^{a_1a_2a_3}\bar w_{1a_1}\bar w_{2a_2}\bar w_{3a_3} \\
&\qquad\int_{[\epsilon,L]^3}\frac{\d^3\vec T}{\epsilon T_1^2T_2^2T_3^2}\bar w_{1\dal_1}k^{T_1}(w_1)\dots \bar w_{4\dal_4}k^\epsilon(w_4)\,.
\end{split} \ee
To make sense of the above we first eliminate the explicit factors of $\bar w$ by rewriting them as derivatives of the product of heat kernels. In order to do so note that
\be \begin{split}
&g^{\vec T,\epsilon}(w_1,w_2,w_3) \coloneqq k^{T_1}(w_1)k^{T_2}(w_2)k^{T_3}(w_3)k^\epsilon(w_4) \\ 
&= \bigg(\frac{\im}{8\pi}\bigg)^{3\times 4}\frac{1}{\epsilon^3 T_1^3T_2^3T_3^3}\exp\bigg(-\frac{1}{4}\bar w_{\ell a}H^{\ell k}w^a_k\bigg)
\end{split} \ee
for
\be (H_{\ell k})_{k,\ell=1}^3 = \begin{pmatrix}
\frac{1}{T_1} + \frac{1}{\epsilon} & \frac{1}{\epsilon} & \frac{1}{\epsilon} \\
\frac{1}{\epsilon} & \frac{1}{T_2} + \frac{1}{\epsilon} & \frac{1}{\epsilon} \\
\frac{1}{\epsilon} & \frac{1}{\epsilon} & \frac{1}{T_3} + \frac{1}{\epsilon}
\end{pmatrix}\,. \ee
Following \cite{Williams:2018ows}, we introduce the vector fields\footnote{These can perhaps be better understood by introducing $\zeta_a = \big(\sum_{i=1}^4T_i)^{-1}\sum_{i=1}^4T_i\p/\p_{w_i^a}$ where we use the shorthand $T_4=\epsilon$. $\zeta$ annihilates $g^{\vec T,\epsilon}$ on the constraint surface $\sum_{i=1}^4w_i=0$, and $\eta_{ia} = \p/\p_{w_i^a} - \zeta_a$ are tangent to this surface. To get the equations \eqref{eq:eta} we have simply eliminated $w_4 = -w_1-w_2-w_3$.}
\be \label{eq:eta} \eta_{4a} \coloneqq - \frac{1}{\epsilon + T_1 + T_2 + T_3}\sum_{k=1}^3 T_k\frac{\p}{\p w_k^a}\,,\qquad \eta_{ka} \coloneqq \frac{\p}{\p w_k^a} + \eta_{4a} \ee
obeying
\be \label{eq:eta-id} \eta_{4a}g^{\vec T,\epsilon} = - \frac{\bar w_{4a}}{4\epsilon}g^{\vec T,\epsilon}\,,\qquad \eta_{ka}g^{\vec T,\epsilon} = -\frac{\bar w_{ka}}{4T_k}g^{\vec T,\epsilon}\,. \ee
Replacing $\bar w_{1a_1}\bar w_{2a_2}\bar w_{3a_3}$ by $\eta_{1a_1}\eta_{2a_2}\eta_{3a_3}$ and integrating by parts to remove the derivatives from $g^{\vec T,\epsilon}$ we get
\be \begin{split}
&\frac{1}{4}\int_{(\C^3)^4}\int_{[\epsilon,L]^3}\frac{\d^3\vec T}{\epsilon T_1T_2T_3}\varepsilon^{a_1a_2a_3}\cL_{\eta_{1a_1}}\cL_{\eta_{2a_2}}\cL_{\eta_{3a_3}}(\d^3w_1\p^{\dal_1}\bh_1\d^3w_2\dots\d^3z_4\p^{\dal_4}\bh_4) \\
&\qquad\d^3\bar w_1\d^3\bar w_2\d^3\bar w_3\,\bar w_{1\dal_1}\bar w_{2\dal_2}\bar w_{3\dal_3}\bar w_{4\dal 4}g^{\vec T,\epsilon}(w_1,w_2,w_3)\,.
\end{split} \ee
Assuming the derivatives with respect to $w_k$ are taken at fixed $z_4$ we have
\be \begin{split} \label{eq:3-eta-id}
&\varepsilon^{a_1a_2a_3}\cL_{\eta_{1a_1}}\cL_{\eta_{2a_2}}\cL_{\eta_{3a_3}} = \frac{\epsilon}{\epsilon + T_1 + T_2 + T_3}\varepsilon^{a_1a_2a_3}\cL_{\p/\p w^{a_1}_1}\cL_{\p/\p w^{a_2}_2}\cL_{\p/\p w^{a_3}_3} \\
&=\frac{\epsilon}{\epsilon + T_1 + T_2 + T_3}\varepsilon^{a_1a_2a_3}\cL_{\p/\p z^{a_1}_1}\cL_{\p/\p z^{a_2}_2}\cL_{\p/\p z^{a_3}_3}\,.
\end{split} \ee
Substituting this back into the expression for the anomaly gives
\be \begin{split}
&\frac{1}{4}\int_{(\C^3)^4}\int_{[\epsilon,L]^3}\frac{\d^3\vec T}{T_1T_2T_3(\epsilon+T_1+T_2+T_3)}\varepsilon^{a_1a_2a_3}\d^3z_1\p_{a_1}\p^{\dal_1}\bh_1\d^3z_2\dots\d^3z_4\p^{\dal_4}\bh_4 \\
&\qquad\d^3\bar w_1\d^3\bar w_2\d^3\bar w_3\,\bar w_{1\dal_1}\bar w_{2\dal_2}\bar w_{3\dal_3}\bar w_{4\dal_4}g^{\vec T,\epsilon}(w_1,w_2,w_3)\,.
\end{split} \ee
Here we've made use of the identity
\be \cL_{\p/\p z^a}(\d^3z\,\omega) = \d^3z\,\cL_{\p/\p z^a}\omega \ee
for $\omega\in\Omega^{0,\bullet}(\C^3)$. Now let's turn our attention to the remaining factors of $\bar w$. Using the identities in equations \eqref{eq:eta-id} and integrating by parts we arrive at
\bea \label{eq:anomaly-calc-3}
&4^3\int_{(\C^3)^4}\int_{[\epsilon,L]^3}\frac{\d^3\vec T\epsilon}{(\epsilon+T_1+T_2+T_3)}\varepsilon^{a_1a_2a_3}\cL_{\eta_{1\dal_1}}\cL_{\eta_{2\dal_2}}\cL_{\eta_{3\dal_3}}\cL_{\eta_{4\dal_4}}(\d^3z_1\p_{a_1}\p^{\dal_1}\bh_1\d^3z_2\\
&\qquad\dots\d^3z_4\p^{\dal_4}\bh_4)\d^3\bar w_1\d^3\bar w_2\d^3\bar w_3\,g^{\vec T,\epsilon}(w_1,w_2,w_3)\,.
\eea
In terms of the original $z_i$ co-ordinates we have
\be \begin{split}
\eta_{1\dal_1} &= \frac{1}{\epsilon + T_1 + T_2 + T_3}\bigg(\epsilon\frac{\p}{\p z^{\dal_1}_1} - (T_2 + T_3)\frac{\p}{\p z^{\dal_1}_2} - T_3\frac{\p}{\p z^{\dal_1}_3}\bigg)\,, \\
\eta_{2\dal_2} &= \frac{1}{\epsilon + T_1 + T_2 + T_3}\bigg(\epsilon\frac{\p}{\p z^{\dal_2}_1} + (\epsilon + T_1)\frac{\p}{\p z^{\dal_2}_2} - T_3\frac{\p}{\p z^{\dal_2}_3}\bigg)\,, \\
\eta_{3\dal_3} &= \frac{1}{\epsilon + T_1 + T_2 + T_3}\bigg(\epsilon\frac{\p}{\p z^{\dal_3}_1} + (\epsilon + T_1)\frac{\p}{\p z^{\dal_3}_2} + (\epsilon + T_1+T_2)\frac{\p}{\p z^{\dal_3}_3}\bigg)\,, \\
\eta_{4\dal_4} &= \frac{1}{\epsilon + T_1 + T_2 + T_3}\bigg(-(T_1 + T_2 + T_3)\frac{\p}{\p z_1^{\dal_4}} - (T_2 + T_3)\frac{\p}{\p z_2^{\dal_4}} - T_3\frac{\p}{\p z_3^{\dal_4}}\bigg)\,.
\end{split} \ee
Substituting the above into equation \eqref{eq:anomaly-calc-3} and exploiting the antisymmetry of the Poisson bracket gives
\be \begin{split}
&4^3\varepsilon^{a_1a_2a_3}\int_{(\C^3)^4}\int_{[\epsilon,L]^3}\frac{\d^3\vec T\epsilon}{(\epsilon+T_1+T_2+T_3)^5}\bigg(-(T_2 + T_3)\frac{\p}{\p z_2^{\dal_1}} - T_3\frac{\p}{\p z_3^{\dal_1}}\bigg) \\
&\qquad\bigg(\epsilon\frac{\p}{\p z_1^{\dal_2}} - T_3\frac{\p}{\p z_3^{\dal_2}}\bigg)\bigg(\epsilon\frac{\p}{\p z^{\dal_3}_1} + (\epsilon + T_1)\frac{\p}{\p z^{\dal_3}_2}\bigg) \\
&\qquad\qquad\bigg(-(T_1 + T_2 + T_3)\frac{\p}{\p z_1^{\dal_4}} - (T_2 + T_3)\frac{\p}{\p z_2^{\dal_4}} - T_3\frac{\p}{\p z_3^{\dal_4}}\bigg) \\
&\qquad\qquad\qquad(\d^3z_1\p^{\dal_1}\p_{a_1}\bh_1\d^3z_2\dots\d^3z_4\p^{\dal_4}\bh_4)\d^3\bar w_1\d^3\bar w_2\d^3\bar w_3g^{\vec T,\epsilon}(w_1,w_2,w_3)\,.
\end{split} \ee
$g^{\vec T,\epsilon}$ is a Gaussian concentrated at $w_1=w_2=w_3=0$, and so we can expand the rest of the integrand in modes around $z_4$. When the $L,\eps\to0$ limits are taken only the zero-mode from the integrand will contribute to the integrals. The lowest moment of the Gaussian is
\be \bigg(\frac{\im}{8\pi}\bigg)^3\frac{1}{\epsilon^3T_1^3T_2^3T_3^3}\bigg(\frac{\epsilon T_1T_2T_3}{\epsilon + T_1 + T_2 + T_3}\bigg)^3 = \bigg(\frac{\im}{8\pi}\bigg)^3\frac{1}{(\epsilon + T_1 + T_2 + T_3)^3}\,. \ee
and therefore the anomaly is
\be \begin{split}
&\bigg(\frac{\im}{2\pi}\bigg)^3\varepsilon^{a_1a_2a_3}\int_{\C^3}\int_{[\epsilon,L]^3}\frac{\d^3\vec T\epsilon}{(\epsilon+T_1+T_2+T_3)^8} \\
&\qquad\bigg(-(T_2 + T_3)\frac{\p}{\p z_2^{\dal_1}} - T_3\frac{\p}{\p z_3^{\dal_1}}\bigg)\bigg(\epsilon\frac{\p}{\p z_1^{\dal_2}} - T_3\frac{\p}{\p z_3^{\dal_2}}\bigg)\bigg(\epsilon\frac{\p}{\p z^{\dal_3}_1} + (\epsilon + T_1)\frac{\p}{\p z^{\dal_3}_2}\bigg) \\
&\qquad\qquad\bigg( - (T_1 + T_2 + T_3)\frac{\p}{\p z_1^{\dal_4}} - (T_2 + T_3)\frac{\p}{\p z_2^{\dal_4}} - T_3\frac{\p}{\p z_3^{\dal_4}}\bigg) \\
&\qquad\qquad\qquad(\p^{\dal_1}\p_{a_1}\bh_1\dots\d^3z_4\p^{\dal_4}\bh_4)|_{z=z_1=z_2=z_3=z_4}\,.
\end{split} \ee
A marginal simplification is obtained by integrating by parts with $\p/\p z_3^{\dal_4}$ in the 4\textsuperscript{th} differential operator and replacing it by $- \p/\p z_1^{\dal_4} - \p/\p z_2^{\dal_4} - \p/\p z_4^{\dal_4}$. We further discard $\p/\p z_4^{\dal_4}$ here by the antisymmetry of the Poisson bracket. Therefore the anomaly is
\be \begin{split}
&\bigg(\frac{\im}{2\pi}\bigg)^3\int_{\C^3}\int_{[\epsilon,L]^3}\frac{\d^3\vec T\epsilon}{(\epsilon+T_1+T_2+T_3)^8}\bigg(-(T_2 + T_3)\frac{\p}{\p z_2^{\dal_1}} - T_3\frac{\p}{\p z_3^{\dal_1}}\bigg) \\
&\qquad\bigg(\epsilon\frac{\p}{\p z_1^{\dal_2}} - T_3\frac{\p}{\p z_3^{\dal_2}}\bigg)\bigg(\epsilon\frac{\p}{\p z^{\dal_3}_1} + (\epsilon + T_1)\frac{\p}{\p z^{\dal_3}_2}\bigg)\bigg(- (T_1 + T_2)\frac{\p}{\p z_1^{\dal_4}} - T_2\frac{\p}{\p z_2^{\dal_4}}\bigg) \\
&\qquad\qquad(\p\p^{\dal_1}\bh_1\p\p^{\dal_2}\bh_2\p\p^{\dal_3}\bh_3\p^{\dal_4}\bh_4)|_{z=z_1=z_2=z_3=z_4}\,.
\end{split} \ee
At this point we run out of tricks, and are forced to expand out the differential operators. Fortunately half of the 16 terms generated are total derivatives. Up to permuting factors of $\bh$, relabelling indices and discarding exact terms the other half are all proportional to a single expression:
\be \label{eq:anomaly-term}
\cA = \p\p^\dd\p_\dal\bh\p\p^\dal\p_\db\bh\p\p^\db\p_\dg\bh\p^\dg\p_\dd\bh\,.
\ee
The coefficient is obtained by performing the integrals over $\vec T$ and then taking $\lim_{\epsilon\to0}$. These steps have been relegated to appendix \ref{app:anomaly-deats}, where we show that the anomaly evaluates to
\be \frac{1}{4\cdot 5!}\bigg(\frac{\im}{2\pi}\bigg)^3\int_{\C^3}\cA = \frac{1}{4\cdot 5!}\bigg(\frac{\im}{2\pi}\bigg)^3\int_{\C^3}\tr(\bs(\p\bs)^3)\,, \ee
where we recall that $\bs^\dal_{~\,\db} = - \p^\dal\p_\db\bh$, and the trace is taken in the fundamental of $\fsl_2(\C)$. It is not hard to see that this does not depend on our trivialization of $\cO(n)$, and so takes the same form on $\PT$
\be \label{eq:BV-PBF-anomaly} \frac{1}{4\cdot 5!}\bigg(\frac{\im}{2\pi}\bigg)^3\int_\PT\tr(\bs(\p\bs)^3)\,. \ee
To see that this is consistent with the anomaly cocycle claimed in equation \eqref{eq:PBF-anomaly} we need to restrict the polyform field $\bh$ into its physical and ghost pieces, $\bh = \chi + h$, whereupon $\bs$ gets similarly restricted to
\be \bs^\db_{~\,\dal} = \psi^\db_{~\,\dal} + s^\db_{~\,\dal} =  - \p^\db\p_\dal\chi - \p^\db\p_\dal h\,. \ee
Substituting into the BV anomaly cocycle gives
\be \frac{1}{5!}\bigg(\frac{\im}{2\pi}\bigg)^3\int_{\PT}\tr(\psi(\p s)^3)\,, \ee
precisely matching the cocycle in equation \ref{eq:PBF-anomaly}. The extra terms present in the BV cocycle \eqref{eq:BV-PBF-anomaly} involving antifields render it BRST invariant off-shell.

%%%%%%%%%%%%%%%%%%%%%%%%%%%%%%%%%%
%%%%%%%%%%%%%%%%%%%%%%%%%%%%%%%%%%

\section{Anomaly cancellation from a Green-Schwarz mechanism on twistor space} 
\label{sec:GS-mechanism}

Since the holomorphic BF and Poisson-BF theories are both anomalous, we must modify these theories if we wish to have a consistent quantum theory on twistor space.\\

One way of achieving this is to couple to appropriate bosonic and fermionic matter so that the coefficients of the anomalies, or equivalently the all-plus amplitudes, vanish on the nose. In the case of sd gravity the anomaly is proportional to the difference between the number of bosonic and fermionic degrees of freedom, and so it can be cancelled by coupling to a single Weyl fermion. The case of sdEYM is a little more involved. Suppose we couple to Weyl fermions and real scalars in the representations $R_f$ and $R_s$ of $\fg$ respectively. Then the conditions for anomaly cancellation are
\bea \label{eq:vanishing-anomalies}
0 &= \dim R_s - 2\dim R_f + 2\dim\fg + 2\,, \\
0 &= \tr_{R_s}(X^2) - 2\tr_{R_f}(X^2) + 2\tr_\ad(X^2)\,, \\
0 &= \tr_{R_s}(X^4) - 2\tr_{R_f}(X^4) + 2\tr_\ad(X^4)
\eea
for all $X\in\fg$. Depending on the dimension of the space of quartic Lie algebra invariants, these impose either 3, 4 or 5 constraints. Of course, the ordinary 4d chiral anomaly must also vanish. Given the considerable freedom in $R_f,R_s$, we expect that in general these identities can be satisfied in many different ways. For example, in supersymmetric theories (even with $\cN=1$ supersymmetry) the fermionic superpartners precisely cancel the loop diagrams and all the anomalies vanish, although we emphasise that the constraints \eqref{eq:vanishing-anomalies} by themselves are far weaker than supersymmetry.

Another, particularly intriguing, anomaly free theory is the holomorphic Poisson-Chern-Simons theory with dynamical field $H\in\Omega^{0,1}(\PT,\oplus_{s\in\Z}\cO(2s-2))$ and action
\be \frac{1}{4\pi\im}\int_\PT\D^3Z\wedge\bigg(H\wedge\dbar H + \frac{1}{3}H\wedge\{H,H\}\bigg)\,. \ee
Here we interpret the integral as extracting the unweighted part of the integrand. The action has the obvious gauge redundancies with parameter $\mathcal{X}\in\Omega^{0}(\PT,\oplus_{s\in\Z}\cO(2s-2))$. On spacetime this theory describes a self-dual higher spin theory whose spectrum includes a scalar, sd Maxwell theory and sd gravity. It suffers from a tower of twistorial anomalies, all of which are proportional to a count of the bosonic degrees of freedom in the theory. Regularised via the Riemann $\zeta$-function, this count gives
\be \frac{1}{2} + \sum_{s\geq1} 1\, \overset{{\rm reg}}{=}\, \frac{1}{2} + \zeta(0) = 0\,. 
\label{zeta_reg_sum}
\ee
Recently a non-commutative variant of this theory was proposed in \cite{Tran:2022tft}, in which the Moyal product determined by the Poisson bracket $\{\ ,\ \}$ is switched on. Note that since the Poisson bracket has weight, this is only possible in the higher spin set-up described above. We expect that the twistorial anomalies in this theory vanish for essentially the same reason. 1-loop $n$-point amplitudes in the spacetime self-dual higher spin theory were computed in~\cite{Skvortsov:2020gpn}; they are also proportional to the sum in~\eqref{zeta_reg_sum}.\\

In this section, we prefer to explore an anomaly cancellation mechanism generalising the results of Costello~\cite{Costello:2021bah} and Costello-Paquette~\cite{Costello:2022wso} on the pure holomorphic BF case. It will therefore be useful to review this case first.

%%%%%%%%%%%%%%%%%%%%%%%%%%%%%%%%%%

\subsection{Anomaly cancellation in holomorphic BF theory} \label{subsec:HBF-cancellation}

To cancel the pure gauge anomaly, Costello uses a six-dimensional version of the Green-Schwarz mechanism, adapted to holomorphic BF theory. To do this, he introduces a new  
 $(2,1)$-form $\eta$ obeying the constraint $\p\eta=0$. That is, $\eta\in\Omega^{0,1}(\PT,\Omega^2_{\rm cl})$ where $\Omega^2_{\rm cl}$ is the sheaf of $\p$-closed $(2,0)$-forms. The kinetic term for $\eta$ and its coupling to holomorphic BF theory are given by\footnote{Although we work in the analytically continued setting, our conventions are chosen so that if we required all fields to be equivariant under Euclidean spinor conjugation then our actions would be real. This induces the standard reality conditions on Euclidean $\R^4$. For this reason we take the integrand in the path integral to be
\[ \exp(\im S^\prime[\eta;b,a]) = \exp(\im S[b,a] - S_\eta[\eta;a])\,. \]
The relative factor of $\im$ here also affects $\delta b$ in equation~\eqref{eq:HBF-eta-transformation}. We adopt analogous conventions for Poisson-BF theory.}
\be \label{eq:HBF-eta-action}
S_\eta[\eta;a] = \frac{1}{4\pi\im}\int_\PT\p^{-1}\eta\,\bar\p\eta 
+ \lambda_\fg\,\eta\,\tr(a\p a)
\ee
where the kinetic term involving $\p^{-1}\eta$ makes sense (at least locally) because $\p\eta=0$. This kinetic terms reveals that $\eta$ is closely related to the closed string mode of the topological B-model. The interaction term involves a constant $\lambda_\fg$ that depends on the Lie algebra $\fg$ of the gauge theory, to be fixed below by the requirement that this interaction allows us to cancel the gauge anomaly. We will sometimes refer to the coupling between $\eta$ and $a$ in~\eqref{eq:HBF-eta-action} as the \emph{gauge counterterm}. 

\begin{figure}[!ht]
\centering
  \includegraphics[scale=0.3]{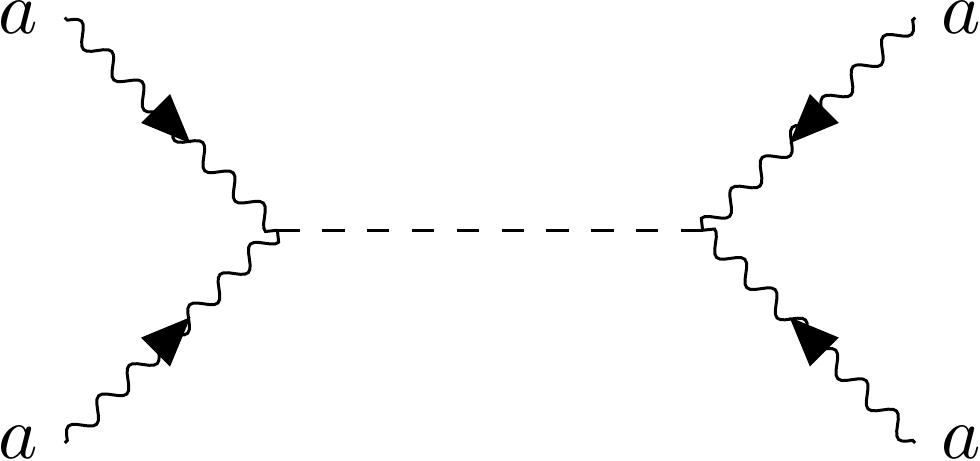}
\caption{\emph{Diagram involving a propagator for the $\eta$ field. The gauge variation of this diagram compensates the holomorphic BF theory anomaly.}} \label{fig:HBF-cancellation}
\end{figure}

The field $\eta$ transforms under BRST transformations of the BF theory, and also has its own BRST transformations with ghost $\theta\in\Omega^2_{\rm cl}$. The presence of $\eta$ also modifies the transformations of $b$. Altogether, we have 
\be \label{eq:HBF-eta-transformation}
\begin{split}
    \delta \eta &= \bar\p\theta - \lambda_\fg\,\tr(\p c\p a)\,, \\
    \delta a & = \bar\p c + [a,c]\\
    \delta b &= \bar\p d + [a,d] - [b,c] - \im\lambda_\fg\,(\theta \p a + \eta \p c)\,,
\end{split} 
\ee
where $c$ are the ghosts for standard gauge transformations while $d$ are the ghosts for the transformations~\eqref{gaugeBFgaugetrans2} of the $b$-field. The modification of the BRST transformation of $b$ is necessary to compensate for the variation of $S_\eta$ under standard BRST transformations of $a$, as well as the $\bar\p\theta$ term in the BRST transformation of $\eta$. However, the interaction in~\eqref{eq:HBF-eta-action} is not invariant under the non-linear part of the transformation $\delta\eta$ in~\eqref{eq:HBF-eta-transformation}. On the support of the classical field equations, one finds that the BRST transformation of the classical action $S^\prime[\eta;b,a]=S[b,a] + \im S_\eta[\eta;a]$ is
\be \label{eq:HBF-eta-variation} 
\delta S^\prime[\eta;b,a]= - \frac{\lambda_\fg^2}{2\pi}\int_\PT\tr(c\p a)\,\tr((\p a)^2)\,.
\ee
Equivalently, we can see the classical failure of gauge invariance~\eqref{eq:HBF-eta-variation} by evaluating the tree-level diagram shown in figure \ref{fig:HBF-cancellation}. This is straightforward using the identity
\be \label{eq:eta-prop}
    \bar\p_1P^\eta_{12} = -2\pi \im\,\p_1\delta_\PT
\ee
 for the $\eta$-propagator $P^\eta_{12}$, where $\delta_\PT$ is the 6-form $\delta$-function on the diagonal\footnote{Technically, we restrict to the $(1,2)$- and $(2,1)$-form parts of this $\delta$-function on the $1^{\rm st}$ and $2^{\rm nd}$ factors of $\PT_1\times\PT_2$, respectively.} of $\PT_1\times\PT_2$.

In~\eqref{eq:HBF-anomaly} we showed that quantum holomorphic BF theory also failed to be gauge invariant, with the phase of the partition function changing under a BRST transformation by
\be
	\frac{1}{3!}\bigg(\frac{\im}{2\pi}\bigg)^3\int_\PT\tr_\ad(c(\p a)^3)\,. 
\ee
In order for the classical failure of BRST invariance~\eqref{eq:HBF-eta-variation} to compensate for this, we must have the identity
\be \label{eq:quartic-trace} 
\tr_{\rm ad}(X^4)= C_\fg\,\tr(X^2)^2 
\ee
for all $X\in\fg$ for some (Lie algebra dependent) constant $C_\fg$. This identity holds when $\fg$ is $\fsl_2(\C)$, $\fsl_3(\C)$ or any of the exceptional Lie algebras as a consequence of the fact that the space of quartic invariants is 1 dimensional for these Lie algebras~\cite{Okubo:1978qe}. It also holds for $\fso_8(\C)$, despite this algebra having a 3 dimensional space of quartic invariants, on account of triality. In these cases the constant $C_\fg$ is given by~\cite{Okubo:1981td}
\be 
	C_\fg = \frac{5(2\bh^\vee)^2}{2(2+\dim\fg)}
\ee
with $\bh^\vee$ the dual Coxeter number of $\fg$. Thus, for these choices of $\fg$, the gauge anomaly~\eqref{eq:HBF-anomaly} is cancelled by the classical variation~\eqref{eq:HBF-eta-transformation}, provided we choose the coupling to be
\be \label{eq:HBF-cancellation} 
	\lambda_\fg^2=\frac{C_\fg}{3!}\bigg(\frac{\im}{2\pi}\bigg)^2\,.
\ee
It is also important to note that, since the theory remains 1-loop exact after coupling to $\eta$, there are no further failures of gauge invariance at higher loop order. Thus, as shown by Costello in~\cite{Costello:2021bah}, for these gauge groups holomorphic BF theory coupled to $\eta$  exists at the quantum level as a theory on twistor space.\\

We remark that, on a Calabi-Yau 3-fold, the 1-loop anomaly in holomorphic BF theory agrees (up to a factor of $1/2$) with the 1-loop anomaly in holomorphic Chern-Simons theory. Similarly, the tree-level diagram cancelling the anomaly coincides with the corresponding diagram in BCOV theory~\cite{Bershadsky:1993cx,Costello:2019jsy} (sometimes referred to as Kodaira-Spencer gravity). In this way, the anomaly cancellation mechanism above is analogous to the Green-Schwarz anomaly cancellation of the topological open-closed B-model, here adapted to a complex 3-fold (twistor space) that is not Calabi-Yau.

%%%%%%%%%%%%%%%%%%%%%%%%%%%%%%%%%%

\subsection{Anomaly cancellation in Poisson-BF theory} 
\label{subsec:PBF-cancellation}

We now show that the same $\eta$ field can cancel the gravitational anomaly in Poisson-BF theory. Analogous to the holomorphic BF case, we couple $\eta$ to Poisson-BF theory using the action
\be \label{eq:PBF-eta-action} 
S^\prime[\eta;g,h]=\frac{1}{2\pi\im}\int_\PT g\wedge T^{2,0}(h)
+ \frac{1}{4\pi}\int_\PT\p^{-1}\eta\,\nbar\eta + \mu\,\eta\,\tr( s\p s)\,. 
\ee
where the coupling $\mu$ will be determined below. As before, $s^{\dot\alpha}_{~\,\dot\beta} = - \p^{\dot\alpha}\p_{\dot\beta}h$ is the partial connection on $T_\cPT$ induced by the Hamiltonian deformation of the complex structure. In the kinetic term for $\eta$  we define
\be
	\nbar\eta = \bar\p\eta + \cL^{1,0}_{\{h,\ \}}\eta\,,
\ee
where the $(1,0)$ Lie derivative is defined by $\cL^{1,0}_V = [V\ip\,,\p\,]$. We use a commutator here since $V = \{h,\ \}$ is a $(0,1)$-form valued vector field. It therefore acts on $\eta$ by $\cL^{1,0}_V\eta = - \p(V\ip\eta)$, allowing us to rewrite the $\eta$ kinetic term as
\be
\frac{1}{4\pi}\int_\PT\p^{-1}\eta\,\nbar\eta = \frac{1}{4\pi}\int_\PT\p^{-1}\eta\,\bar\p\eta + \eta\,\{h,\ \}\ip\eta\,. \ee
From now on we will suppress the $(1,0)$ Lie derivative and just write the vector field so that, {\it e.g.}, $\nbar\eta = \bar\p\eta + \{h,\eta\}$. The fact that the $\eta$ kinetic term must be covariantized in the presence of $h$ can be viewed as an example of the twistor version of the statement that all fields couple to gravity. We will sometimes refer to the final term in equation \eqref{eq:PBF-eta-action} as the \emph{Poisson counterterm}.

Just as in the gauge theory case, the coupling to $\eta$ introduces a modification of the BRST transformation of $g$, as well as introducing its own ghost $\theta$. We have the BRST transformations
\bea \label{eq:PBF-eta-transformation}
    \delta h &= \nbar\chi = \bar\p\chi + \{h,\chi\}\,,\\
    \delta g &= \nbar\phi - \{g,\chi\} + \frac{\im}{2}\p_\dal(\theta\p^\dal\ip\eta + \eta\p^\dal\ip\theta) + \im\mu\,\Big((\p^\db\p_\dal\eta)(\p\psi^\dal_{~\,\db}) +(\p^\db\p_\dal\theta)(\p s^\dal_{~\,\db})\Big)\\
    \delta\eta &= \nbar\theta - \mu\,\tr(\p\psi\p s)
\eea
where $\psi^\dal_{~\,\db} = - \p^\dal\p_\db\chi$. Since $\eta$ and $\theta$ are $\p$-closed, the $(1,0)$ Lie derivatives in \eqref{eq:PBF-eta-transformation} act as
\be \p^\dal\p_\db\eta = \cL^{1,0}_{\p^\dal}\cL^{1,0}_{\p_\db}\eta = \p(\p^\dal\ip\p(\p_\db\ip\eta))\,. \ee At the quantum level, the action should also include ghost and gauge fixing terms, and the corresponding ghosts, antighosts and Nakanishi-Lautrup fields have their own BRST transformations. These are best described using the BV formulation of the theory, discussed in subsection~\ref{subsec:BV-eta}.

\begin{figure}[t!]
\centering
  \includegraphics[scale=0.3]{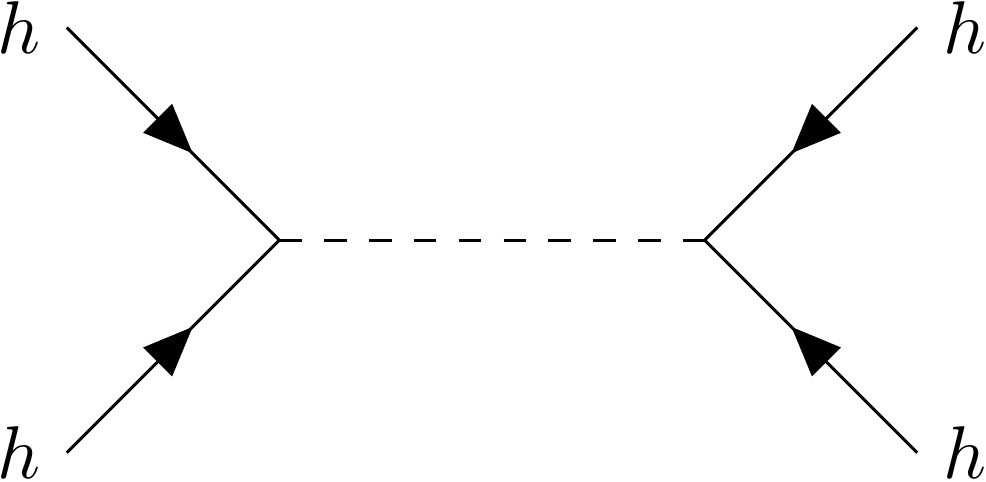}
\caption{\emph{Diagram involving a propagator for the $\eta$ field. The variation of this diagram under Hamiltonian diffeomorphisms $\delta h = \nbar\chi$ compensates the diffeomorphism anomaly of holomorphic Poisson-BF theory.}} \label{fig:PBF-cancellation}
\end{figure}

As in the gauge theory, the classical action~\eqref{eq:PBF-eta-action} is not quite invariant under these BRST transformations. On-shell, one finds
\be \label{eq:PBF-eta-variation}
	\delta S^\prime[\eta;g,h] = - \frac{\mu^2}{2\pi}\int_\PT\tr(\psi\p s)\,\tr((\p s)^2)\,. 
\ee
Again, this can also be viewed as failure of the tree-level diagram in figure \ref{fig:PBF-cancellation} to be invariant under the linear part of $\delta h = \nbar\chi$. However, in contrast to holomorphic BF theory, the fact that the kinetic term for $\eta$ also couples to $h$ modifies the anomaly in the partition function, because $\eta$ itself can now run around the loop, as in figure~\ref{fig:eta-anomaly}. This loop diagram is evaluated in appendix~\ref{app:eta-box}, but it's easy to understand the result via a heuristic argument. Since the $\eta$ propagators are not directed, the anomaly diagram has a $\Z_2$ reflection symmetry. This suppresses its contribution by a factor of $1/2$ compared to the Poisson anomaly diagram~\ref{fig:eta-anomaly}. On the other hand, as a $(2,1)$-form with values in $\Omega^2_{\rm cl}$, $\eta$ carries two $(0,1)$-form degrees of freedom: three for a generic $(2,1)$-form, less one for the $\p\eta=0$ constraint. This factor of 2 cancels the symmetry factor of $1/2$. We therefore expect that the BRST variation of the diagram in figure~\ref{fig:eta-anomaly} to be exactly equal to the Poisson anomaly given in equation~\eqref{eq:PBF-anomaly} in the absence of the gauge contribution. This is borne out by our explicit calculation.

\begin{figure}[!ht]
\centering
  \includegraphics[scale=0.3]{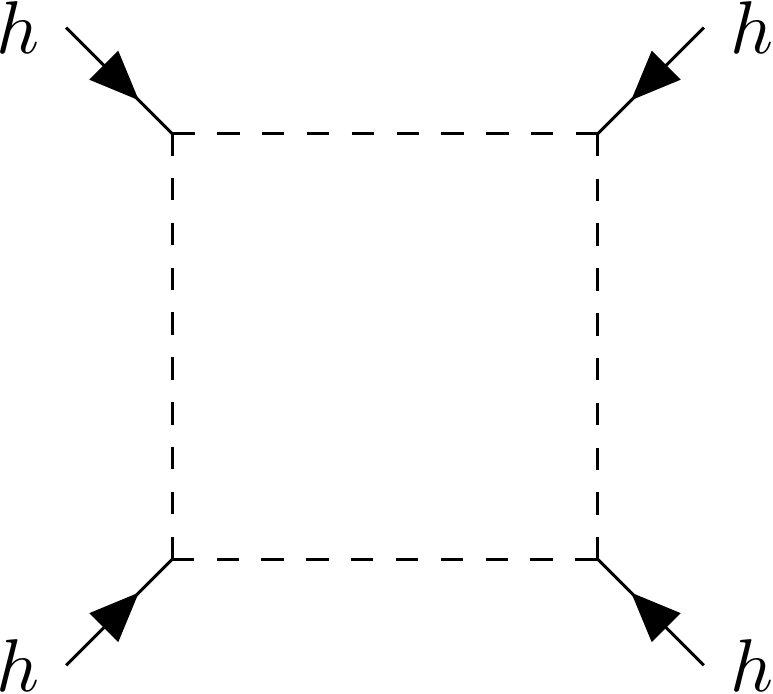}
\caption{\emph{In holomorphic Poisson-BF theory, $\eta$ also contributes to the 1-loop anomaly.}} \label{fig:eta-anomaly}
\end{figure}

The total 1-loop anomaly in Poisson-BF theory coupled to $\eta$ is thus twice the result of the gravitational anomaly in Poisson-BF theory itself. It can be written as
\be 
	\frac{2}{5!}\bigg(\frac{\im}{2\pi}\bigg)^3\int_\PT\tr(\psi(\p s)^3) = 
	\frac{1}{5!}\bigg(\frac{\im}{2\pi}\bigg)^3\int_\PT\tr(\psi\p s)\,\tr((\p s)^2)\,, 
\ee
where the second expression uses the $\fsl_2$ trace identity $\tr(X^4) = \frac{1}{2}\tr(X^2)^2$. This total loop anomaly can be cancelled by the exchange diagram in figure~\ref{fig:PBF-cancellation}, or equivalently the classical variation~\eqref{eq:PBF-eta-variation}  by taking
\be 
	\mu^2 = \frac{1}{5!}\bigg(\frac{\im}{2\pi}\bigg)^2\,. 
\ee
It is important to notice that the counterterm coupling of $\eta$ cannot appear in any loop diagram, so that this theory remains 1-loop exact. Thus there are no higher loop anomalies, and we conclude that holomorphic Poisson-BF theory coupled to $\eta$ as in~\eqref{eq:PBF-eta-action} exists as a quantum theory on twistor space.

Given that the anomaly cancellation in holomorphic BF theory can be interpreted as a Green-Schwarz mechanism, it is natural to speculate that the cancellation of the Poisson-BF anomaly also has a string theory counterpart. We leave this possibility for future work.

%%%%%%%%%%%%%%%%%%%%%%%%%%%%%%%%%%

\subsection{Anomaly cancellation in the coupled gauge-gravitational theory} 
\label{subsec:PGBF-cancellation}

Finally, we consider including $\eta$ in the coupled gauge-gravitational theory. The complete action is
\bea \label{eq:PGBF-eta-action} 
S^\prime[\eta;b,a;g,h]&=\frac{1}{2\pi\im}\int_\PT g\wedge T^{0,2}(h) + b\wedge\cF^{0,2}(a)
\\
&\qquad+\frac{1}{4\pi}\int_\PT\p^{-1}\eta\,\nbar\eta
+ \lambda_\fg\,\eta\,\tr(a\p a) + \mu\,\eta\,\tr( s\p s)\,,
\eea
where the first line  describes holomorphic BF theory coupled to holomorphic Poisson-BF theory as above (involving $\nbar a= \bar\p a + \{h,a\}$), and where the second line consists of the sum of the $\eta$ interaction we met in the holomorphic and Poisson-BF theories separately. 
The full BRST transformations of the physical fields in this theory are
\begin{subequations}
\bea \label{eq:PGBF-eta-transformation-1}
     \delta a &= \nbar c + [a,c] + \{a,\chi\} = \bar\p c + \{h,c\} + [a,c] + \{a,\chi\}\,,\\
    \delta b &= \nbar d + [a,d] - \{b,\chi\} - [b,c] - \im\lambda_\fg\,(\theta\p a+\eta\p c) \\
\eea
for the fields of holomorphic BF theory,
\bea \label{eq:PGBF-eta-transformation-2}
    \delta h &= \bar\p\chi + \{h,\chi\}\,,\\
    \delta g &= \nbar\phi - \{g,\chi\} - \{b,c\} + \{a,d\}\\
    &\qquad  + \frac{\im}{2}\p_\dal(\theta\p^\dal\ip\eta + \eta\p^\dal\ip\theta) + \im\mu\,\Big((\p^\db\p_\dal\eta)(\p \psi^\dal_{~\,\db}) +(\p^\db\p_\dal\theta)(\p s^\dal_{~\,\db})\Big)
\eea
for the gravitational fields of Poisson-BF theory, and
\be\label{eq:PGBF-eta-transformation-3}
   \delta\eta = \bar\p\theta - \lambda_\fg\,\tr(\p c\p a) - \mu\,\tr(\p\psi\p s)
\ee
\end{subequations}
for $\eta$. These are just the BRST transformations of $(b,a;g,h)$ in the coupled gauge-Poisson-BF theory, together with the sum of the $\eta$ terms introduced above. Once again, the BRST transformations of these fields and the associated ghosts and antighosts are best understood in the BV formalism, to be discussed in~\ref{subsec:BV-eta}.\\

The coupling to gravitational fields does not affect the pure gauge anomaly, so to cancel this using the $\eta$ exchange diagram we must again choose $\fg=\fsl_2$, $\fsl_3$, $\fso_8$ or any of the exceptional Lie algebras and tune
\be
	\lambda_\fg^2 = \frac{C_\fg}{3!}\bigg(\frac{\im}{2\pi}\bigg)^2\,.
\ee
Similarly, the coupling $\mu$ is fixed by requiring that BRST variation of the tree-level diagram in figure~\ref{fig:PBF-cancellation} cancels the pure gravitational anomaly. In the theory~\eqref{eq:PGBF-eta-action} this anomaly receives contributions from any of the gravitational, gauge or $\eta$ fields running around the loop. Thus, we must now choose 
the coupling $\mu$ to obey
\be \label{eq:PBF-cancellation} 
	\mu^2 = \frac{2+\dim\fg}{2\cdot5!}\bigg(\frac{\im}{2\pi}\bigg)^2\,. 
\ee
to cancel the pure gravitational anomaly in coupled gauge-Poisson-BF theory. There is no more freedom to tune couplings in this model.

\begin{figure}[t!]
\centering
  \includegraphics[scale=0.3]{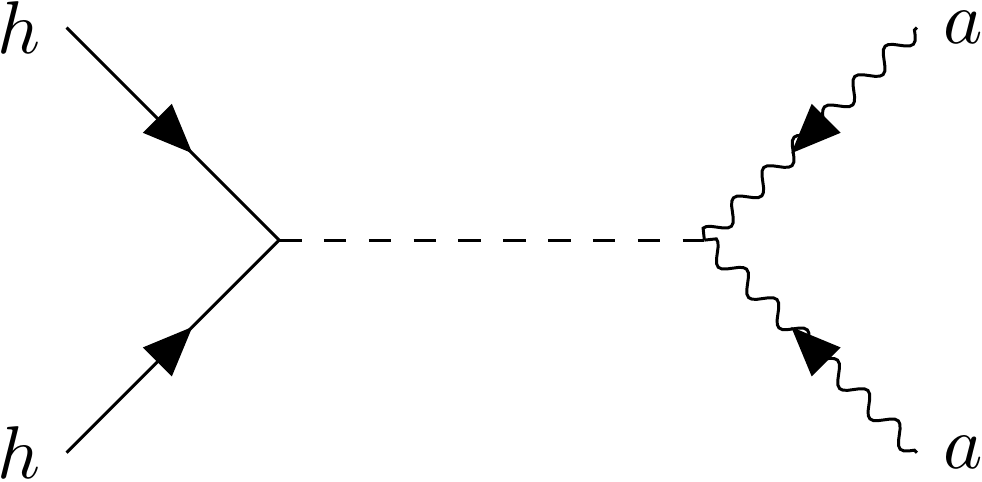}
\caption{\emph{Exchange diagram compensating the mixed gauge-gravitational anomaly.}} \label{fig:PGBF-mixed-cancellation}
\end{figure}

Remarkably, exactly these values of the couplings also cancel the mixed gauge-gravitational anomaly. The BRST variation of the tree-level diagram shown in figure~\ref{fig:PGBF-mixed-cancellation} is readily seen to be
\be \label{eq:PGBF-eta-variation}
-\frac{\lambda_\fg\mu}{2\pi}\int_\PT\tr_\ad(c\p a)\,\tr((\p s)^2) 
-\frac{\lambda_\fg\mu}{2\pi}\int_\PT \tr_\ad((\p a)^2)\,\tr(\psi\p s)\,.
\ee
Comparing to the mixed anomaly~\eqref{eq:PGBF-mixed-cocycle} we can see that the condition for cancellation of the mixed Poisson-gauge anomaly is
\be \label{eq:PGBF-mixed-cancellation} 
	\lambda_\fg\mu = -\frac{2\bh^\vee}{4!}\bigg(\frac{\im}{2\pi}\bigg)^2\,. 
\ee
Consistency of equations \eqref{eq:HBF-cancellation}, \eqref{eq:PBF-cancellation} and \eqref{eq:PGBF-mixed-cancellation} now requires that
\be
\frac{C_\fg}{3!}\bigg(\frac{2+\dim\fg}{2\cdot 5!}\bigg)\bigg(\frac{\im}{2\pi}\bigg)^4 = \lambda_\fg^2\mu^2 = (\lambda_\fg\mu)^2 = \bigg(-\frac{2\bh^\vee}{4!}\bigg)^2\bigg(\frac{\im}{2\pi}\bigg)^4\,.
\ee
This indeed holds as a consequence of the condition
$C_\fg = 5(2\bh^\vee)^2/2(2+\dim\fg)$ appearing in the trace identity $\tr_{\rm ad}(X^4) = C_\fg\,\tr(X^2)^2$ that we met in cancelling the pure gauge anomaly~\cite{Okubo:1978qe}. 

It is very striking that a single $\eta$ field, with just two free couplings, is able to  simultaneously cancel the pure gauge, gravitational and mixed anomalies on twistor space, for any choice of $\fg$ where the gauge anomaly itself cancels. We emphasise that the fact that the kinetic term of $\eta$ necessarily couples to gravity was crucial for this to work: This covariant coupling ensures that, on-shell, $\eta$ represents an element of $H^1(\mathcal{PT},\Omega^2_{\rm cl})$ even on a curved twistor background $\mathcal{PT}$. Had we na{\"i}vely  used the kinetic term $\int_\PT \p^{-1}\eta\,\bar\p\eta$ without coupling to the background complex structure, $\eta$ would not contribute to the 1-loop gravitational anomaly. This in turn would have caused a mismatch between the values of $\mu$ and $\lambda_\fg$ required
to cancel the pure gauge and pure gravitational anomalies, and those required to cancel the mixed anomaly. The mismatch would be a factor of $(1+\dim\fg)/(2+\dim\fg)$ which is non-trivial for any finite dimensional gauge group.

%%%%%%%%%%%%%%%%%%%%%%%%%%%%%%%%%%

\subsection{The classical BV description of \texorpdfstring{$\eta$}{eta}}
\label{subsec:BV-eta}

For many purposes, it is also useful to describe $\eta$ in the BV framework. To do this, we promote it to a BV superfield $\bseta\in\Omega^{0,\bullet}(\PT,\Omega^2_{\rm cl})[1]$. In addition to~\eqref{eq:BV-PGBF-action}, the BV action for the coupled theory including $\bseta$ contains
\be \label{eq:BV-PGBF-eta-action}
S_\eta[\boldsymbol{\eta};\ba;\bh] 
= \frac{1}{4\pi\im}\int_\PT\p^{-1}\bseta\,\nbar\bseta + \lambda_\fg\,\bseta\,\tr(\ba\p\ba) + \mu\,\bseta\,\tr(\bs\p\bs)\,,
\ee
where $\bs^\dal_{~\,\db} = - \p^\dal\p_\db\bh$. We extend the odd symplectic form by
\be
\omega = \frac{1}{4\pi\im}\int_\PT \delta\bseta\,\p^{-1}(\delta\bseta)\,.
\ee
This leads to the same antibracket for $\bseta$ as used in the closed string sector of the B-model~\cite{Costello:2019jsy}. As in section \ref{sec:GS-mechanism}, the new terms in the action modify the BRST transformations of $\bb,\bg$ to
\bea \label{eq:BV-PGBF-eta-transform-1}
    \delta\bb &= \{S^\prime,\bb\}_{\rm BV} = \bar\p\bb + [\ba,\bb] - \im\lambda_\fg\,\boldsymbol{\eta}\p\ba\,, \\
    \delta\bg &= \{S^\prime,\bg\}_{\rm BV} \\ &=\bar\p\bg + \{\bh,\bg\} + \{\ba,\bb\} + \frac{i}{2}\p_\dal(\bseta\p^\dal\ip\bseta) + \im\mu\,\big(\p^\db\p_\dal\bseta\big)\big(\p\bs^\dal_{~\,\db}\big)\,.
\eea
Here, as explained in subsection \ref{subsec:PBF-cancellation}, $(1,0)$ vector fields act by $(1,0)$ Lie derivative, $\cL^{1,0}_V = [V\ip\,,\p\,]$. We also have BRST transformations of $\bseta$
\be \label{eq:BV-PGBF-eta-transform-2}
    \delta\bseta = \{S^\prime,\bseta\}_\mathrm{BV} = \nbar\bseta - \frac{\lambda_\fg}{2}\,\tr(\p\ba^2) - \frac{\mu}{2}\,\tr(\p\bs^2)\,.
\ee
The transformations of $\ba$ and $\bh$ are unaltered (because the new couplings do not involve $\bb$ or $\bg$). The failure of the combined action $S^\prime[\bseta;\bb,\ba;\bg,\bh] = S[\bb,\ba;\bg,\bh] + \im S_\eta[\bseta;\ba;\bh]$ to be classically BRST invariant is then
\bea \label{eq:BV-PGBF-eta-variation}
&\delta S^\prime = \{S^\prime,S^\prime\}_\mathrm{BV} \\
&= - \frac{1}{8\pi}\int_\PT\big(\lambda_\fg^2\,\tr(\ba\p\ba)\tr((\p\ba)^2) + 2\lambda_\fg\mu\,\tr(\ba\p\ba)\tr((\p\bs)^2) + \mu^2\,\tr(\bs\p\bs)\tr((\p\bs)^2)\big)\,.
\eea
Restricting to the physical and ghost fields $\ba = a + c$, $\bs = \psi + s$ we recover equations \eqref{eq:HBF-eta-variation}, \eqref{eq:PBF-eta-variation} and \eqref{eq:PGBF-eta-variation}. Note that, in contrast to what we found in subsections \ref{subsec:HBF-cancellation}, \ref{subsec:PBF-cancellation} and \ref{subsec:PGBF-cancellation}, equation \eqref{eq:BV-PGBF-eta-variation} holds off-shell.

%%%%%%%%%%%%%%%%%%%%%%%%%%%%%%%%%%
%%%%%%%%%%%%%%%%%%%%%%%%%%%%%%%%%%

\section{Reduction to space-time}
\label{sec:space-time-theory}

In this section we argue that Poisson-BF theory coupled to $\eta$ descends to self-dual gravity on space-time coupled to a scalar field $\rho$ with a 4$^{\rm th}$-order kinetic term, that plays the role of a gravitational axion. 
%\be 
%S[e,\Gamma;\rho] = \frac{1}{2}\int_{\R^4}\big(\Gamma_{\al\beta}\wedge\d(e^{\dal\al}\wedge e_{\dal}^{~\,\beta}) + \vol_g(\Delta_g\rho)^2 + \mu^\prime\rho{\tilde R}^\dal_{~\,\db}\wedge \tilde R^\db_{~\,\dal}\big)\,,
%\ee
%where $e^{\al\dal}$ is the vierbein 1-form and $\Gamma_{\alpha\beta}$ is a Lagrange multiplier.  $\tilde R$ is the curvature of the $\fsl_2(\C)_+$ part of the spin connection $\omega$
%\be {\tilde R}^{\dot\alpha}_{~\,\dot\beta} = \d\omega^{\dot\alpha}_{~\,\dot\beta} + \omega^{\dot\alpha}_{~\,\dot\gamma}\wedge\omega^{\dot\gamma}_{~\,\dot\beta} \ee
%where $\omega$ is defined by
%\be \d e^{\dal\al} + \omega^\dal_{~\,\db}\wedge e^{\db\alpha} + \omega^\al_{~\,\beta}\wedge e^{\dal\beta} = 0\,. \ee
%We further determine the value of the coupling $\mu^\prime$ by requiring the all-plus 4-pt 1-loop graviton amplitude vanishes, and then argue that for generic external momenta the theory has trivial amplitudes in all signatures.

%%%%%%%%%%%%%%%%%%%%%%%%%%%%%%%%%%

\subsection{Poisson-BF theory on twistor space as self-dual gravity on space-time} \label{subsec:PBF-to-sdGR}

In this section we show how, at the classical level, Poisson-BF theory on twistor space can be reduced to an action for self-dual gravity on space-time. This reduction has been performed before in~\cite{Sharma:2021pkl}, but the discussion here provides an alternative and more explicit presentation.\\

%{\color{blue} Please check all signs in the following!} 
Consider the deformed almost complex structure $\bar{\nabla}=\bar\p + \{h,\ \}$ that, off-shell, will not be integrable. We construct a set  $\{e^0,\theta^{\dal}\}$ of weighted 1-forms on twistor space that obey
\be
    \nbar\,\lrcorner\, e^0 =0\,,\qquad\qquad\nbar\,\lrcorner\,\theta^{\dal}=0\,.
\ee
Specifically, since the deformation preserves the fibration over $\CP^1$, we can choose 
\begin{subequations}
\label{pseudo_basis}
\begin{equation}
    e^0 = \la\lambda\d\lambda\ra
\end{equation}
to be the $(1,0)$-form of weight 2 on $\CP^1$, while
\begin{equation}
    \label{deformed_hol_theta}
    \theta^{\dal} = \d\mu^{\dal} - \p^{\dal}h = \d\mu^{\dal} -\epsilon^{\dal{\dot\beta}}\frac{\p h}{\p\mu^{\dot\beta}}
\end{equation}
\end{subequations}
are 1-forms of weight 1 on the fibres of $\CPT\to\CP^1$. At each point of twistor space, these forms form a basis of (weighted) forms that are of type $(1,0)$ in the deformed almost complex structure.

The off-shell space-time $M$ is the moduli space of pseudo-holomorphic curves in the deformed almost complex structure that are invariant under a prescribed antiholomorphic involution on $\CPT$.\footnote{As we work analytically on space-time, we will not need to make this involution explicit.} These curves generate a diffeomorphism $p:M\times\CP^1\to\CPT$ that satisfies \cite{McDuff:2012}
\be
\bar\p_{\sigma}\ip p^*\theta^{\dal} = 0\,,\qquad\bar\p_{\sigma}\ip p^*e^0=0\,,
\ee
where $\bar\p_\sigma$ is the antiholomorphic vector tangent to each curve. The second of these conditions just says that the twistor co-ordinate $\lambda_\alpha(x,\sigma,\hat{\sigma})$ must be holomorphic  along each curve, and since it has homogeneity 1, Liouville's theorem fixes $\lambda_\alpha=\sigma_\alpha$, up to a possible $\GL(2,\C)$ transformation. (In general, this $\GL(2,\C)$ transformation can be $x$-dependent. It amounts to a choice of frame for the undotted spin bundle on $M$.) %{\color{blue} Fill in why can fix this frame as trivial}.) 
Thus, $\bar\partial_\sigma\ip p^*e^0=0$ amounts to saying that we can use $\lambda$ instead of $\sigma$ as the curve parameter. With this understood, the pseudo-holomorphic curves are determined by the non-linear pde
\be
\bar\p_0\ip p^*\theta^{\dal} = 0\,,
\ee
where ${\bar\p}_0=\la\lambda\hat\lambda\ra\,\lambda_\alpha\,\p / \p\hat{\lambda}_{\alpha}$ is an antiholomorphic vector along the base of $\CPT\to\CP^1$.\\

To obtain an action on $M$, we wish to reduce the twistor space action~\eqref{PoissonBFaction} along these pseudo-holomorphic curves. This is the analogue of fixing a gauge along the curves $\mu^{\dal}=x^{\dal\alpha}\lambda_\alpha$ of flat twistor space in the case of self-dual Yang-Mills theory. However, it is important to note that the diffeomorphism $p$ is neither holomorphic, nor will it preserve the Poisson structure. Thus, to make use of these curves we must first re-express~\eqref{PoissonBFaction} in terms of quantities that do not explicitly refer to this Poisson structure. It is remarkable that this can be achieved.\\

To do so, consider the 1-form of weight 2
\be
\label{symplectic_pot_def}
    \vartheta = \frac{1}{2}[\mu\,\d\mu] - h\,.
\ee
This is the symplectic potential for the twistor sigma models of \cite{Adamo:2021bej}. Note that it is \emph{not} of type $(1,0)$ in the deformed almost complex structure (i.e., is not in the span of $e^0,\theta^{\dal}$), so only provides a locally defined potential. It obeys the structure equation
\be
\label{vartheta_structure}
\begin{split}
    &\d\vartheta = \frac{1}{2}\d\mu^{\dal}\wedge\d\mu_{\dal} - \p h - \,\dbar h\\
    &= \frac{1}{2}\left(\d\mu^{\dal}-\p^{\dal}h\right)\wedge\left(\d\mu_{\dal}-\p_{\dal}h\right)-T -e^0\wedge\p_0 h \\
    &= \frac{1}{2}\theta^{\dal}\wedge\theta_{\dal} -T -e^0\wedge\p_0h\,,
\end{split}
\ee
where $T=\bar\p h+ \frac{1}{2}\p^{\dal}h\wedge\p_{\dal}h$ as before. We conclude that
\be
\label{dzdrho}
 e^0\wedge\d\vartheta = e^0\wedge\left(\Sigma-\,T\right)\,,
\ee
where we have introduced the weight 2 form
\be
\label{symplectic_form_def}
    \Sigma = \frac{1}{2}\theta^{\dal}\wedge\theta_{\dal}
\ee
that has type (2,0) in the deformed almost complex structure.  Alternatively, \eqref{dzdrho} may be phrased globally on $\CPT$ as 
\be
\d(\Sigma-T) = 0\qquad\text{mod}\ e^0\,,
\ee
which is a type of Bianchi identity. This form can also be derived by subtracting twice the Bianchi identity $\dbar T + \{h,T\} = 0$ for $T$ from the structure equation for $\Sigma$. Thus, even off-shell, $\Sigma-T$ is closed modulo $e^0$, so forms a symplectic form of weight 2 on the fibres of $\CPT\to\CP^1$, with $\vartheta$ the associated symplectic potential, defined up to an exact form. However, $\Sigma-T$ is not of type $(2,0)$, and $(\Sigma-T)^2\neq0$ unless the Poisson-BF equations of motion $T=0$ hold.

To make use of this, note that since the Poisson-BF action depends only on $g\wedge T$, we are free to make the replacement
\be
\label{g_expand}
 g = \frac{1}{2}e^0\wedge\d\mu^\dal\wedge\d\mu_\dal\wedge\tilde{g} \to e^0\wedge\Sigma\wedge\tilde{g}\,,
\ee
for some $\tilde g\in\Omega^{0,1}(\PT,\cO(-6))$. This gives
\be
\label{BF_action_manipulate}
   S[g,h] = \frac{1}{2\pi\im}\int_{\CPT} e^0\wedge \tilde g\wedge\Sigma\wedge T = -\frac{1}{4\pi\im}\int_{\CPT}e^0\wedge\tilde{g}\wedge(\Sigma-T)\wedge(\Sigma-T)\,.
\ee
The terms we have added in the second going to the second line both vanish: $\Sigma\wedge\Sigma=0$ as it would be a $(4,0)$-form in the deformed almost complex structure and $T\wedge T=0$ since it would be a $(0,4)$-form in the background complex structure. The form of the action given in~\eqref{BF_action_manipulate} makes contact with the discussion above; we see that the equations of motion for $g$ are equivalent to the statement $(\Sigma-T)^2=0$ mod $e^0$. The purpose of recasting the action this way is that, identifiying $\Sigma-T= \d\vartheta$ as above, the Lagrangian in~\eqref{BF_action_manipulate} is built purely from geometric forms and in particular does not refer to the Poisson structure. Thus
\begin{equation}
\label{covariant_twistor_action}
S[\tilde g,\vartheta]=-\frac{1}{4\pi\im}\int\limits_{\CPT} e^0\wedge\tilde{g}\wedge \d\vartheta\wedge\d\vartheta = -\frac{1}{4\pi\im}\int\limits_{M\times\CP^1} e^0\wedge p^*\left(\tilde{g}\wedge \d\vartheta\wedge\d\vartheta \right)\,,
\end{equation}
where $p:M\times\CP^1\to \CPT$ is our non-Poisson diffeomorphism and we have used the fact that $p^*e^0=e^0$.

We are now in position to reduce the action. Decompose
\be
p^*\tilde{g} = \bar{e}^0\tilde{g}_{0} + \tilde{g}_{\rm vert}\,,\qquad\text{where}\qquad \bar\p_{0}\ip  \tilde{g}_{\rm vert}=0\,.
\ee
Integrating out $\tilde{g}_{\rm vert}$ imposes 
\begin{equation}
\label{T_partial_zero}
\begin{aligned} 
0 &= e^0\wedge \bar\p_0\ip p^*(\d\vartheta\wedge\d\vartheta) = 2e^0\wedge  p^*(\d\vartheta)\wedge \bar\p_0\ip p^*(\d\vartheta)\\
&= 2e^0\wedge p^*(\rd\vartheta) \wedge\left(\cL_0 (p^*\vartheta) - \d(\bar\p_0\ip p^*\vartheta)\right)
\end{aligned} 
\end{equation}
as a constraint on $\d\vartheta$, where in the final line $\cL_0$ is the Lie derivative along $\bar\p_0$. In other words, the $\tilde{g}_{\rm vert}$ equation of motion imposes
\begin{equation}
\label{spdrag}
\cL_0 (p^*\vartheta) = \d(\bar\p_0\ip p^*\vartheta) \qquad\qquad\text{mod $e^0$}
\end{equation}
so that, taken modulo $e^0$, $\vartheta$ is invariant along the curves up to an exact form.  We can solve~\eqref{spdrag} to find
\begin{subequations}
\begin{equation}
\label{p*v}
p^*\vartheta = q^{\alpha\beta}(x)\lambda_\alpha\lambda_\beta + \d f\qquad\text{mod $e^0$}\,,
\end{equation}
for some triple of 1-forms $q^{\alpha\beta}$ on $M$, and where formally
\begin{equation}
 f = \bar\p_{0}^{-1}(\bar\p_{0}\ip p^*\vartheta)\,,
\end{equation}
\end{subequations}
to be understood using the Cauchy kernel of $\bar\p_{0}$. With this $p^*\vartheta$, we have $p^*\d\vartheta = \d q^{\alpha\beta}\,\lambda_\alpha\lambda_\beta$ mod $e^0$, and we note that the freedom in inverting $\bar\p_{0}$ does not affect $p^*\d\vartheta$. \\

The remaining terms in the action may be written as 
\begin{equation}
S[\tilde{g}_0,\vartheta]= -\frac{1}{4\pi\im}\int\limits_{M\times\CP^1} e^0\wedge\bar{e}^0\,\tilde{g}_0\,\lambda_\alpha\lambda_\beta\lambda_\gamma\lambda_\delta\wedge\left(\d q^{\alpha\beta}\wedge\d q^{\gamma\delta}\right)\,,
\end{equation}
and all the $\CP^1$ dependence is either explicit or in $\tilde{g}_0$. Thus, if we let
\be
\label{Psi_gravity_def}
\Psi_{\al\beta\gamma\delta}(x) = -\frac{1}{2\pi\im}\int_{\CP^1} e^0\wedge \bar{e}^0 \,\tilde{g}_0\,\lambda_\al\lambda_\beta\lambda_\gamma\lambda_\delta
\ee
be the Penrose transform of $\tilde{g}_0$, the Poisson-BF action on twistor space reduces to 
\begin{equation}
\label{Kirill_action}
S[\Psi,q]=\frac{1}{2}\int_M \Psi_{\alpha\beta\gamma\delta}\,\d q^{\alpha\beta}\wedge \d q^{\gamma\delta}
\end{equation}
as an action on $M$. This is an action for self-dual gravity with vanishing cosmological constant, in a form first presented by~\cite{Krasnov:2021cva}. To put it in a somewhat more familiar form, we replace $\d q^{\alpha\beta}$ by a generic 2-form $\Sigma^{\alpha\beta}$ on $M$, and impose $\d\Sigma^{\alpha\beta}=0$ by a Lagrange multiplier. Thus, at least on an open patch of $M$, \eqref{Kirill_action} is equivalent to
\be
S[\Psi,\Sigma,\Gamma] = \int_M\left(\Gamma_{\al\beta}\wedge\d\Sigma^{\al\beta}+\frac{1}{2}\,\Psi_{\al\beta\gamma\delta}\,\Sigma^{\al\beta}\wedge\Sigma^{\gamma\delta}\right)\,,
\ee
which was the self-dual gravity action discussed in section~\ref{sec:sdG}.\\

We remark that recasting the Poisson-BF action on twistor space as~\eqref{covariant_twistor_action} is analogous to the covariant form of twistor action for self-dual gravity at $\Lambda\neq0$ given in equation~(4.5) of~\cite{Mason:2007ct}. The field equations for (all of) $\tilde {g}$ from~\eqref{covariant_twistor_action} implies that $e^0\wedge\d\vartheta$ has rank 3, so defines an almost complex structure which turns out to be integrable automatically. It would clearly be interesting to investigate this covariant form of the twistor action further.

%%%%%%%%%%%%%%%%%%%%%%%%%%%%%%%%%%

\subsection{\texorpdfstring{$\eta$}{eta} on twistor space as a \texorpdfstring{4\textsuperscript{th}}{4th}-order gravitational axion on space-time} \label{subsec:eta-to-axion}

Next we consider the  action for the space-time field corresponding to $\eta$. On shell, at the linearised level on the basic twistor space $\PT$, $\eta$ represents an element of $H^{0,1}(\PT,\Omega^2_{\rm cl})$. The Penrose transform of such a field was first computed by Mason in~\cite{Mason:FAv1} who showed that it gives a scalar field $\rho$ of conformal weight 0, obeying the 4\textsuperscript{th}-order field equation $(\Delta)^2\rho=0$, where $\Delta$ is the Laplacian on $\R^4$. Explicitly, the Penrose transform may be constructed as 
\be 
	\rho = \frac{1}{2\pi\im}\int_{L_x}\p^{-1}\eta\,, 
\ee
while the kinetic term for $\eta$ was argued in~\cite{Costello:2021bah} to descend to the quadratic action
\be 
	\frac{1}{2}\int_{\R^4}(\Delta\rho)^2 
\ee
on $\R^4$. The ambiguity in defining $\p^{-1}\eta$ introduces a shift redundancy $\rho\sim\rho + C$ for constant $C$, implying that the $\R^4$ theory can depend on $\rho$ only through $\d\rho$. Indeed, the undifferentiated $\rho$ should not be allowed even as an operator in the theory. For example, it is argued in~\cite{Costello:2021bah} that theories on $\R^4$ with twistor progenitors have correlation functions which are analytic functions of their arguments. This is clearly not true of the $\rho$-$\rho$ propagator, which is logarithmic.\\

Now let's review what happens in the gauge theory case if we switch on the coupling $\lambda_\fg$, introducing the gauge counterterm in the twistor action~\eqref{eq:HBF-eta-action}. It was argued in \cite{Costello:2021bah} that on space-time this leads to a coupling of $\rho$ to sdYM theory through the term
\be \label{eq:gauge-axion}
	\sqrt{2}\lambda_\fg^\prime\int_{\R^4}\rho\,\tr(F\wedge F) 
\ee
which is compatible with the shift redundancy. This coupling reveals $\rho$ to be a type of axion, and henceforth we shall refer to it as a 4\textsuperscript{th}-order axion. We demonstrate in section \ref{sec:amplitudes} that $\lambda^\prime_\fg$ can be identified with the coefficient of the gauge counterterm, $\lambda_\fg$, justifying the rather unusual normalization. We remark that 4\textsuperscript{th}-order axions of exactly this type appear in $\cN=4$ conformal supergravity~\cite{Fradkin:1985am} and in particular in the twistor string~\cite{Berkovits:2004jj}.\footnote{It is interesting to note that any term appearing in the space-time action involving $F^-$ can be eliminated by a field redefinition of $B^-$. We can therefore equivalently think of this interaction as
\be
\frac{\lambda_\fg^\prime}{2}\int_{\R^4}\rho\,\tr(F^+\wedge F^+)\,,
\ee
though in this case the symmetry $\rho\sim\rho+C$ must be accompanied by a shift of $B^-$ proportional to $F^-$. Given that a 4\textsuperscript{th}-order scalar behaves somewhat like two 2\textsuperscript{nd}-order scalars, it is natural to speculate that $\rho$ is simultaneously playing the r\^{o}le of both the axion and dilaton appearing in `$\cN=0$ Einstein-Yang-Mills theory' \cite{Chiodaroli:2017ngp,Faller:2018vdz}. This theory is notable as it appears as the double copy of Yang-Mills with Yang-Mills coupled to a biadjoint scalar.}\\

Next let's turn our attention to the couplings between Poisson-BF theory and $\eta$ on twistor space \eqref{eq:PBF-eta-action}, which correspond to interactions between the metric and $\rho$ on space-time. To start with we take $\mu=0$, so that the only interaction lies in the covariant kinetic term
\be \frac{1}{4\pi\im}\int_\PT\p^{-1}\eta\,\nbar\eta\,. \ee
Classically this is invariant under Poisson diffeomorphisms, and so it should be diffeomorphism invariant on space-time. Furthermore, since the $\eta$ equation of motion is linear, it should descend to an expression quadratic in $\rho$. Since $\rho$ has dimension 0, a dimension 4 term can have at most 4 derivatives, or at most two curvatures. The only Lorentz invariant terms compatible with these observations are 
\begin{itemize}
    \item[-] $\rho \,\Delta_g^2\rho$ with no powers of the curvature, 
    \item[-] $\rho\, R\Delta_g\rho$, $\rho\, R^{\mu\nu}\nabla_\mu\nabla_\nu\rho$ and $R\,\nabla_\mu\rho\nabla^\mu\rho$, each of which is linear in the curvature, or
    \item[-] $R^2\,\rho^2$, $R_{\mu\nu}R^{\mu\nu}\rho^2$, $(\ast R)_{\mu\nu\rho\sigma}R^{\mu\nu\rho\sigma}\rho^2$ and $(\ast R\ast)_{\mu\nu\rho\sigma}R^{\mu\nu\rho\sigma}\rho^2$, all of which are quadratic.
\end{itemize}
On the support of the sd Einstein equations all terms involving $R_{\mu\nu}$ or $R$ vanish. These can therefore be removed by a field redefinition of the Lagrange multiplier field $\Gamma$. Furthermore, none of the terms involving quadratic curvature invariants respect the shift symmetry $\rho\sim\rho+C$. We conclude that, modulo field redefinitions, the covariant $\eta$ kinetic term must descend to
\be \label{eq:rho-cov-kin} \frac{1}{2}\int_{\R^4}\vol_g\,(\Delta_g\rho)^2\,,
\ee 
%= \frac{1}{2}\int_{\R^4}\rho\wedge\d(\ast\d)^3\rho\,. \ee
using the curved space Laplacian and measure.\\

Next, suppose we switch on the coupling $\mu$, introducing the Poisson counterterm
\be \label{eq:Poisson-counter} \frac{\mu}{4\pi\im}\int_\PT\eta\wedge\p^\dal\p_\db h\wedge\p^\db\p_\dal\p h
\ee
on twistor space.
Although this leads to a classical failure of gauge invariance on $\PT$, it does not do so on space-time. This is for the same reason that the twistorial gravitational anomaly does not correspond to a space-time anomaly. By considering the effect of the term \eqref{eq:Poisson-counter} on the $\eta,g$ equations of motion, we find that it leads to at worst terms linear in $\rho$ on space-time. Furthermore, integrating by parts we can remove all derivatives from $\rho$. The only dimension 4 possibilities are either
\begin{itemize}
    \item[-] linear in $\rho$: $\rho\,R^2$, $\rho\,R_{\mu\nu}R^{\mu\nu}$, $\rho\,R_{\mu\nu\rho\sigma}(\ast R)^{\mu\nu\rho\sigma}$, $\rho\,R_{\mu\nu\rho\sigma}(\ast R\ast)^{\mu\nu\rho\sigma}$, $\rho\,\nabla_\mu\nabla^\mu R$
    \item[-] or independent of $\rho$: $R^2$, $R_{\mu\nu}R^{\mu\nu}$, $R_{\mu\nu\rho\sigma}(\ast R)^{\mu\nu\rho\sigma}$, $R_{\mu\nu\rho\sigma}(\ast R\ast)^{\mu\nu\rho\sigma}$, $\nabla_\mu\nabla^\mu R$.
\end{itemize}
On-shell $R^2,R_{\mu\nu}R^{\mu\nu},\nabla_\mu\nabla^\mu R$ vanish, and the Euler scalar
\be R_{\mu\nu\rho\sigma}(\ast R\ast)^{\mu\nu\rho\sigma} \ee
becomes proportional to the Chern-Pontryagin scalar
\be R_{\mu\nu\rho\sigma}(\ast R)^{\mu\nu\rho\sigma}\,. \ee
Both the Euler and Chern-Pontryagin scalars are topological, and their integrals give the Euler character and gravitational instanton number respectively. We therefore conclude that switching on $\mu$ turns $\rho$ into a kind of 4\textsuperscript{th}-order `gravitational axion'. Viewing the Riemann tensor as a 2-form with values in endomorphisms of the tangent bundle we can write down the coupling as 
\be \label{eq:grav-axion} \frac{\mu^\prime}{\sqrt{2}}\int_{\R^4}\rho\,R^\mu_{~\,\nu}\wedge R^\mu_{~\,\nu}\,. \ee
We emphasise that this respects the shift symmetry $\rho\sim\rho+C$ on account of the exactness of the Chern-Pontryagin class. As noted in the gauge theory case, there is some freedom in how we choose to represent this coupling. In section \ref{sec:amplitudes} we show that $\mu^\prime$ can be identified with the coefficient of the Poisson counterterm on $\PT$, $\mu$.

Putting this all together, we find that Poisson BF-theory coupled to $\eta$ through a covariant kinetic term and Poisson counterterm on $\PT$ descends to sd gravity coupled to a 4\textsuperscript{th}-order gravitational axion on space-time, with action
\be S^\prime[\rho;e,\Gamma] = \int_{\R^4}\bigg(e^{\alpha\dal}\wedge e^\beta_{~\,\dal}\wedge\d\Gamma_{\alpha\beta} + \frac{\im}{2}\vol_g\,(\Delta_g\rho)^2 + \frac{\im\mu^\prime}{\sqrt{2}}\rho\,R^\mu_{~\,\nu}\wedge R^\nu_{~\,\mu}\bigg)\,. \ee
Here we have chosen to write the sd gravity part of the action explicitly in terms of vielbeins, which determine the metric $g$ in the usual way.\\

Finally, essentially identical arguments show that holomorphic gauge-Poisson-BF theory coupled to $\eta$ though a covariant kinetic term and with both gauge and Poisson counterterms descends to the sum of sdEYM as described in subsection \ref{subsec:PBF-to-sdGR} together with the covariant $\rho$ kinetic term \eqref{eq:rho-cov-kin}, and both the gauge \eqref{eq:gauge-axion} and gravitational \eqref{eq:grav-axion} axion couplings on space-time.

%%%%%%%%%%%%%%%%%%%%%%%%%%%%%%%%%%
%%%%%%%%%%%%%%%%%%%%%%%%%%%%%%%%%%

\section{Axions and quantum integrability of self-dual Einstein-Yang-Mills} \label{sec:amplitudes}

It is important to recognise that the gauge, diffeomorphism and mixed anomalies investigated in the previous sections were anomalies of the theory on \emph{twistor} space. No such anomalies are possible in 4 real dimensions, for any semi-simple $\fg$, since the only non-trivial representation of $\fg$ the theory involves is the adjoint. What then is the consequence of this twistorial anomaly cancellation for the theory on $\R^4$?

The philosophy of this paper (as in~\cite{Costello:2021bah}) is that the self-dual Yang-Mills and self-dual vacuum Einstein equations are integrable \emph{because} they come from twistor space. The beauty of the Penrose-Ward and non-linear graviton constructions is that they generate solutions to these equations purely from the holomorphic structure of twistor space. If the twistor space theory is anomalous, then we should expect integrability of the theory on $\R^4$ to be lost at the quantum level. In fact, it is well known that sdYM and sd gravity possess non-trivial 1-loop scattering amplitudes~\cite{Bern:1998xc,Cangemi:1996rx,Chalmers:1996rq}, describing the interaction of arbitrary numbers of positive helicity gluons or gravitons. Such non-trivial interactions within the self-dual sector are anathema to integrability.

In this section, we recall the 4-pt 1-loop gluon, graviton and mixed amplitudes in sdEYM. The coefficient of the graviton amplitude is modified when we couple to a 4\textsuperscript{th}-order scalar, which we evaluate by recasting it as a pair of 2\textsuperscript{nd}-order scalars as standard. Switching on gauge and gravitational axion interactions and tuning the corresponding couplings $\lambda_\fg^\prime$ and $\mu^\prime$ we show that tree level axion exchange can simultaneously cancel all of the 4-pt 1-loop amplitudes.

We then sketch a recursive argument demonstrating that if the 1-loop 4-pt all-plus amplitudes vanish, then all higher point 1-loop amplitudes also vanish. From this we infer that, at least for generic external momenta, sdEYM coupled to a 4\textsuperscript{th}-order gauge/gravitational axion $\rho$ (for appropriate values of the couplings) has trivial amplitudes.

%%%%%%%%%%%%%%%%%%%%%%%%%%%%%%%%%%

\subsection{The 4-point 1-loop amplitudes in self-dual Einstein-Yang-Mills} \label{subsec:4-pt-amplitudes}

We first recall the 4-pt 1-loop amplitudes in sdEYM.\\

The 1-loop 4-gluon amplitude is identical to that in pure sdYM \cite{Mahlon:1993fe,Bern:1993qk}
\be \label{eq:4-gluon-amp} \cA^\mathrm{1-loop}_\mathrm{sdYM}\big(1^+_\sfa,2^+_\sfb,3^+_\sfc,4^+_\sfd\big) = - \frac{\im}{(4\pi)^2}\frac{4}{3}\frac{[12][34]}{\la12\ra\la34\ra}\tr_\ad(\mathrm{t}_\sfa\mathrm{t}_\sfb\mathrm{t}_\sfc\mathrm{t}_\sfd)\,. \ee
Although this coincides with the 1-loop 4-gluon all-plus amplitude in full (non self-dual) Yang-Mills, in full Einstein-Yang-Mills it acquires corrections: the above expression coincides with the order $g_\mathrm{YM}^4$ piece, but the full amplitude receives further contributions at order $g_\mathrm{YM}^2\kappa^2$, $\kappa^4$ \cite{Nandan:2018ody,Faller:2018vdz}. These subleading contributions involve gravitons running through the loop, which cannot arise in the self-dual theory on account of the BF-type form of the action.\\

There are in principle two non-vanishing mixed amplitudes, the first is the 2-gluon, 2-graviton amplitude, which takes the form \cite{Nandan:2018ody,Faller:2018vdz}
\be \label{eq:2-gluon-2-graviton-amp}
\cA^\mathrm{1-loop}_\mathrm{sdEYM}\big(1^+_\sfa,2^+_\sfb,3^{++},4^{++}\big) = - \frac{\im}{(4\pi)^2}\frac{s}{6}\frac{[12][34]^2}{\la12\ra\la34\ra^2}\tr_\ad(\mathrm{t}_\sfa\mathrm{t}_\sfb)\,.
\ee
Here $s=\la12\ra[12]$ is the familiar Mandelstam variable. Again, this does not coincide with the 2-gluon, 2-graviton 1-loop all-plus amplitude in full EYM, which receives subleading corrections of order $\kappa^4$.

We might also have expected a 3-gluon, 1-graviton mixed amplitude, but in sdEYM this vanishes. It does not vanish in full EYM, but is of order $g_\mathrm{YM}\kappa^3$. This is consistent with our findings in subsection \ref{subsec:space-time-anomalies}, where we saw that the term in the anomaly polynomial cubic in the twistorial field $a$ involved the background complex structure on twistor space rather than its fluctuations.\\

Finally, consider the 1-loop 4-graviton amplitude in sdEYM. This receives contributions from both gluons and gravitons running around the loop, and these contributions are proportional. The full amplitude is \cite{Grisaru:1979re,Dunbar:1994bn}
\be \label{eq:4-graviton-amp-EYM} \cA^\mathrm{1-loop}_\mathrm{sdEYM}(1^{++},2^{++},3^{++},4^{++}) = - \frac{\im}{(4\pi)^2}\frac{s^2+t^2+u^2}{120}\frac{[12]^2[34]^2}{\la12\ra^2\la34\ra^2}(1+\dim\fg)\,, \ee
where $t=\la23\ra[23]$ and $u=\la13\ra[13]$ are the remaining Mandelstam variables. This coincides with the 4-graviton  all-plus amplitude in full EYM at 1-loop. More generally, it's well known that helicity conservation in supersymmetric theories necessitates that all-plus graviton amplitudes are proportional to the difference between the number of bosonic and fermionic degrees of freedom \cite{Grisaru:1979re}. In the present work this follows from the fact that a field with values in the vector bundle $V$ contributes to the twistorial gravitational anomaly polynomial in the form $\todd_4(T_\cPT)\mathrm{rk}(V)$.\\

The all-plus gravity amplitude receives a further contribution when we introduce the 4\textsuperscript{th}-order scalar $\rho$ on space-time, even before we turn on the gauge and gravitational axion couplings  $\lambda_\fg^\prime,\mu^\prime$ introduced in section \ref{subsec:eta-to-axion}. From the discussion in the previous paragraph and the computation in appendix \ref{app:eta-box}, we expect that a 4\textsuperscript{th}-order scalar carries twice the degrees of freedom of a 2\textsuperscript{nd}-order scalar. A trick for seeing this directly is to consider the kinetic term for $\rho$:
\be
\frac{1}{2}\int_{\R^4}\vol_g\,(\Delta_g\rho)^2\,.
\ee
We can rewrite this by introducing a new scalar field $\tau$ as
\be \label{eq:4th-to-2nd-order} \frac{1}{2}\int_{\R^4}\vol_g\,(2\tau\Delta_g\rho - \tau^2)\,. \ee
Although the $\tau^2$ term does depend on $g$ through the volume from, on the support of the sd Einstein equations it's possible to fix
\be \label{eq:det-g} \det{g} = 1\,. \ee Indeed, Plebanski's 1\textsuperscript{st} equation, which is a partially on-shell gauge fixing of the sd Einstein equations, is actually equivalent to equation \eqref{eq:det-g} for a K\"{a}hler metric $g$. In Plebanski's 2\textsuperscript{nd} formulation it's a consequence of the partially on-shell gauge fixing \cite{Plebanski:1975wn,Siegel:1992wd}. This means that when coupling to sd gravity, we can ignore the $g$ dependence in the $\tau^2$ term.

It's then not difficult to see that the 1-loop graviton amplitudes built from the action \eqref{eq:4th-to-2nd-order} do not involve any insertions of the $\tau^2$ term, which we treat as a vertex. The field $\rho$, which is a `real' 4\textsuperscript{th}-order scalar, therefore contributes to all-plus graviton amplitudes in the same way as a `complex' 2\textsuperscript{nd}-order scalar, exactly as expected.\footnote{Note, however, that if we were to diagonalise the $\rho,\tau$ kinetic term we'd find that the two terms would have opposite sign. The field with the wrong sign kinetic term is known as an Ostrogradsky ghost.} Hence, the 4-graviton 1-loop amplitude in sdEYM coupled to a 4\textsuperscript{th}-order scalar is
\be \label{eq:4-graviton-amp} \cA^\mathrm{1-loop}_{\mathrm{sdEYM}+\rho}(1^{++},2^{++},3^{++},4^{++}) = - \frac{\im}{(4\pi)^2}\frac{s^2+t^2+u^2}{120}\frac{[12]^2[34]^2}{\la12\ra^2\la34\ra^2}(2+\dim\fg)\,. \ee
Notably, the coefficient $2+\dim\,\fg$ is the same one that appears in the 4-graviton all-plus 1-loop amplitude in $\cN=0$ EYM, viewed as a double copy~\cite{Faller:2018vdz}.

%%%%%%%%%%%%%%%%%%%%%%%%%%%%%%%%%%

\subsection{Cancelling the 4-point 1-loop amplitudes with axion exchange} \label{subsec:4-pt-cancellation}

Now we show that by tuning the couplings $\lambda_\fg^\prime$, $\mu^\prime$ it's possible to simultaneously cancel all of the 4-pt 1-loop amplitudes from subsection \ref{subsec:4-pt-amplitudes}.\\

First, we review the cancellation of the 4-gluon 1-loop amplitude from equation \eqref{eq:4-gluon-amp}, which was demonstrated in \cite{Costello:2022wso}. For a positive helicity gluon momentum eigenstate with eigenvalues $k_{\alpha\dal} = \kappa_\alpha\tilde\kappa_\dal$, we have
\be F = \frac{1}{2}\mathrm{t}_\sfa\,\tilde\kappa_\dal\tilde\kappa_\db\,\d x^{\alpha\dal}\wedge\d x_\alpha^{~\,\db}\, e^{\im\la\kappa|x|\tilde\kappa]}\,. \ee
Here $\la\kappa|x|\tilde\kappa] = x^{\alpha\dal}\kappa_\alpha\tilde\kappa_\dal = x\cdot k$. Recalling the interaction vertex \eqref{eq:gauge-axion} and noting that the $\rho$ propagator is
\be G_\rho(x,y) = \int\frac{\d^4p}{(2\pi)^4}\,\frac{\e^{\im(x-y)\cdot p}}{p^4} \ee
we can see that axion exchange contributes
\bea
&2{\lambda^\prime_\fg}^2\int_{\R^4_x\times\R^4_y}\tr(F_1(x)\wedge F_2(x))\,G_\rho(x,y)\,\tr(F_3(y)\wedge F_4(y)) \\
&=2{\lambda^\prime_\fg}^2\frac{[12][34]}{\la12\ra\la34\ra}\tr(\mathrm{t}_{\sfa_1}\mathrm{t}_{\sfa_2})\tr(\mathrm{t}_{\sfa_3}\mathrm{t}_{\sfa_4})(2\pi)^4\delta^{(4)}(k_1+k_2+k_3+k_4)\,.
\eea
The combination $[12][34]/\la12\ra\la34\ra$ is totally symmetric on account of momentum conservation. Stripping off the momentum conserving $\delta$-function, and summing over the 3 possible channels gives a total amplitude
\be \label{eq:gluon-counter-amp} - 6\im{\lambda^\prime_\fg}^2\frac{[12][34]}{\la12\ra\la34\ra}\tr(\mathrm{t}_{(\sfa_1}\mathrm{t}_{\sfa_2})\tr(\mathrm{t}_{\sfa_3}\mathrm{t}_{\sfa_4)})\,. \ee
This compensates the 4-gluon 1-loop amplitude \eqref{eq:4-gluon-amp}, which itself needs to be summed over configurations which are not equivalent under cyclic permutations of the external legs. Since this is an amplitude in full Yang-Mills, for which gluon propagators are undirected, we should also quotient by reflections.
\be \sum_{\sigma\in S_4/D_4}\cA^\mathrm{1-loop}_\mathrm{YM}\big(1^+_{\sfa_{\sigma(1)}},2^+_{\sfa_{\sigma(2)}},3^+_{\sfa_{\sigma(3)}},4^+_{\sfa_{\sigma(4)}}\big) = - \frac{\im}{4\pi^2}\frac{[12][34]}{\la12\ra\la34\ra}\tr_\ad(\mathrm{t}_{(\sfa_1}\mathrm{t}_{\sfa_2}\mathrm{t}_{\sfa_3}\mathrm{t}_{\sfa_4)})\,. \ee
Using the trace identity $\tr_\fg(X^4) = C_\fg\tr(X^2)^2$ this is equal to
\be - C_\fg\frac{\im}{4\pi^2}\frac{[12][34]}{\la12\ra\la34\ra}\tr(\mathrm{t}_{(\sfa_1}\mathrm{t}_{\sfa_2})\tr(\mathrm{t}_{\sfa_3}\mathrm{t}_{\sfa_4)})\,. \ee
In order for this to cancel \eqref{eq:gluon-counter-amp} we should therefore take
\be \label{eq:tune-lambda-prime} {\lambda_\fg^\prime}^2 = \frac{C_\fg}{3!}\bigg(\frac{\im}{2\pi}\bigg)^2\,. \ee

Next, let us turn our attention to the 2-gluon, 2-graviton 1-loop amplitude. We anticipate that this should be cancelled by tree level exchange of an axion where at one end of the propagator we insert the gauge axion vertex, and at the other the gravitational axion vertex. The linearized curvature of a positive helicity graviton momentum eigenstate on flat space is purely self-dual and is given by
\be
R_{\mu\nu} = \frac{1}{2}R_{\mu\nu\rho\sigma}\,\d x^\rho\wedge\d x^\sigma = \frac{1}{2}\eps_{\alpha\beta}\,\tilde\kappa_{\dal}\tilde\kappa_{\db}\tilde\kappa_{\dot\gamma}\tilde\kappa_{\dot\delta}\,\d x^{\gamma\dot\gamma}\wedge\d x_\gamma^{~\,\dot\delta}\,\e^{\im\la\kappa|x|\tilde\kappa]}\,.
\ee
Axion exchange therefore contributes
\bea
&- \lambda^\prime_\fg\mu^\prime\int_{\R^4_x\times\R^4_y}\tr(F_1(x)\wedge F_2(x))\,G_\rho(x,y)\,R_{3\mu\nu}(y)\wedge R_4^{\mu\nu}(y) \\
&= - \lambda_\fg^\prime\mu^\prime s\frac{[12][34]^2}{\la12\ra\la34\ra^2}\tr(\mathrm{t}_{\sfa_1}\mathrm{t}_{\sfa_2})(2\pi)^4\delta^{(4)}(k_1+k_2+k_3+k_4)\,,
\eea
and stripping off the momentum conserving $\delta$-function this is
\be \im\lambda_\fg^\prime\mu^\prime s\frac{[12][34]^2}{\la12\ra\la34\ra^2}\tr(\mathrm{t}_{\sfa_1}\mathrm{t}_{\sfa_2})\,. \ee
We can therefore see that in order to cancel the 2-gluon, 2-graviton 1-loop amplitude from equation \eqref{eq:2-gluon-2-graviton-amp} we should take
\be \label{eq:tune-lambda-mu-prime} \lambda_\fg^\prime\mu^\prime = - \frac{2\bh^\vee}{4!}\bigg(\frac{\im}{2\pi}\bigg)^2\,. \ee

Finally let's attempt to simultaneously cancel the 4-graviton 1-loop amplitude. The contribution of axion exchange with insertions of the gravitational axion vertex at both ends of the propagator is
\bea
&\frac{{\mu^\prime}^2}{2}\int_{\R^4_x\times\R^4_y}R_{1\mu\nu}(x)R_2^{\mu\nu}(x)\,G_\rho(x,y)\,R_{3\rho\sigma}(y)R_4^{\rho\sigma}(y) \\
&= \frac{{\mu^\prime}^2}{2}\frac{[12]^4[34]^2}{\la34\ra^2}(2\pi)^4\delta^{(4)}(k_1+k_2+k_3+k_4) \\
&= \frac{{\mu^\prime}^2s^2}{2}\frac{[12]^2[34]^2}{\la12\ra^2\la34\ra^2}(2\pi)^4\delta^{(4)}(k_1+k_2+k_3+k_4)\,.
\eea
Stripping off the momentum conserving delta function and summing over the three possible channels we arrive at the following formula for the amplitude
\be - \frac{\im{\mu^\prime}^2(s^2+t^2+u^2)}{2}\frac{[12]^2[34]^2}{\la12\ra^2\la34\ra^2}\,. \ee
In order to cancel the 4-graviton 1-loop amplitude \eqref{eq:4-graviton-amp} we must take
\be \label{eq:tune-mu-prime} {\mu^\prime}^2 = \frac{2+\dim\fg}{2\cdot 5!}\bigg(\frac{\im}{2\pi}\bigg)^2\,. \ee
Consistency of equations \eqref{eq:tune-lambda-prime}, \eqref{eq:tune-lambda-mu-prime} and \eqref{eq:tune-mu-prime} therefore requires that
\be \frac{C_\fg}{3!}\frac{2+\dim\fg}{2\cdot 5!}\bigg(\frac{\im}{2\pi}\bigg)^4 = {\lambda_\fg^\prime}^2{\mu^\prime}^2 = (\lambda^\prime_\fg\mu^\prime)^2 = \bigg(-\frac{2\bh^\vee}{4!}\bigg)^2\bigg(\frac{\im}{2\pi}\bigg)^4\,. \ee
This is only possible if
\be C_\fg = \frac{5(2\bh^\vee)^2}{2(2+\dim\fg)}\,. \ee
As we saw in subsection \ref{sec:GS-mechanism}, this holds for a simple Lie algebra $\fg$ whenever we have the trace identity $\tr_\ad(X^4)\propto\tr(X^2)^2$. Comparing equations \eqref{eq:tune-lambda-prime}, \eqref{eq:tune-lambda-mu-prime} and \eqref{eq:tune-mu-prime} to the anomaly cancellation conditions \eqref{eq:HBF-cancellation}, \eqref{eq:PGBF-mixed-cancellation} and \eqref{eq:PBF-cancellation} it's clear that we can identify
\be \lambda_\fg^\prime = \lambda_\fg\,,\qquad\qquad \mu^\prime = \mu \ee
as claimed in subsection \ref{subsec:eta-to-axion}. For these values of the couplings all 4-pt 1-loop amplitudes in the twistorial theory described in section \ref{sec:space-time-theory} vanish. This is consistent with our expectation that there should be no amplitudes in a 4d quantum integrable theory.

%%%%%%%%%%%%%%%%%%%%%%%%%%%%%%%%%%

\subsection{A sketch of the vanishing of higher point amplitudes} \label{subsec:n-pt-cancellation}

We've seen that all 4-pt 1-loop amplitudes in sdEYM coupled to a mixed gauge/gravitational 4\textsuperscript{th}-order axion vanish. Now we sketch a proof that leverages this result to show the vanishing of all higher point amplitudes, focusing on the case of sd gravity coupled to a 4\textsuperscript{th}-order gravitational axion for brevity.\footnote{Note that the vanishing of higher point amplitudes in the case of sdYM coupled to a 4\textsuperscript{th}-order axion follows from the equivalence between form factors and 2d correlators proven in \cite{Costello:2022wso}.} We work perturbatively around the flat background, taking $e^{\alpha\dal} = \d x^{\alpha\dal} + \delta e^{\alpha\dal}$.\\

The first step is to characterise the tree amplitudes in the theory. It's well known that, with generic external kinematics, all $n$-pt trees in sd gravity vanish for $n>3$. See {\it e.g.}~\cite{Dixon:2013uaa} for the analogous argument in sdYM, which generalises straightforwardly. Upon coupling to the 4\textsuperscript{th}-order gravitational axion, there are three new classes of tree amplitudes. The first involve an internal axion exchange, but no external axion states. These are generically non-vanishing, and are expected to cancel the 1-loop all-plus amplitudes. The second involve a single external axion state, and the third correct the axion 2-pt function.

\begin{figure}[!ht]
\centering
  \includegraphics[scale=0.3]{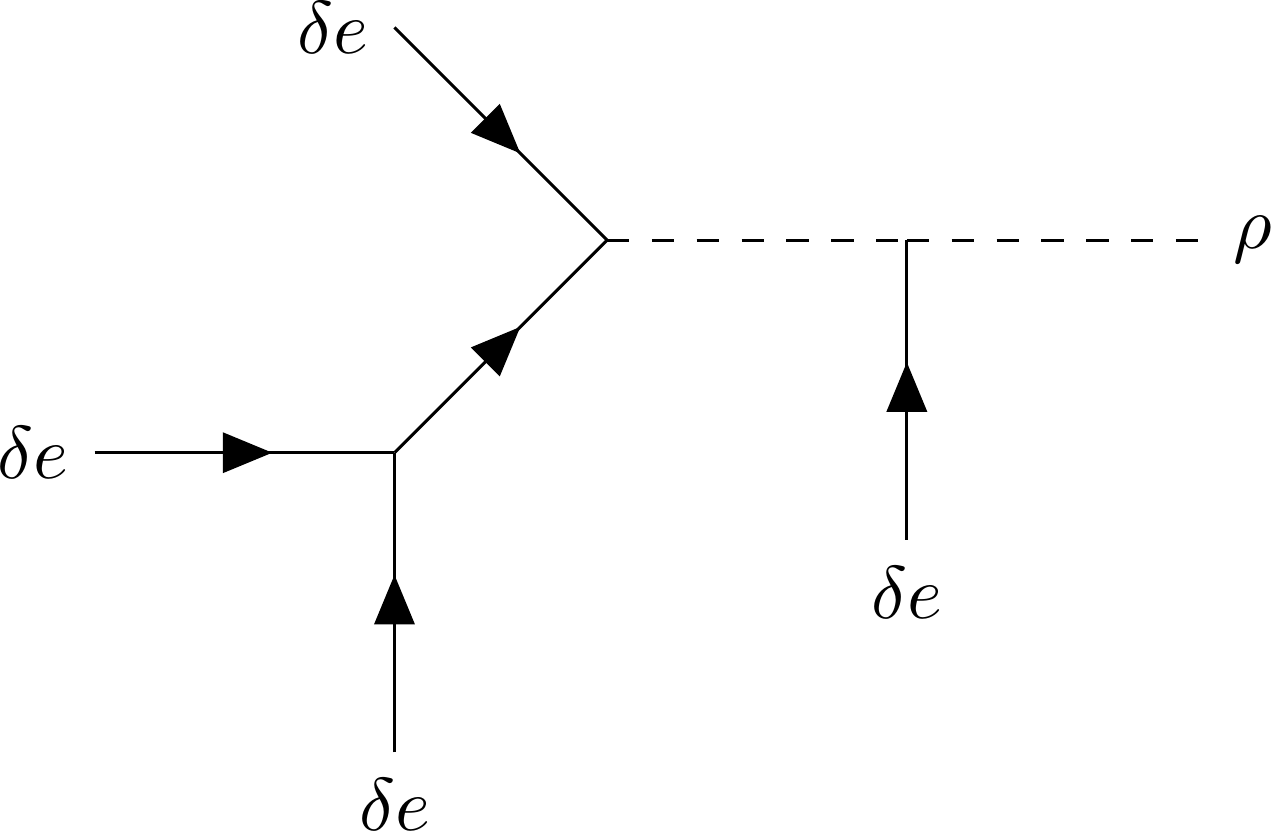}
\caption{\emph{An example of a tree diagram in sd gravity coupled to a 4\textsuperscript{th}-order gravitational axion with one axion leg.}} \label{fig:axion-tree}
\end{figure}

Let's concentrate on the amplitudes with one external axion. The remaining $n-1$ external legs correspond to positive helicity gravitons. An example of such a diagram with $n=5$ is illustrated in figure \ref{fig:axion-tree}.

For $n>3$ these diagrams all vanish. To show this we adopt the following strategy. Each positive helicity graviton state is accompanied by a pair of left-handed polarization spinors $\varepsilon_i,\varepsilon^\prime_i$, however, diffeomorphism invariance requires that the amplitude is actually independent of this data. By counting the number of uncontracted spinor indices in the internal vertices and propagators, we find that in every diagram contributing to the amplitude a pair of polarizations must contract, or a polarization must contract with the left-handed momentum spinor of the external axion $\kappa_n$. By choosing the polarizations so that $\varepsilon_i=\varepsilon^\prime_i=\kappa_n$ for $i=1,\dots,n-1$, all of these contractions vanish. Therefore the amplitude is zero.

The most dangerous interaction vertices, {\it i.e.} those most likely to undermine the above argument, are those with many derivatives but which are of low order in $\delta e$. This is because space-time derivatives introduce spinor indices into which polarizations can contract, whereas every extra graviton leg is accompanied by two further polarizations. Exploiting this observation allows us to dramatically simplify the vertices we need to consider.

The momentum space sd gravity propagator involves one momentum vector with free indices, and the sd gravity interaction vertex involves one derivative with free indices.

Turning our attention to the 4\textsuperscript{th}-order axion, we employ the same trick used in subsection \ref{subsec:4-pt-amplitudes} and split it into a pair of 2\textsuperscript{nd}-order scalars with kinetic term 
\be \label{eq:4th-to-2nd-order-again} \frac{1}{2}\int_{\R^4}\vol_g\,(2\tau\Delta_g\rho - \tau^2)\,. \ee
The one axion trees involve no insertions of $\tau^2$, which we treat as vertex. The diagram in figure \ref{fig:axion-tree}, for example, is replaced by a similar diagram with directed quadratic scalar propagators. Since the axion kinetic term is now 2\textsuperscript{nd}-order, the quadratic axion vertices also get modified. The flat space $\rho$-$\tau$ propagator has no free indices. The most dangerous interaction generated by covariant kinetic term in \eqref{eq:4th-to-2nd-order-again} is
\be \label{eq:kinetic-vertex} -2\int_{\R^4}\vol_\delta\,\delta e_{\alpha\beta\dal\db}\p^{\alpha\dal}\tau\p^{\beta\db}\rho\,. \ee
It introduces two derivatives with free indices, and involves $\delta e$ to linear order. Similarly, the most dangerous part of the counterterm vertex is
\be \label{eq:counterterm-vertex} \frac{\mu}{\sqrt{2}}\int_{\R^4}\vol_\delta\,\rho\,\p^\gamma_{~\,(\dg|}\p^\alpha_{~\,(\dal}\delta e_{\db)|\dd)\alpha\gamma}\p^{\delta\dd}\p^{\beta\db}\delta e^{\dal\db}_{~\,~\,\beta\delta}\,, \ee
which introduces four derivatives with free indices and is quadratic in $\delta e$.

Consider a general $n$-pt 1-axion tree diagram built from these worst case vertices. It will necessarily involve one insertion of the counterterm vertex \eqref{eq:counterterm-vertex}, and let $k$ be the number of vertices of the form \eqref{eq:kinetic-vertex}. The overall number of polarization spinors is $2(n-1)$, and the number of free left-handed spinor indices in the amplitude is readily computed to be $4+2(n-k-3)+2k = 2(n-1)$. Finally note that the external axion state only ever appears in the amplitude in the form $\d\rho$, as is necessitated by the shift symmetry $\rho\sim\rho+C$. One of the internal momenta in the amplitude is therefore $k_n^{\alpha\dal} = \kappa_n^\alpha\tilde\kappa_n^\dal$. We conclude that in every diagram either two polarization spinors contract, or a polarization spinor contracts with $\kappa_n$. Setting all polarizations equal to $\kappa_n$ gives zero.

One important feature of the above argument is that it cannot be applied to axion exchange diagrams, which we know cannot vanish since they must compensate the all-plus amplitudes. It can, however, be straightforwardly adapted to show that the two axion amplitudes vanish.\\

Now we've characterised all trees, we can move on to the $n$-pt all-plus amplitudes. As before, we restrict to the case of sd gravity coupled to a 4\textsuperscript{th}-order gravitational axion. We've already seen in subsection \ref{subsec:4-pt-cancellation} that all 4-pt all-plus amplitudes in the theory vanish. Our strategy will be to induct on $n$.

The $n$-pt all-plus amplitudes receive two different classes of contributions. The first are from loop diagrams, with either gravitons or axions running through the loop, and second are from axion exchange diagrams.

We know the sum over the first class of diagrams will be twice to the standard $n$-pt 1-loop all-plus graviton amplitudes, since the axion contributes as if it were a complex scalar. The collinear properties of these amplitudes are well understood, in particular sending $k_2\to t/(1-t)k_1$ and writing $K = k_1 + k_2$ we have
\be \cA^\mathrm{1-loop}_n(1^{++},\dots,n^{++}) \xrightarrow{\parallel} \mathrm{Split}^\mathrm{tree}_{--}(1^{++},2^{++};t)\cA^\mathrm{1-loop}_{n-1}(K^{++},3^{++},\dots,n^{++}) \ee
for $\mathrm{Split}^\mathrm{tree}_{--}(1^{++},2^{++};t)$ the standard tree graviton splitting function \cite{Bern:1998xc}. Actually, a somewhat stronger version of this statement is true: the above decomposition also holds in the holomorphic collinear limit, as defined in \cite{Ball:2021tmb}. Here we take $\kappa_2\mapsto\sqrt{t/(1-t)}\kappa_1$, whilst keeping $\tilde\kappa_1,\tilde\kappa_2$ fixed. The momentum $K$ is determined by
\be \kappa_K=\frac{1}{\sqrt{(1-t)}}\kappa_1=\frac{1}{\sqrt{t}}\kappa_2\,,\qquad \tilde\kappa_K = \sqrt{(1-t)}\tilde\kappa_1 + \sqrt{t}\tilde\kappa_2\,. \ee
Roughly speaking, in the holomorphic collinear limit we take $\la12\ra\to0$ whilst keeping $[12]$ fixed.

We can also characterise the holomorphic collinear limits of the sum of the axion exchange diagrams. Since these are tree diagrams holomorphic poles can only arise when the momenta becoming collinear are attached via a trivalent vertex to a graviton or axion propagator. Let $\cA^\mathrm{exchange}_n$ be the sum of the $n$-pt axion exchange diagrams, and $\cA^\mathrm{axion}_n$ the one axion, $n$-pt tree amplitude. In the holomorphic collinear limit we have
\bea
\cA^\mathrm{exchange}_n(1^{++},\dots,n^{++})\xrightarrow{\parallel}&\mathrm{Split}^\mathrm{tree}_{--}(1^{++},2^{++};t)\cA^\mathrm{exchange}_{n-1}(K^{++},3^{++},\dots,n^{++}) \\
&\qquad+\mathrm{Split}^\mathrm{axion}(1^{++},2^{++};t)\cA^\mathrm{axion}_{n-1}(K,3^{++},\dots,n^{++})
\eea
for $\mathrm{Split}^\mathrm{axion}(1^{++},2^{++};t)$ the two graviton to one axion splitting function. However, we have already seen that the one axion, $n$-pt tree vanishes, and so we infer that $\cA^\mathrm{exchange}_n$ has exactly the same holomorphic collinear behaviour as $\cA^\mathrm{tree}_n$.

Now we are ready to induct. Suppose that the $n$-pt all-plus amplitude in sd gravity coupled to the 4\textsuperscript{th}-order axion vanishes, {\it i.e.},
\be 2\cA^\mathrm{1-loop}_n(1^{++},\dots,n^{++}) + \cA^\mathrm{exchange}_n(1^{++},\dots,n^{++}) = 0\,. \ee
Then all holomorphic collinear limits of the $(n+1)$-pt all-plus amplitude must vanish. On the grounds of dimension and helicity the $(n+1)$-pt all-plus amplitude must have weight $2+n$ in square brackets, and $2-n$ in angle brackets. For $n>4$ it must therefore have holomorphic collinear poles, unless it vanishes. Since we've already seen that the $4$-pt all-plus amplitude is zero, by induction all $n$-pt all-plus amplitudes are also zero.

Showing that the all-plus amplitudes vanish in the case of sdEYM coupled to a 4\textsuperscript{th}-order mixed gauge/gravitational axion is no more difficult.

%%%%%%%%%%%%%%%%%%%%%%%%%%%%%%%%%%
%%%%%%%%%%%%%%%%%%%%%%%%%%%%%%%%%%

\section{Discussion}

In this note, we've computed the anomalies in the twistor formulations of both sd gravity and sdEYM. As in the case of sdYM \cite{Costello:2021bah}, these anomalies do not represent a failure of diffeomorphism or gauge invariance on space-time, but instead should be viewed as obstructing integrability. Indeed, we've found that they can be identified with the non-vanishing 4-pt all-plus graviton and gluon amplitudes, which have long been conjectured to violate integrability \cite{Bardeen:1995gk}.

We have also explored a number of ways of cancelling these anomalies. One possibility is to couple to appropriate bosonic or fermionic matter so that the coefficients of the anomalies, or equivalently the all-plus amplitudes, vanish on the nose. A particular instance of this is the case of supersymmetric theories, for which the twistorial anomalies necessarily vanish. However, we again emphasise that cancellation (and hence quantum integrability) is a far weaker condition than supersymmetry. The twistorial anomalies also cancel in a self-dual higher spin theory since the potential anomaly is proportional to a count of the degrees of freedom in the theory: appropriately regularised, this count gives zero.

The method for cancelling the anomalies we have devoted the majority of our effort towards is by coupling to a single field $\eta\in\Omega^{2,1}_\mathrm{cl}$ on twistor space describing a 4\textsuperscript{th}-order scalar on space-time. The motivation to look for such a cancellation was provided by Costello in \cite{Costello:2021bah}. Therein, he used this method to cancel the gauge anomaly in the twistor formulation of sdYM. The field $\eta$ plays a role analogous to that of the closed string field in the topological string: in particular tree level exchanges of $\eta$ cancel the anomaly through a kind of Green-Schwarz mechanism. For this to be possible in the case of sdYM it's necessary that the trace identity $\tr_\ad(X^4)\propto\tr(X^2)^2$ should hold. We've seen that a similar mechanism can be applied in the case of sd gravity. Furthermore, in the case of sdEYM we found that if it's possible to cancel the twistorial gauge anomaly, then the mixed and gravitational twistorial anomalies can also be eliminated. This remarkable cancellation demands a stringy interpretation. It is intriguing to note that the field $\eta$ naturally arises as the holomorphic twist of the 6d $(1,0)$ tensor multiplet \cite{Saberi:2020pmw}, potentially hinting at a similar Green-Schwarz mechanism in 6d supersymmetric theories. \\

There are a number of natural questions left open by this work:

\begin{itemize}
\item[-] Once the twistorial anomalies are cancelled, the all-plus amplitudes obstructing integrability vanish. Indeed, we verified that the only non-zero amplitudes in sdEYM coupled to a 4\textsuperscript{th}-order mixed gauge/gravitational axion are at three points. This may cause the reader some distress. In a gravitational theory there are no local operators, and the natural observable available to us is the S-matrix. We have found that in order for a theory of sd gravity to be integrable its S-matrix must be trivial. What, then, can we compute in such a theory? We conjecture that the natural observables are not line operators, but surface operators. Indeed, it was argued in \cite{Costello:2022wso} that the chiral algebra living on a holomorphic surface defect is isomorphic to the celestial chiral algebra of sd gravity.

\item[-] In \cite{Costello:2022wso,Bu:2022dis} (see also \cite{Costello:2022upu}) a correspondence between the conformal blocks of a particular 2d chiral algebra and local operators in sdYM coupled to the 4\textsuperscript{th}-order axion was obtained by exploiting the twistor formulation of the latter. Furthermore, correlators of the chiral algebra evaluated in a conformal block are related to amplitudes in the presence of the corresponding operator. This allowed the authors to recover the MHV, NMHV and 1-loop all-plus amplitudes in YM. The gravitational case was also briefly considered, owing to its close connections with the $w_{1+\infty}$ symmetry of celestial holography \cite{Guevara:2021abz, Strominger:2021lvk} and self-dual gravity \cite{Adamo:2021lrv,Adamo:2021zpw,Ball:2021tmb}. Although without an understanding of its anomalies, the discussion was necessarily classical. It would be fascinating to develop this proposal in the gravitational case, and we expect that our analysis of the twistorial anomaly will be essential in obtaining loop-level results.

\item[-] Although we have found that 4d quantum integrability necessitates the vanishing of all non-trivial flat space amplitudes, we expect that in more general (asymptotically flat) backgrounds, perhaps those with a non-zero value for the Lagrange multiplier field, this will no longer be the case. See \cite{Costello:2022jpg} for an example of a 4d quantum integrable theory with non-vanishing amplitudes.

\item[-] sd gravity has also been proposed as the target space description of the $\cN=2$ string \cite{Ooguri:1990ww,Ooguri:1991fp}. Curiously, it's believed that all but the 3-pt amplitudes in this theory vanish, suggesting that its string field theory has an anomaly free description on twistor space. However, the $\cN=2$ string depends on a choice of K\"{a}hler structure on the target, so it cannot describe the theory of sd gravity we've studied in this work. We expect it's related to a holomorphic Poisson-Chern-Simons theory on twistor space, which can be defined using a choice of K\"{a}hler structure on space-time along the lines of holomorphic Chern-Simons theory \cite{Bittleston:2020hfv,Costello:2021bah}. This theory suffers from a 1-loop anomaly equal to $1/2$ of that in holomorphic Poisson-BF theory. If this really is the $\cN=2$ string, then the anomaly must be cancelled, but how?
\end{itemize}

\acknowledgments

It is a pleasure to thank Kevin Costello and Natalie Paquette for very helpful comments on a draft version of this paper, and Kasia Budzik, Kevin Costello, Maciej Dunajski, Davide Gaiotto, Lionel Mason, Natalie Paquette and Jingxiang Wu for useful discussions.

\section*{Declarations}

Research at Perimeter Institute is supported in part by the Government of Canada through the Department of Innovation, Science and Economic Development and by the Province of Ontario through the Ministry of Colleges and Universities. AS has been supported by a Mathematical Institute Studentship, Oxford and by the ERC grant GALOP ID: 724638. The work of DS has been partly funded by STFC HEP Theory Consolidated grant ST/T000694/1. The authors have no competing interests to declare that are relevant to the content of this article. Data sharing not applicable to this article as no datasets were generated or analysed during the current study. 

%\newpage

%%%%%%%%%%%%%%%%%%%%%%%%%%%%%%%%%%
%%%%%%%%%%%%%%%%%%%%%%%%%%%%%%%%%%
%%%%%%%%%%%%%%%%%%%%%%%%%%%%%%%%%%

\begin{appendix}

%%%%%%%%%%%%%%%%%%%%%%%%%%%%%%%%%%
%%%%%%%%%%%%%%%%%%%%%%%%%%%%%%%%%%

\section{Direct computation of the mixed anomaly} \label{app:mixed-anomaly-calc}

In this appendix we compute the mixed Poisson-gauge anomaly directly using the same approach as in section \ref{sec:anomaly-calc}, {\it i.e.}, we restrict to the patch $\{\lambda\neq\hat\alpha\}\subset\PT^3$ and use the BV formalism. Recall that in addition to extending $h,g$ to polyform fields $\bh,\bg$, $a,b$ are also expanded to
\be 
\mathbf{a}\in\Omega^{0,\bullet}(\PT,\fg)[1]\,,\qquad \mathbf{b}\in\Omega^{3,\bullet}(\PT,\fg^\vee)[1]\,. \ee
We use an identical gauge fixing and regularisation as in subsection \ref{subsec:BV-quantum}, so that the $\bh$-$\bg$ propagator is the pullback of the $(0,2)$-form propagator from equation \eqref{eq:reg-prop}. The $\ba$-$\bb$ propagator takes the same form, though is accompanied by the identity $\mathrm{id.}\in\fg\otimes\fg^\vee$.\\

We wish to evaluate the BRST variations of both of the diagrams appearing in figure \ref{fig:PG-mixed-anomaly}. These can be computed as a sum over all ways of replacing an internal propagator by a heat kernel. Unfortunately there are inequivalent ways of specifying an internal edge for both diagrams.

\begin{figure}[!ht]
\centering
  \begin{subfigure}[t]{0.49\textwidth}
  \centering
    \includegraphics[scale=0.3]{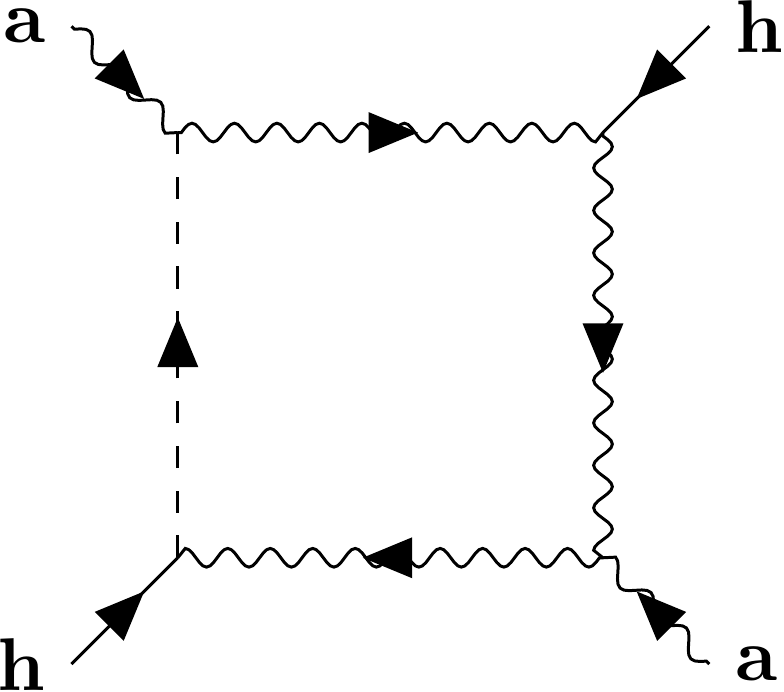}
  \caption*{(O1)}
  \end{subfigure}
\caption{\emph{Gauge variation of mixed anomaly diagram (O).}} \label{fig:O1-mixed-anomaly}
\end{figure}

For the diagram (O) there are 2 ways of doing so up to isomorphism. Representing the distinguished edge by a dotted line, one of these possibilities is illustrated in figure \ref{fig:O1-mixed-anomaly}. The other is obtained by reversing the direction of the internal propagators. Neither has a symmetry factor, and they contribute equally to the anomaly.

\begin{figure}[!ht]
\centering
  \begin{subfigure}[t]{0.32\textwidth}
  \centering
    \includegraphics[scale=0.3]{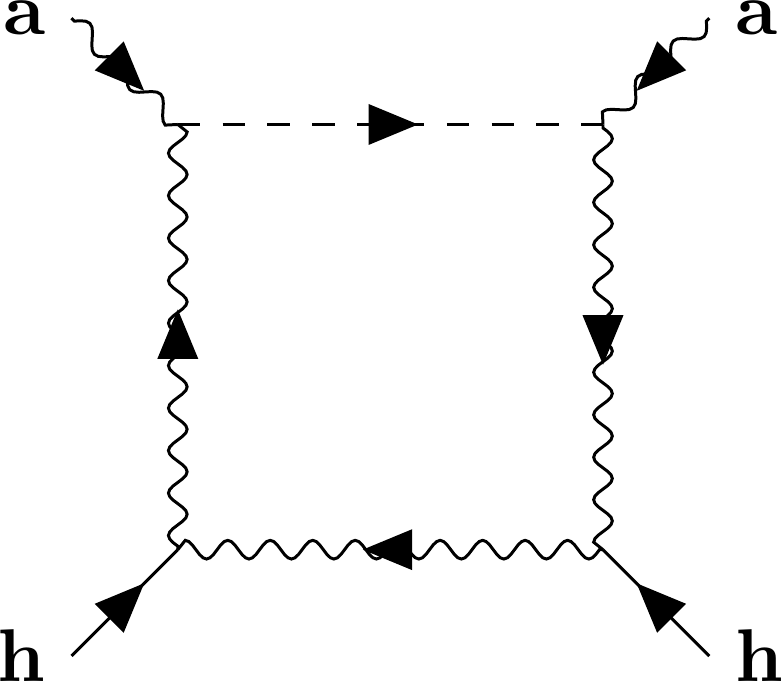}
  \caption*{(A1)}
  \end{subfigure}
  \begin{subfigure}[t]{0.32\textwidth}
  \centering
    \includegraphics[scale=0.3]{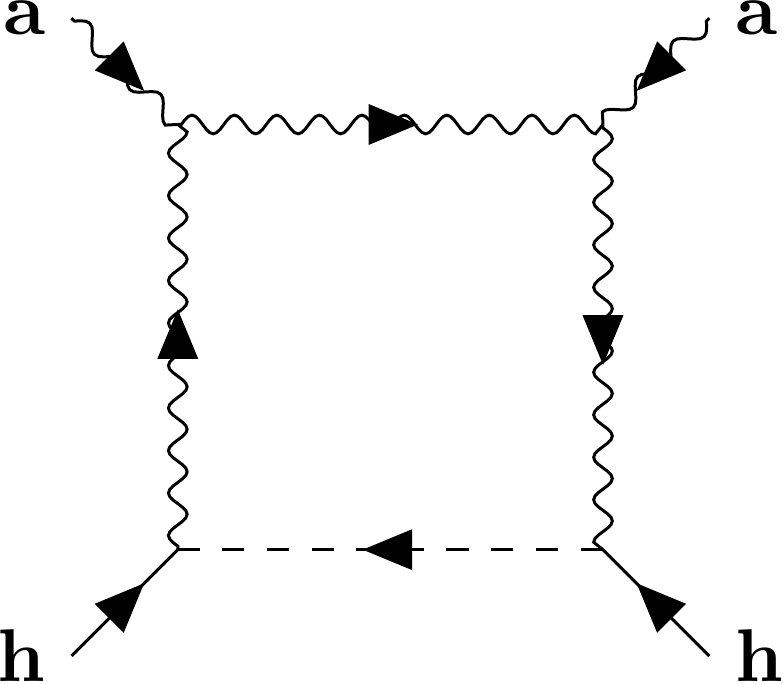}
  \caption*{(A2)}
  \end{subfigure}
  \begin{subfigure}[t]{0.32\textwidth}
  \centering
    \includegraphics[scale=0.3]{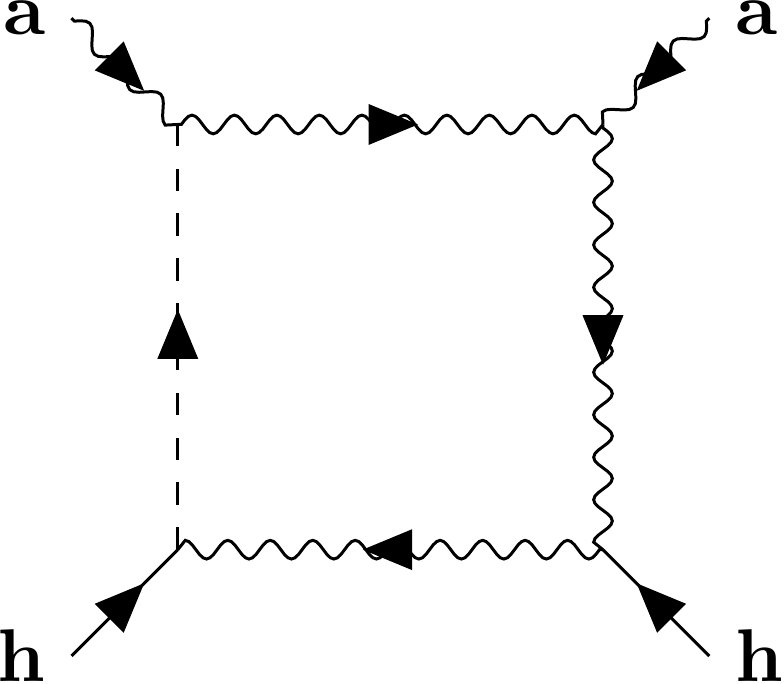}
  \caption*{(A3)}
  \end{subfigure}
\caption{\emph{Gauge variation of mixed anomaly diagram (A)}} \label{fig:A123-mixed-anomaly}
\end{figure}

On the other hand, for the diagram (A) there are 4 distinct ways of specifying an internal edge up to isomorphism. 3 options are illustrated in figure \ref{fig:A123-mixed-anomaly}. The final diagram is obtained by reversing the orientation of the internal propagators in (A3). Its contribution is equal to that of (A3). All diagrams are unique up to isomorphism, and so they are not accompanied by symmetry factors.

We therefore only need to evaluate the diagrams (O1), (A1), (A2) and (A3). Since these computations are similar to those performed in evaluating the box diagram in section \ref{sec:anomaly-calc} and appendix \ref{app:anomaly-deats} we will be brief, emphasising only the major differences. 

Let's begin by considering the 2 diagrams of type (O1). Together they contribute
\be \label{eq:O1-anomaly} -2f_{\sfd\sfa}^{~~\sfc}f_{\sfc\sfb}^{~~\sfd}\bigg(\frac{1}{2\pi}\bigg)^4\int_{(\C^3)^4}\d^3z_1\ba^\sfa_1P^{[\epsilon,L]}_{12}\d^3z_2\{\bh_2,P^{[\epsilon,L]}_{23}\}_2\d^3z_3\ba_3^\sfb\d^3z_4\{\bh_4,\overline K^\epsilon_{41}\}_4\,. \ee
We always take the distinguished edge to connect the vertices located at $z_3,z_4$. Substituting in the explicit forms of the regulated propagator and heat kernel, and applying the identity \eqref{eq:epsilon-identity} gives
\be
\begin{split}
&\frac{2\cdot2\mathbf{h}^\vee \kappa_{\sfa\sfb}}{4^2}\varepsilon^{a_1a_2a_3}\int_{(\C^3)^4}\d^3z_1\ba_1^\sfa\d^3z_2\dots\d^3z_4\p^{\dal_4}\bh_4\d^3\bar z_{41}\,\varepsilon^{a_1a_2a_3}{\bar z}_{12a_1}{\bar z}_{23a_2}{\bar z}_{34a_3} \\
&\qquad\bar z_{23\dal_2}\bar z_{41\dal_4}\int_{[\epsilon,L]^3}\frac{\d^3\vec T}{\epsilon T_1T_2^2T_3}g^{\vec T,\epsilon}(z_{12},z_{23},z_{34})\,.
\end{split}
\ee
Here we've used the fact $f_{\sfd\sfa}^{~~\sfc}f_{\sfc\sfb}^{~~\sfd} = 2\bh^\vee\kappa_{\sfa\sfb}$ for $\bh^\vee$ the dual Coxeter number and $\kappa_{\sfa\sfb}=\tr(\mathrm{t}_\sfa\mathrm{t}_\sfb)$, which follows from of normalization of the $\fg$-invariant bilinear $\tr$. The only substantial difference with the pure Poisson-BF case is that there are fewer explicit factors of
\be \frac{\bar z_{i(i+1)}}{4T_M} \ee
on account of the fact that the holomorphic BF theory interaction involves no derivatives. Using the identities \eqref{eq:eta-id} these can be replaced with copies of $\eta_{i(i+1)}$ acting on $g^{\vec T,\epsilon}$. Subsequently integrating by parts and employing equation \eqref{eq:3-eta-id} gives
\be \begin{split}
&2\cdot4^3\cdot2\bh^\vee \kappa_{\sfa\sfb}\varepsilon^{a_1a_2a_3}\int_{[\epsilon,L]^3}\frac{\d^3\vec T\epsilon}{(\epsilon+T_1+T_2+T_3)}\int_{(\C^3)^4}\cL_{\eta^2_{\dal_2}}\cL_{\eta_{4\dal_4}}(\d^3z_1\p_{a_1}\ba_1^\sfa\dots\\
&\qquad\d^3z_4\p^{\dal_4}\bh_4)\d^3\bar w_1\d^3\bar w_2\d^3\bar w_3\,g^{\vec T,\epsilon}(w_1,w_2,w_3)\,.
\end{split} \ee
Expanding the object multiplying the Gaussian $g^{\vec T,\epsilon}$ in modes around $w_1=w_2=w_3=0$, only the zero-mode contributes in the limit $\lim_{L\to0}\lim_{\epsilon\to0}$. This allows us to perform the integrals over $w_1$, $w_2$ and $w_3$. Substituting in the explicit form of the vector fields $\eta_{2\dal_2},\eta_{4\dal_4}$ we can then simplify using antisymmetry of the bi-vector and integration by parts in $\eta_{4\dal_4}$. We arrive at the following expression
\be \label{eq:O1-calc} \begin{split}
&2\cdot2\bh^\vee\kappa_{\sfa\sfb}\bigg(\frac{\im}{2\pi}\bigg)^3\int_{[\epsilon,L]^3}\frac{\d^3\vec T\epsilon}{(\epsilon+T_1+T_2+T_3)^6}\int_{\C^3}\bigg(\epsilon\frac{\p}{\p z_1^{\dal_2}} - T_3\frac{\p}{\p z_3^{\dal_2}}\bigg) \\
&\qquad\bigg(-(T_1+T_2)\frac{\p}{\p z_1^{\dal_4}} - T_2\frac{\p}{\p z_2^{\dal_4}}\bigg)(\p\ba_1^\sfa\p^{\dal_2}\p\bh_2\p \ba_3^\sfb\p^{\dal_4}\bh_4)|_{z=z_1=z_2=z_3=z_4}\,.
\end{split} \ee
At this point we simply expand out the differential operators, perform the integration over $\vec T$ and then take $\lim_{\epsilon\to0}$. These steps have been relegated to appendix \ref{app:anomaly-deats}. After permuting factors and relabelling indices we find that the result can be expressed as a sum of 3 terms
\be \label{eq:O1-anomaly-2}
\frac{4\cdot2\bh^\vee}{5!\cdot 2^5}\bigg(\frac{\im}{2\pi}\bigg)^3\int_{\C^3}(- 3\cB_1 + 13\cB_2 + 5\cB_3)\,,
\ee
where
\be \begin{split} \label{eq:O1-terms}
\cB_1 &= \tr(\p_\dal\p_\db\p\ba\p\ba)\p^\dal\p\bh\p^\db\bh\,, \\
\cB_2 &= \tr(\p_\dal\p\ba\p_\db\p\ba)\p^\dal\p\bh\p^\db\bh\,, \\
\cB_3 &= \tr(\p_\dal\p\ba\p \ba)\p_\db\p^\dal\p\bh\p^\db\bh\,.
\end{split} \ee

Now let's turn our attention to the remaining diagrams. The diagram (A1) contributes
\be -f_{\sfd\sfa}^{~~\sfc}f_{\sfc\sfb}^{~~\sfd}\int_{(\C^3)^4}\d^3z_1\ba^\sfa_1P^{[\epsilon,L]}_{12}\d^3z_2\{\bh_2,P^{[\epsilon,L]}_{23}\}_2\d^3z_3\{\bh_3,P^{[\epsilon,L]}_{34}\}_3\d^3z_4\ba_4^\sfb\overline K^\epsilon_{41}\,. \ee
Repeating the procedure outlined above in the limit $\lim_{L\to0}\lim_{\epsilon\to0}$ this is equal to 
\be \begin{split} \label{eq:A1-calc}
&2\bh^\vee\kappa_{\sfa\sfb}\bigg(\frac{\im}{2\pi}\bigg)^3\int_{[\epsilon,L]^3}\frac{\d^3\vec T\epsilon}{(\epsilon+T_1+T_2+T_3)^6}\int_{\C^3}\bigg(\epsilon\frac{\p}{\p z_1^{\dal_2}} - T_3\frac{\p}{\p z_3^{\dal_2}}\bigg) \\
&\qquad\bigg(\epsilon\frac{\p}{\p z_1^{\dal_3}} + (\epsilon+T_1)\frac{\p}{\p z_2^{\dal_3}}\bigg)(\p\ba_1^\sfa\p^{\dal_2}\p\bh_2\p^{\dal_3}\p\bh_3\ba_4^\sfb)|_{z=z_1=z_2=z_3=z_4}\,.
\end{split} \ee
Aside from reordering the factors of $\ba,\bh$, the differences between this and equation \eqref{eq:O1-calc} are in the replacement $\eta_{4\dal_4}\mapsto\eta_{3\dal_3}$. In appendix \ref{app:anomaly-deats} we show that expanding out the differential operators, performing the integral over $\vec T$ and taking the limit $\lim_{\epsilon\to0}$ gives
\be \label{eq:A1-anomaly}
\frac{2\bh^\vee}{5!\cdot 2^5}\bigg(\frac{\im}{2\pi}\bigg)^3\int_{\C^3}(\cB_1 + \cB_3 - 16\cB_4)
\ee
for $\cB_1,\cB_3$ as above and
\be \label{eq:A1-term}
\cB_4 = \tr(\p\ba\p\ba)\p^\db\p_\dal\p\bh\p^\dal\p_\db\bh\,.
\ee

Similar calculations, also relegated to appendix \ref{app:anomaly-deats}, show that diagram (A2) contributes
\be \label{eq:A2-anomaly}
\frac{2\bh^\vee}{5!\cdot 2^5}\bigg(\frac{\im}{2\pi}\bigg)^3\int_{\C^3}(34\cB_1+13\cB_2+86\cB_3)
\ee
and the diagram (A3) contributes
\be \label{eq:A3-anomaly}
-\frac{2\cdot2\bh^\vee}{5!\cdot2^5}\bigg(\frac{\im}{2\pi}\bigg)^3\int_{\C^3}(22\cB_1+43\cB_2)\,.
\ee

Collecting terms from equations \eqref{eq:O1-anomaly}, \eqref{eq:A1-anomaly}, \eqref{eq:A2-anomaly} and \eqref{eq:A3-anomaly}, the total mixed anomaly is
\be \begin{split} \label{eq:mixed-anomaly-total}
&-\frac{2\bh^\vee}{5!\cdot 2^5}\bigg(\frac{\im}{2\pi}\bigg)^3\int_{\C^3}(21\cB_1 + 21\cB_2 - 107\cB_3 + 16\cB_4)\,.
\end{split} \ee
This expression can be dramatically simplified by integrating by parts. For example
\be \cB_1 + \cB_2 = \p_\db(\tr(\p_\dal\p\ba\p\ba)\p^\dal\p\bh\p^\db\bh) - \cB_3\,. \ee
Since $\cB_1,\cB_2$ have the same coefficient in equation \eqref{eq:mixed-anomaly-total}, this allows us to eliminate both in favour of $\cB_3$. Furthermore
\be \cB_3 = \p_\dal(\tr(\p\ba\p\ba)\p_\db\p^\dal\p\bh\p^\db\bh) - \cB_3 - \cB_4\,, \ee
allowing us to eliminate $\cB_3$ in favour of $\cB_4$. We therefore find that the total anomaly is
\be -\frac{21+107+32}{2^9\cdot3\cdot5}2\mathbf{h}^\vee\bigg(\frac{\im}{2\pi}\bigg)^3\int_{\C^3}\cB_4 = -\frac{2\bh^\vee}{48}\bigg(\frac{\im}{2\pi}\bigg)^3\int_{\C^3}\cB_4\,. \ee
Writing this out explicitly, the mixed Poisson-gauge anomaly is
\be -\frac{2\bh^\vee}{48}\bigg(\frac{\im}{2\pi}\bigg)^3\int_{\C^3}\tr(\p\ba\p\ba)\p^\db\p_\dal\p\bh\p^\dal\p_\db\bh\,, \ee
or in terms of $\bs^\dal_{~\,\db} = - \p^\dal\p_\db\bh$
\be \label{eq:BV-mixed-cocycle} -\frac{2\bh^\vee}{48}\bigg(\frac{\im}{2\pi}\bigg)^3\int_{\C^3}\tr(\p\ba\p\ba)\tr(\bs\p\bs)\,. \ee
It is easy to verify that this is independent of our choice of trivialization of $\cO(n)$, and so takes the same form on $\PT$.

In order to recover the mixed anomaly cocycles in subsection \ref{subsec:anom-box} we restrict $\ba,\bs$ to their physical and ghost parts
\be \begin{split}
&\ba = c + a\,,\qquad\bs^\db_{~\,\dal} = \psi^\db_{~\,\dal} + s^\db_{~\,\dal} = - \p^\db\p_\dal\bh = - \p^\db\p_\dal\chi - \p^\db\p_\dal h\,.
\end{split} \ee
Substituting into \eqref{eq:BV-mixed-cocycle} we get
\be -\frac{2\bh^\vee}{24}\int_{\C^3}\tr(c\p a)\tr(\p s^2) - \frac{2\bh^\vee}{24}\int_{\C^3}\tr(\chi\p s)\tr(\p a^2)\,, \ee
which coincides with equation \eqref{eq:PGBF-mixed-cocycle}. The extra terms present in the mixed BV cocycle render it BRST invariant off-shell.

%%%%%%%%%%%%%%%%%%%%%%%%%%%%%%%%%%
%%%%%%%%%%%%%%%%%%%%%%%%%%%%%%%%%%

\section{Evaluating the contribution of \texorpdfstring{$\eta$}{eta} to the anomaly} \label{app:eta-box}

In section \ref{subsec:PBF-cancellation} we saw that in order to consistently couple the field $\eta$ to Poisson-BF theory we needed to introduce a tree level coupling between $\eta$ and $h$. One consequence of this is that $\eta$ itself contributes to the Poisson anomaly by running through the loop of the box diagram, as illustrated in figure \ref{fig:eta-anomaly}. In this appendix we evaluate its contribution by explicitly calculating the BRST variation of this diagram using the BV techniques employed in section \ref{sec:anomaly-calc} and appendix \ref{app:mixed-anomaly-calc}. We refer the reader to subsection \ref{subsec:BV-eta} for a review of the classical BV description of $\eta$, which involves replacing it with the polyform field $\bseta$. Recall that when performing these explicit computations we work in a patch of $\PT$ biholomorphic to $\C^3$.

First need to construct the $\bseta$ propagator, which obeys the polyform counterpart of equation \eqref{eq:eta-prop}:
\be (\bar\p_1+\bar\p_2)P^{\bseta}_{12} = -2\pi\im\p_1\delta^6_{12} \ee
Here $\delta^6$ is the 6-form $\delta$-function obeying
\be \int_{\C^3}\delta^6 = 1\,, \ee
and we restrict $\delta^6_{12}$ to its $(1,\bullet)$- and $(2,\bullet)$-form parts on the $1\textsuperscript{st}$ and $2\textsuperscript{nd}$ factors of $(\C^3)_1\times(\C^3)_2$ respectively. Note that before performing this restriction we may write $\delta^6 = -\d^3Z\bar\delta^3$, so that
\be \delta^6_{12} = -\frac{1}{2}\eps_{\dal\db}(\d z_1\d v^\dal_2\d v^\db_2 + 2\d z_2\d v^\dal_1\d v^\db_2)\bar\delta^3_{12} = -\d^3Z_{12}^{1,2}\bar\delta^3_{12}\,. \ee
Therefore
\be (\bar\p_1+\bar\p_2)P_{12}^{\bseta} = -2\pi\im(\d^3Z^{1,2}_{12}\p_1)\bar\delta^3_{12}\,. \ee
Recalling that the $\bg$-$\bh$ polyform propagator obeys
\be \bar\p P = -2\pi\bar\delta^3\,, \ee
we can obtain the $\bseta$ propagator as
\be P_{12}^{\bseta} = \im(\d^3Z^{1,2}_{12}\p_1)P_{12}\,. \ee
To avoid divergences in the loop diagram this needs to be regulated. Fortunately we can employ exactly the same heat kernel regularisation as used in the case of Poisson-BF theory. We simply replace $P_{12}$ with its mollified counterpart $P^{[\epsilon,L]}_{12}$.

The final ingredient we need is the interaction vertex, given by
\be -\frac{1}{4\pi\im}\int_\PT\bseta\,\{\bh,\ \}\ip\bseta = -\frac{1}{4\pi\im}\int_\PT\bseta\,\p^\dal \bh\,\p_\dal\ip\bseta\,. \ee
It takes exactly the same form upon restriction to $\C^3$.
%The factor of $1/2$ in the above is the expected symmetry factor for a vertex quadratic in $\eta$.

Now let's evaluate the BRST variation of the box diagram illustrated in figure \ref{fig:eta-anomaly}, which is equal to
\be \label{eq:axion-anomaly-formula} -\frac{1}{2}\bigg(\frac{1}{2\pi}\bigg)^4\int_{(\C^3)^4}\{\bh_1,\ \}_1\ip(\d^3Z^{1,2}_{12}\p_1)P^{[\epsilon,L]}_{12}\dots\{\bh_4,\ \}_4\ip(\d^3Z^{1,2}_{41}\p_4)\overline K^\epsilon_{41}\,. \ee
The overall minus sign appears because a Grassmann odd field is running through the loop, and the factor of $1/2$ is due the $\Z_2$ symmetry of the box diagram. This symmetry is present because $\bseta$ propagators are undirected, in contrast to BF-type theory propagators.

Our strategy will be to massage the expression \eqref{eq:axion-anomaly-formula} into a multiple of equation \eqref{eq:anomaly-calc-1}, which we've already evaluated. We can expand
\be \begin{aligned} &\d^3Z^{1,2}_{12}\p_1 = \frac{1}{2}\eps_{\dal\db}(\d z_1\d v^\dal_2\d v^\db_2 + 2\d z_2\d v^\dal_1\d v^\db_2)\p_1 \\
&= \frac{1}{2}\eps_{\dal\db}(\d z_1\d v_1^\dg\d v_2^\dal\d v_2^\db\p_{v_1^\dg} - \d z_2\d v_1^\dal\d v_1^\db\d v_2^\dg\p_{v_1^\dg} - 2\d z_1\d z_2\d v_1^\dal\d v_2^\db\p_{z_1})\,,
\end{aligned} \ee
so that
%\be \begin{aligned}
%\p_{v^1_1}\ip\D^3Z^{1,2}_{12}\p_1 &= -\d z_1\d z_2\d v^2_2\p_{z_1} - (\d z_1\d v^1_2\d v^2_2 + \d v^2_1\d z_2\d v^1_2)\p_{v^1_1} - \d v^2_1\d z_2\d v^2_2\p_{v^1_2}\,, \\
%\p_{v^2_1}\ip\D^3Z^{1,2}_{12}\p_1 &= \d z_1\d z_2\d v^1_2\p_{z_1} +\d v^1_1 \d z_2\d v^1_2\p_{v^1_1} + (\d z_1\d v^1_2\d v^2_2 + \d v^1_1\d z_2\d v^2_2)\p_{v^1_2}\,, \\
%\end{aligned} \ee
%or more succinctly,
\be \p_{v^\dal_1}\ip\d^3Z^{1,2}_{12} = \frac{1}{2}\eps_{\dal\db}(2\d v^\db_1\d z_2\d v^\dg_2\p_{v^\dg_1} - \d z_1\d z_2\d v^\db_2\p_{z_1}) - \frac{1}{2}\eps_{\db\dg}\d z_1\d v^\db_2\d v_2^\dg\p_{v^\dal_1}\,. \ee
The term involving $\d z_1\d z_2$ cannot contribute in \eqref{eq:axion-anomaly-formula}. This is because each copy of $\p_{v_i^\dal}\ip\D^3Z_{ij}^{1,2}\p_i$ must contribute at least one $\d z_k$ with either $k=i,j$, but a maximum of four can appear before the integral is saturated. We can therefore replace
\be \p_{v^\dal_1}\ip\d^3Z^{1,2}_{12} \mapsto\frac{1}{2}(2\eps_{\dal\db}\d v_1^\db\d z_2\d v_2^\dg - \delta_\dal^{~\,\dg}\eps_{\db\dd}\d z_1\d v_2^\db\d v^\dd_2)\p_{v_1^\dg} = \eps_{\dal\db}\Omega^{\db\dg}_{12}\p_{v_1^\dg}\,. \ee
Substituting this into equation \eqref{eq:axion-anomaly-formula} gives
\be \label{eq:axion-anomaly-1} -\frac{1}{2}\bigg(\frac{1}{2\pi}\bigg)^4\int_{(\C^3)^4}\Omega^{\dal_1\db_1}_{12}\Omega^{\dal_2\db_2}_{23}\Omega^{\dal_3\db_3}_{34}\Omega^{\dal_4\db_4}_{41}\p_{v_1^{\dal_1}}\bh_1\p_{v^{\db_1}_1}P^{[\epsilon,L]}_{12}\dots\p_{v^{\dal_4}_4}\bh_4\p_{v^{\db_4}_4}\overline K^\epsilon_{41}\,, \ee
which can be simplified by expanding
\be %\begin{aligned}
\Omega^{\dal_1\db_1}_{12}\Omega^{\dal_2\db_2}_{23}\Omega^{\dal_3\db_3}_{34}\Omega^{\dal_4\db_4}_{41}
= \d^3Z_1\d^3Z_2\d^3Z_3\d^3Z_4(\eps^{\dal_1\db_1}\eps^{\dal_2\db_2}\eps^{\dal_3\db_3}\eps^{\dal_4\db_4} + \eps^{\dal_1\db_4}\eps^{\dal_2\db_1}\eps^{\dal_3\db_2}\eps^{\dal_4\db_3})
%\end{aligned}
\ee
to obtain
\be \begin{aligned} \label{eq:axion-anomaly-2}
&-\frac{1}{2}\bigg(\frac{1}{2\pi}\bigg)^4\int_{(\C^3)^4}\big(\d^3Z_1\p_{v_1^{\dal_1}}\bh_1\p_{v_{1\dal_1}}P^{[\epsilon,L]}_{12}\dots\d^3Z_4\p_{v_4^{\dal_4}}\bh_4\p_{v_{4\dal_4}}\overline K^\epsilon_{41} \\
&+ \d^3Z_1\p_{v_1^{\dal_1}}
\bh_1\p_{v_{1\dal_2}}P^{[\epsilon,L]}_{12}\d^3Z_2\p_{v_2^{\dal_2}}\bh_2\p_{v_{2\dal_3}}P^{[\epsilon,L]}_{23}\d^3Z_3\p_{v_3^{\dal_3}}\bh_3\p_{v_{3\dal_4}}P^{[\epsilon,L]}_{34}\d^3Z_4\p_{v_4^{\dal_4}}\bh_4\p_{v_{4\dal_1}}\overline K^\epsilon_{41}\big)\,.
\end{aligned} \ee
The first term is exactly $1/2$ of equation \eqref{eq:anomaly-calc-1}, so let's concentrate on the second. By translational invariance of the propagator it's equal to
\bea
&-\frac{1}{2}\bigg(\frac{1}{2\pi}\bigg)^4\int_{(\C^3)^4}\p_{v_{1\dal_1}}\overline K^\epsilon_{41}\p_{v_1^{\dal_1}}
\bh_1\d^3Z_1\p_{v_{2\dal_2}}P^{[\epsilon,L]}_{12}\p_{v_2^{\dal_2}}\bh_2\d^3Z_2 \\
&\p_{v_{3\dal_3}}P^{[\epsilon,L]}_{23}\p_{v_3^{\dal_3}}\bh_3\d^3Z_3\p_{v_{4\dal_4}}P^{[\epsilon,L]}_{34}\p_{v_4^{\dal_4}}\bh_4\d^3Z_4 \\
&= -\frac{1}{2}\bigg(\frac{1}{2\pi}\bigg)^4\int_{(\C^3)^4}\{\overline K_{41}^\epsilon,\bh_1\}_1\d^3Z_1\{P^{[\epsilon,L]}_{12},\bh_2\}_2\d^3Z_2\{P^{[\epsilon,L]}_{23},\bh_3\}\d^3Z_3\{P^{[\epsilon,L]}_{34},\bh_4\}\d^3Z_3\,.
\eea
Up to an overall factor of $1/2$ this is precisely the expression we'd get from evaluating diagram \ref{fig:PBF-anomaly-BRST} if the direction of all internal propagators were reversed. However, the value of the diagram is independent of the orientation of these propagators.

We therefore conclude that both terms in equation \eqref{eq:axion-anomaly-2} equate to $1/2$ of equation \eqref{eq:anomaly-calc-1}, and the overall contribution of $\bseta$ to the anomaly when running through the loop of the box diagram the same as for Poisson-BF fields. This was computed in subsection \ref{subsec:loopsec} to be
\be \frac{1}{4\cdot5!}\bigg(\frac{\im}{2\pi}\bigg)^3\int_{\C^3}\tr(\bs(\p\bs)^3)\,. \ee
We use this result in subsections \ref{subsec:PBF-cancellation} and \ref{subsec:PGBF-cancellation} when cancelling anomalies using $\bseta$. It plays a particularly important role in the latter, where it is responsible for the shift $\dim(\fg)+1\mapsto\dim(\fg)+2$ leading to simultaneous cancellation of the gauge, mixed and gravitational twistorial anomalies.

%%%%%%%%%%%%%%%%%%%%%%%%%%%%%%%%%%
%%%%%%%%%%%%%%%%%%%%%%%%%%%%%%%%%%

\section{Anomaly calculation details} \label{app:anomaly-deats}

In this appendix we evaluate integrals appearing in the calculation of anomalies in subsections \ref{subsec:loopsec} and \ref{app:mixed-anomaly-calc}.\\

We begin by evaluating the expression
\bea \label{eq:anomaly-deats-1}
&\lim_{\epsilon\to0}\bigg(\frac{\im}{2\pi}\bigg)^3\int_{\C^3}\int_{[\epsilon,L]^3}\frac{\d^3\vec T\epsilon}{(\epsilon+T_1+T_2+T_3)^8}\bigg(-(T_2 + T_3)\frac{\p}{\p z_2^{\dal_1}} - T_3\frac{\p}{\p z_3^{\dal_1}}\bigg) \\
&\qquad\bigg(\epsilon\frac{\p}{\p z_1^{\dal_2}} - T_3\frac{\p}{\p z_3^{\dal_2}}\bigg)\bigg(\epsilon\frac{\p}{\p z^{\dal_3}_1} + (\epsilon + T_1)\frac{\p}{\p z^{\dal_3}_2}\bigg)\bigg( - (T_1 + T_2)\frac{\p}{\p z_1^{\dal_4}} - T_2\frac{\p}{\p z_2^{\dal_4}}\bigg) \\
&\qquad\qquad(\p^{\dal_1}\p\bh_1\p^{\dal_2}\p \bh_2\p^{\dal_3}\p \bh_3\p^{\dal_4}\bh_4)|_{z=z_1=z_2=z_3=z_4}
\eea
for the anomaly in pure Poisson-BF theory, which we obtained in subsection \ref{subsec:loopsec}. We proceed by expanding out the product of differential operators. Up to permuting factors of $\bh$ and relabelling indices we get 4 distinct terms.
\be \begin{split} \label{eq:expand-diff-ops}
&\bigg(-(T_2 + T_3)\frac{\p}{\p z_2^{\dal_1}} - T_3\frac{\p}{\p z_3^{\dal_1}}\bigg)\bigg(\epsilon\frac{\p}{\p z_1^{\dal_2}} - T_3\frac{\p}{\p z_3^{\dal_2}}\bigg)\bigg(\epsilon\frac{\p}{\p z^{\dal_3}_1} + (\epsilon + T_1)\frac{\p}{\p z^{\dal_3}_2}\bigg) \\
&\qquad\bigg(-(T_1 + T_2)\frac{\p}{\p z_1^{\dal_4}} - T_2\frac{\p}{\p z_2^{\dal_4}}\bigg)(\p^{\dal_1}\p \bh_1\dots\p^{\dal_4}\bh_4)|_{z=z_1=z_2=z_3=z_4} \\
&= (-(\epsilon+T_1)(T_1+T_2)(T_2+T_3)T_3 - (\epsilon+T_1)(T_1+T_2)T_3^2 + \epsilon^2T_2T_3 - \epsilon T_2T_3^2)\cA_1 \\
&+ (-\epsilon(T_1+T_2)T_3^2 + \epsilon(\epsilon + T_1)(T_1+T_2)(T_2+T_3) + \epsilon^2T_2(T_2+T_3) - (\epsilon+T_1)T_2T_3^2)\cA_2 \\
&+ (-\epsilon(T_1+T_2)(T_2+T_3)T_3 + \epsilon(\epsilon + T_1)(T_1+T_2)T_3 - \epsilon T_2(T_2+T_3)T_3 + \epsilon(\epsilon + T_1)T_2T_3)\cA_3 \\
&+ (\epsilon^2(T_1+T_2)(T_2+T_3) + \epsilon^2(T_1+T_2)T_3 + \epsilon(\epsilon + T_1)T_2(T_2+T_3) - (\epsilon+T_1)T_2(T_2+T_3)T_3)\cA_4\,,
\end{split} \ee
where
\be
\begin{split}
\cA_1 &= \p_\db\p_\dg\p^\dal\p\bh\p_\dd\p^\db\p\bh\p_\dal\p^\dg\p\bh\p^\dd\bh\,, \\
\cA_2 & = \p_\db\p_\dg\p^\dal\p\bh\p_\dal\p_\dd\p^\db\p\bh\p^\dg\p\bh\p^\dd\bh\,, \\
\cA_3 &= \p_\dg\p_\dd\p^\dal\p\bh\p_\dal\p^\db\p\bh\p_\db\p^\dg\p\bh\p^\dd\bh\,, \\
\cA_4 &= \p_\db\p_\dg\p_\dd\p^\dal\p\bh\p_\dal\p^\db\p\bh\p^\dg\p\bh\p^\dd\bh\,. \\
\end{split}
\ee
Fortunately $\cA_3$ vanishes
\be \begin{split}
&\cA_3 = \p_\dd\p_\dg\p^\dal\p\bh\p_\dal\p^\db\p\bh\p_\db\p^\dg\p\bh\p^\dd\bh = - \p_\dd\p^\dg\p_\dal\p\bh\p^\dal\p_\db\p\bh\p^\db\p_\dg\p\bh\p^\dd\bh \\
&= -\p_\dd\p_\dal\p^\dg\p\bh\p_\dg\p^\db\p\bh\p_\db\p^\dal\p\bh\p^\dd\bh = -\cA_3
\end{split} \ee
and $\cA_4$ is a total derivative
\be \begin{split}
&\cA_4 = \p_\db(\p_\dg\p_\dd\p^\dal\p\bh\p_\dal\p^\db\p\bh\p^\dg\p\bh\p^\dd\bh) - \p_\dg\p_\dd\p^\dal\p\bh\p_\dal\p^\db\p\bh\p_\db\p^\dg\p\bh\p^\dd\bh \\
&- \p_\dg\p_\dd\p^\dal\p\bh\p_\dal\p^\db\p\bh\p^\dg\p\bh\p_\db\p^\dd\bh = \p_\db(\p_\dg\p_\dd\p^\dal\p\bh\p_\dal\p^\db\p\bh\p^\dg\p\bh\p^\dd\bh) \\
&- \p(\p_\dg\p_\dd\p^\dal\p\bh\p_\dal\p^\db\bh\p^\dg\p\bh\p_\db\p^\dd\bh) - 2\cA_3\,.
\end{split} \ee
Furthermore
\be \begin{split}
&\cA_1 + \cA_2 = \p_\dal(\p_\db\p_\dg\p^\dal\p\bh\p_\dd\p^\db\p\bh\p^\dg\p\bh\p^\dd\bh) - \p_\db\p_\dg\p^\dal\p\bh\p_\dd\p^\db\p\bh\p^\dg\p\bh\p_\dal\p^\dd\bh \\
&= \p_\dal(\p_\db\p_\dg\p^\dal\p\bh\p_\dd\p^\db\p\bh\p^\dg\p\bh\p^\dd\bh) + \p(\p_\db\p_\dg\p^\dal\p\bh\p_\dal\p^\dd\bh\p_\dd\p^\db\p\bh\p^\dg\bh) - \cA_3\,,
\end{split} \ee
so that modulo exact terms $\cA_2 \sim - \cA_1$. Finally, we can massage $\cA_1$ into a slightly nicer form as follows
\bea &\cA_1 = \p_\dg(\p_\db\p^\dal\p\bh\p_\dd\p^\db\p\bh\p_\dal\p^\dg\p\bh\p^\dd\bh) - \p_\db\p^\dal\p\bh\p_\dd\p^\db\p\bh\p_\dal\p^\dg\p\bh\p_\dg\p^\dd\bh -\cA_3 \\ 
&= \p_\dg(\p_\db\p^\dal\p\bh\p_\dd\p^\db\p\bh\p_\dal\p^\dg\p\bh\p^\dd\bh) - \p^\dd\p_\dal\p\bh\p^\dal\p_\db\p\bh\p^\db\p_\dg\p\bh\p^\dg\p_\dd\bh\,.
\eea
Working up to total derivatives, equation \eqref{eq:expand-diff-ops} can therefore be expressed in terms of $\bs^\db_{~\,\dal} = - \p^\db\p_\dal\bh$ using the identities
\be \cA\coloneqq\tr(\bs\p\bs^3)\sim-\cA_1\sim\cA_2\,,\qquad\cA_3=0\,,\qquad\cA_4\sim0\,, \ee
as
\bea
&\big(T_1^2T_2T_3+ T_1T_2^2T_3+T_1T_2T_3^2+2T_1^2T_3^2 \\ 
&\qquad + (T_1T_2^2+T_1^2T_2+T_2T_3^2+T_2^2T_3+T_3T_1^2+T_3^2T_1+2T_1T_2T_3)\epsilon \\
&\qquad\qquad+ (T_1T_2+T_2T_3+T_3T_1+2T_2^2)\epsilon^2\big)\cA\,. \eea
At this point we are in a position to integrate over $\vec T$. This operation is invariant under permutations of $\{T_1,T_2,T_3\}$, so there are only 6 independent integrals to evaluate
\bea
&\lim_{\epsilon\to0}\int_{[\epsilon,L]^3}\frac{\d^3\vec{T}\,\epsilon T_1^2T_2T_3}{(T_1+T_2+T_3+\epsilon)^8} = \int_{[1,\infty)^3}\frac{\d^3\vec{t}\,t_1^2t_2t_3}{(t_1+t_2+t_3+1)^8} = \frac{5\cdot31}{7!\cdot2^7}\,, \\
&\lim_{\epsilon\to0}\int_{[\epsilon,L]^3}\frac{\d^3\vec{T}\,\epsilon T_1^2T_2^2}{(T_1+T_2+T_3+\epsilon)^8} = \int_{[1,\infty)^3}\frac{\d^3\vec{t}\,t_1^2t_2^2}{(t_1+t_2+t_3+1)^8} = \frac{239}{7!\cdot2^7}\,, \\
&\lim_{\epsilon\to0}\int_{[\epsilon,L]^3}\frac{\d^3\vec{T}\,\epsilon^2 T_1^2T_2}{(T_1+T_2+T_3+\epsilon)^8} = \int_{[1,\infty)^3}\frac{\d^3\vec{t}\,t_1^2t_2}{(t_1+t_2+t_3+1)^8} = \frac{2^2\cdot11}{7!\cdot2^7}\,, \\
&\lim_{\epsilon\to0}\int_{[\epsilon,L]^3}\frac{\d^3\vec{T}\,\epsilon^2 T_1T_2T_3}{(T_1+T_2+T_3+\epsilon)^8} = \int_{[1,\infty)^3}\frac{\d^3\vec{t}\,t_1t_2t_3}{(t_1+t_2+t_3+1)^8} = \frac{2^5}{7!\cdot2^7}\,, \\
&\lim_{\epsilon\to0}\int_{[\epsilon,L]^3}\frac{\d^3\vec{T}\,\epsilon^3 T_1T_2}{(T_1+T_2+T_3+\epsilon)^8} = \int_{[1,\infty)^3}\frac{\d^3\vec{t}\,t_1t_2}{(t_1+t_2+t_3+1)^8} = \frac{2^3\cdot13}{7!\cdot2^7}\,, \\
&\lim_{\epsilon\to0}\int_{[\epsilon,L]^3}\frac{\d^3\vec{T}\,\epsilon^3 T_1^2}{(T_1+T_2+T_3+\epsilon)^8} = \int_{[1,\infty)^3}\frac{\d^3\vec{t}\,t_1t_2}{(t_1+t_2+t_3+1)^8} = \frac{17}{7!\cdot2^7}\,. \\
\eea
We therefore find that the pure Poisson-BF anomaly is \eqref{eq:anomaly-deats-1}
\bea
&\frac{(3\cdot5\cdot31 + 2\cdot239 + 6\cdot2^2\cdot11 + 2\cdot2^5 + 3\cdot13 + 2\cdot17)}{7!\cdot 2^7}\bigg(\frac{\im}{2\pi}\bigg)^3\int_{\C^3}\cA \\
&=\frac{1}{4\cdot5!}\bigg(\frac{\im}{2\pi}\bigg)^3\int_{\C^3}\tr(\bs(\p\bs)^3)
\eea
This is the form used in subsection \ref{subsec:loopsec}.\\

In appendix \ref{app:mixed-anomaly-calc} we are required to evaluate 4 separate anomaly integrals.

The first, associated to the mixed Poisson-gauge anomaly diagram (O1), is \eqref{eq:O1-calc}
\be \begin{split} \label{eq:O1-calc-app}
&\lim_{\epsilon\to0}2\cdot2\bh^\vee\bigg(\frac{\im}{2\pi}\bigg)^3\int_{\C^3}\int_{[\epsilon,L]^3}\frac{\d^3\vec T\epsilon}{(\epsilon+T_1+T_2+T_3)^6}\bigg(\epsilon\frac{\p}{\p z_1^{\dal_2}} - T_3\frac{\p}{\p z_3^{\dal_2}}\bigg) \\
&\qquad\bigg(-(T_1+T_2)\frac{\p}{\p z_1^{\dal_4}}-T_2\frac{\p}{\p z_2^{\dal_4}}\bigg)\tr(\p\ba_1\p^{\dal_2}\p\bh_2\p\ba_3\p^{\dal_4}\bh_4)|_{z=z_1=z_2=z_3=z_4}\,.
\end{split} \ee
We proceed by expanding out
\be \begin{split}
&\bigg(\epsilon\frac{\p}{\p z_1^{\dal_2}} - T_3\frac{\p}{\p z_3^{\dal_2}}\bigg)\bigg(-(T_1+T_2)\frac{\p}{\p z_1^{\dal_4}}- T_2\frac{\p}{\p z_2^{\dal_4}}\bigg)\tr(\p\ba_1\p^{\dal_2}\p\bh_2\p\ba_3\p^{\dal_4}\bh_4)|_{z=z_1=z_2=z_3=z_4} \\
&= - \epsilon(T_1+T_2)\tr(\p_{\dal_2}\p_{\dal_4}\p\ba\p\ba)\p^{\dal_2}\p\bh\p^{\dal_4}\bh + T_3(T_1+T_2)\tr(\p_{\dal_4}\p\ba\p_{\dal_2}\p\ba)\p^{\dal_2}\p\bh\p^{\dal_4}\bh \\
&- \epsilon T_2\tr(\p_{\dal_2}\p\ba\p\ba)\p_{\dal_4}\p^{\dal_2}\p\bh\p^{\dal_4}\bh + T_2T_3\tr(\p\ba\p_{\dal_2}\p\ba)\p_{\dal_4}\p^{\dal_2}\p\bh\p^{\dal_4}\p\bh\,.
\end{split} \ee
The final two terms are equal by the cyclicity of the trace, so there are only two independent integrals to perform. These evaluate to
\bea \label{eq:mixed-integrals-1}
&\lim_{\epsilon\to0}\int_{[\epsilon,L]^3}\frac{\d^3\vec{T}\,\epsilon^2T_1}{(T_1+T_2+T_3+\epsilon)^6} = \int_{[1,\infty)^3}\frac{\d \vec{t}\,t_1}{(t_1+t_2+t_3+1)^6} = \frac{3}{5!\cdot 2^5}\,, \\
&\lim_{\epsilon\to0}\int_{[\epsilon,L]^3}\frac{\d^3\vec{T}\,\epsilon T_1T_2}{(T_1+T_2+T_3+\epsilon)^6} = \int_{[1,\infty)^3}\frac{\d \vec{t}\,t_1t_2}{(t_1+t_2+t_3+1)^6} = \frac{13}{5!\cdot 2^5}\,.
\eea
Therefore equation \eqref{eq:O1-calc-app} is equal to
\be \label{eq:O1-anomaly-app} \frac{4\cdot2\bh^\vee}{5!\cdot 2^5}\bigg(\frac{\im}{2\pi}\bigg)^3\int_{\C^3}(-3\cB_1+13\cB_2+5\cB_3) \ee
for
\be \begin{split} \label{eq:O1-terms-app}
\cB_1 &= \tr(\p_\dal\p_\db\p\ba\p\ba)\p^\dal\p\bh\p^\db\bh\,, \\
\cB_2 &= \tr(\p_\dal\p\ba\p_\db\p\ba)\p^\dal\p\bh\p^\db\bh\,, \\
\cB_3 &= \tr(\p_\dal\p\ba\p\ba)\p_\db\p^\dal\p\bh\p^\db\bh\,.
\end{split} \ee
This is the form assumed in equation \eqref{eq:O1-anomaly-2}.

The contribution from the diagram (A1) is \eqref{eq:A1-calc}
\be \begin{split} \label{eq:A1-calc-app}
&\lim_{\epsilon\to0}2\bh^\vee\bigg(\frac{\im}{2\pi}\bigg)^3\int_{\C^3}\int_{[\epsilon,L]^3}\frac{\d^3\vec T\epsilon}{(\epsilon+T_1+T_2+T_3)^6}\bigg(\epsilon\frac{\p}{\p z_1^{\dal_2}} - T_3\frac{\p}{\p z_3^{\dal_2}}\bigg) \\
&\qquad\bigg(\epsilon\frac{\p}{\p z_1^{\dal_3}} + (\epsilon+T_1)\frac{\p}{\p z_2^{\dal_3}}\bigg)\tr(\p\ba_1\p^{\dal_2}\p\bh_2\p^{\dal_3}\p\bh_3\ba_4)|_{z=z_1=z_2=z_3=z_4}\,.
\end{split} \ee
Expanding out
\be \begin{split}
&\bigg(\epsilon\frac{\p}{\p z_1^{\dal_2}} - T_3\frac{\p}{\p z_3^{\dal_2}}\bigg)\bigg(\epsilon\frac{\p}{\p z_1^{\dal_3}} + (\epsilon+T_1)\frac{\p}{\p z_2^{\dal_3}}\bigg)\tr(\p\ba_1\p^{\dal_2}\p \bh_2\p^{\dal_3}\p\bh_3\ba_4)|_{z=z_1=z_2=z_3=z_4} \\
&= \epsilon^2\tr(\p_{\dal_2}\p_{\dal_3}\p\ba\ba)\p^{\dal_2}\p\bh\p^{\dal_3}\p\bh - \epsilon T_3\tr(\p_{\dal_3}\p\ba\ba)\p^{\dal_2}\p\bh\p_{\dal_2}\p^{\dal_3}\p\bh \\
&+\epsilon(\epsilon+T_1)\tr(\p_{\dal_2}\p\ba\ba)\p_{\dal_3}\p^{\dal_2}\p\bh\p^{\dal_3}\p\bh - (\epsilon+T_1)T_3\tr(\p\ba\ba)\p_{\dal_3}\p^{\dal_2}\p\bh\p_{\dal_2}\p^{\dal_3}\p\bh\,,
\end{split} \ee
the middle 2 terms can be seen to be equal by relabelling dummy indices. There is only one integral independent of those evaluated in \eqref{eq:mixed-integrals-1}:
\be \label{eq:mixed-integrals-2}
\lim_{\epsilon\to0}\int_{[\epsilon,L]^3}\frac{\d^3\vec{T}\,\epsilon^3}{(T_1+T_2+T_3+\epsilon)^6} = \int_{[1,\infty)^3}\frac{\d^3\vec{t}}{(t_1+t_2+t_3+1)^6} = \frac{1}{5!\cdot 2^5}\,.
\ee
Equation \eqref{eq:A1-calc} therefore evaluates to
\be \frac{2\bh^\vee}{5!\cdot2^5}\bigg(\frac{\im}{2\pi}\bigg)^3\int_{\C^3}(\cB_1+\cB_3-16\cB_4) \ee
for $\cB_1,\cB_3$ as above and
\be \label{eq:A1-term-app}
\cB_4 = \tr(\p\ba\p\ba)\p^\db\p_\dal\p\bh\p^\dal\p_\db\bh\,.
\ee
This is the form assumed in equation \eqref{eq:A1-anomaly}.

The mixed Poisson-gauge anomaly diagram (A2) evaluates to
\be \begin{split} \label{eq:A2-calc-app}
&\lim_{\epsilon\to0}2\bh^\vee\bigg(\frac{\im}{2\pi}\bigg)^3\int_{\C^3}\int_{[\epsilon,L]^3}\frac{\d^3\vec T\epsilon}{(\epsilon+T_1+T_2+T_3)^6}\bigg(-(T_2+T_3)\frac{\p}{\p z_2^{\dal_1}} - T_3\frac{\p}{\p z_3^{\dal_1}}\bigg) \\
&\qquad\bigg(-(T_1+T_2)\frac{\p}{\p z_1^{\dal_4}} - T_2\frac{\p}{\p z_2^{\dal_4}}\bigg)\tr(\p^{\dal_1}\p\bh_1\p\ba_2\p\ba_3\p^{\dal_4}\bh_4)|_{z=z_1=z_2=z_3=z_4}\,,
\end{split} \ee
which we seek to evaluate in the limit $\epsilon\to0$. Expanding out the brackets gives
\be \begin{split} &\bigg((T_2+T_3)\frac{\p}{\p z_2^{\dal_1}} + T_3\frac{\p}{\p z_3^{\dal_1}}\bigg)\bigg((T_1+T_2)\frac{\p}{\p z_1^{\dal_4}} + T_2\frac{\p}{\p z_2^{\dal_4}}\bigg) \\
&\qquad\tr(\p^{\dal_1}\p\bh_1\p\ba_2\p\ba_3\p^{\dal_4}\bh_4)|_{z=z_1=z_2=z_3=z_4} \\
&= T_2(T_2+T_3)\tr(\p_{\dal_1}\p_{\dal_4}\p\ba\p\ba)\p^{\dal_1}\p\bh\p^{\dal_4}\bh+ T_2T_3\tr(\p_{\dal_4}\p\ba\p_{\dal_1}\p\ba)\p^{\dal_1}\p\bh\p^{\dal_4}\bh \\ 
& + (T_1+T_2)(T_2+T_3)\tr(\p_{\dal_1}\p\ba\p\ba)\p_{\dal_4}\p^{\dal_1}\p\bh\p^{\dal_4}\bh \\ 
&+ (T_1+T_2)T_3\tr(\p\ba\p_{\dal_1}\p\ba)\p_{\dal_4}\p^{\dal_1}\p\bh\p^{\dal_4}\bh\,.
\end{split} \ee
The final two terms are equal by cyclicity of the trace. There is only one new integral to evaluate:
\be \label{eq:mixed-integrals-3} \lim_{\epsilon\to0}\int_{[\epsilon,L]^3}\frac{\d^3\vec{T}\,\epsilon T_1^2}{(T_1+T_2+T_3+\epsilon)^6} = \int_{[1,\infty)^3}\frac{\d^3\vec{t}\, t_1^2}{(t_1+t_2+t_3+\epsilon)^6} = \frac{21}{5!\cdot2^5}\,. \ee
Equation \eqref{eq:A2-calc-app} is therefore equal to
\be \frac{2\bh^\vee}{5!\cdot 2^5}\bigg(\frac{\im}{2\pi}\bigg)^3\int_{\C^3}(34\cB_1 + 13\cB_2 + 86\cB_3) \ee
for $\cB_1,\cB_2,\cB_3$ as above. This matches equation \eqref{eq:A2-anomaly}.

Finally, the mixed Poisson-gauge anomaly diagram (A3) is
\be \begin{split} \label{eq:A3-calc-app}
&\lim_{\epsilon\to0}2\cdot 2\bh^\vee\bigg(\frac{\im}{2\pi}\bigg)^3\int_{\C^3}\int_{[\epsilon,L]^3}\frac{\d^3\vec T\epsilon}{(\epsilon+T_1+T_2+T_3)^6}\bigg(\epsilon\frac{\p}{\p z_1^{\dal_3}} + (\epsilon+T_1)\frac{\p}{\p z_2^{\dal_3}}\bigg) \\
&\qquad\bigg(-(T_1+T_2)\frac{\p}{\p z_1^{\dal_4}} - T_2\frac{\p}{\p z_2^{\dal_4}}\bigg)\tr(\p\ba_1\p\ba_2\p^{\dal_3}\p\bh\p^{\dal_4}\bh)|_{z=z_1=z_2=z_3=z_4}
\end{split} \ee
\end{appendix}
in the limit $\epsilon\to0$. Expanding out the brackets gives
\be \begin{split}
&\bigg(\epsilon\frac{\p}{\p z_1^{\dal_3}} + (\epsilon+T_1)\frac{\p}{\p z_2^{\dal_3}}\bigg)\bigg(-(T_1+T_2)\frac{\p}{\p z_1^{\dal_4}} - T_2\frac{\p}{\p z_2^{\dal_4}}\bigg)\tr(\p\ba_1\p\ba_2\p^{\dal_3}\p\bh\p^{\dal_4}\bh)|_{z=z_1=z_2=z_3=z_4} \\
&= -\epsilon(T_1+T_2)\tr(\p_{\dal_3}\p_{\dal_4}\p\ba\p\ba)\p^{\dal_3}\p\bh\p^{\dal_4}\bh - (\epsilon+T_1)T_2\tr(\p\ba\p_{\dal_3}\p_{\dal_4}\p\ba)\p^{\dal_3}\p\bh\p^{\dal_4}\bh \\
&- \epsilon T_2\tr(\p_{\dal_3}\p\ba\p_{\dal_4}\p\ba)\p^{\dal_3}\p\bh\p^{\dal_4}\bh - (\epsilon+T_1)(T_1+T_2)\tr(\p_{\dal_4}\p\ba\p_{\dal_3}\p\ba)\p^{\dal_3}\p\bh\p^{\dal_4}\bh\,.
\end{split} \ee
The first two terms and last two terms are equal by cyclicity of the trace. Employing the integrals in equations \eqref{eq:mixed-integrals-1}, \eqref{eq:mixed-integrals-3} we find that \eqref{eq:A3-calc-app} evaluates to
\be -\frac{2\cdot 2\bh^\vee}{5!\cdot2^5}\bigg(\frac{\im}{2\pi}\bigg)^3\int_{\C^3}(22\cB_1+43\cB_2)\,. \ee
This is the form quoted in equation \eqref{eq:A3-anomaly}.

%%%%%%%%%%%%%%%%%%%%%%%%%%%%%%%%%%
%%%%%%%%%%%%%%%%%%%%%%%%%%%%%%%%%%
%%%%%%%%%%%%%%%%%%%%%%%%%%%%%%%%%%

\newpage

\bibliographystyle{JHEP}
\bibliography{grint}

\providecommand{\href}[2]{#2}\begingroup\raggedright\begin{thebibliography}{10}

\bibitem{Plebanski:1975wn}
J.~F. Plebanski, {\it {Some solutions of complex {Einstein} equations}},  {\em
  J. Math. Phys.} {\bf 16} (1975) 2395--2402.

\bibitem{Dunajski:2000iq}
M.~Dunajski and L.~Mason, {\it {Hyperk\"{a}hler hierarchies and their twistor
  theory}},  {\em Commun. Math. Phys.} {\bf 213} (2000) 641--672,
  [\href{http://arxiv.org/abs/math/0001008}{{\tt math/0001008}}].

\bibitem{Ward:1990qt}
R.~S. Ward, {\it {Einstein-Weyl spaces and $\mathrm{SU}(\infty)$ Toda fields}},
   {\em Class. Quant. Grav.} {\bf 7} (1990) L95--L98.

\bibitem{Dunajski:2000rf}
M.~Dunajski, L.~J. Mason, and P.~Tod, {\it {Einstein-Weyl geometry, the dKP
  equation and twistor theory}},  {\em J. Geom. Phys.} {\bf 37} (2001) 63--93,
  [\href{http://arxiv.org/abs/math/0004031}{{\tt math/0004031}}].

\bibitem{Park:1989vq}
Q.-H. Park, {\it {Selfdual gravity as a large $N$ limit of the two-dimensional
  nonlinear $\sigma$ model}},  {\em Phys. Lett. B} {\bf 238} (1990) 287--290.

\bibitem{Ooguri:1991fp}
H.~Ooguri and C.~Vafa, {\it {Geometry of $\mathcal{N}=2$ strings}},  {\em Nucl.
  Phys. B} {\bf 361} (1991) 469--518.

\bibitem{Penrose:1976js}
R.~Penrose, {\it {Nonlinear gravitons and curved twistor theory}},  {\em Gen.
  Rel. Grav.} {\bf 7} (1976) 31--52.

\bibitem{Atiyah:1978wi}
M.~F. Atiyah, N.~J. Hitchin, and I.~M. Singer, {\it {Selfduality in
  four-dimensional {Riemannian} geometry}},  {\em Proc. Roy. Soc. Lond. A} {\bf
  362} (1978) 425--461.

\bibitem{Mason:1991rf}
L.~J. Mason and N.~M.~J. Woodhouse, {\em {Integrability, selfduality, and
  twistor theory}}.
\newblock Clarendon Press, 1991.

\bibitem{Bern:1998xc}
Z.~Bern, L.~J. Dixon, M.~Perelstein, and J.~S. Rozowsky, {\it {One loop {$n$}
  point helicity amplitudes in (self-dual) gravity}},  {\em Phys. Lett. B} {\bf
  444} (1998) 273--283, [\href{http://arxiv.org/abs/hep-th/9809160}{{\tt
  hep-th/9809160}}].

\bibitem{Bern:1998sv}
Z.~Bern, L.~J. Dixon, M.~Perelstein, and J.~S. Rozowsky, {\it {Multileg one
  loop gravity amplitudes from gauge theory}},  {\em Nucl. Phys. B} {\bf 546}
  (1999) 423--479, [\href{http://arxiv.org/abs/hep-th/9811140}{{\tt
  hep-th/9811140}}].

\bibitem{Bern:1993qk}
Z.~Bern, G.~Chalmers, L.~J. Dixon, and D.~A. Kosower, {\it {One-loop {$N$}
  gluon amplitudes with maximal helicity violation via collinear limits}},
  {\em Phys. Rev. Lett.} {\bf 72} (1994) 2134--2137,
  [\href{http://arxiv.org/abs/hep-ph/9312333}{{\tt hep-ph/9312333}}].

\bibitem{Mahlon:1993fe}
G.~Mahlon, {\it {One loop multi-photon helicity amplitudes}},  {\em Phys. Rev.
  D} {\bf 49} (1994) 2197--2210,
  [\href{http://arxiv.org/abs/hep-ph/9311213}{{\tt hep-ph/9311213}}].

\bibitem{Bardeen:1995gk}
W.~A. Bardeen, {\it {Selfdual Yang-Mills theory, integrability and multiparton
  amplitudes}},  {\em Prog. Theor. Phys. Suppl.} {\bf 123} (1996) 1--8.

\bibitem{Costello:2021bah}
K.~J. Costello, {\it {Quantizing local holomorphic field theories on twistor
  space}},  \href{http://arxiv.org/abs/2111.08879}{{\tt arXiv:2111.08879}}.

\bibitem{Mason:2007ct}
L.~Mason and M.~Wolf, {\it {Twistor actions for self-dual supergravities}},
  {\em Commun. Math. Phys.} {\bf 288} (2009) 97--123,
  [\href{http://arxiv.org/abs/0706.1941}{{\tt arXiv:0706.1941}}].

\bibitem{Williams:2018ows}
B.~R. Williams, {\it {Renormalization for holomorphic field theories}},  {\em
  Commun. Math. Phys.} {\bf 374} (2020), no.~3 1693--1742,
  [\href{http://arxiv.org/abs/1809.02661}{{\tt arXiv:1809.02661}}].

\bibitem{Costello:2007ei}
K.~J. Costello, {\it {Renormalisation and the Batalin-Vilkovisky formalism}},
  \href{http://arxiv.org/abs/0706.1533}{{\tt arXiv:0706.1533}}.

\bibitem{Ooguri:1990ww}
H.~Ooguri and C.~Vafa, {\it {Selfduality and $\mathcal{N}=2$ String magic}},
  {\em Mod. Phys. Lett. A} {\bf 5} (1990) 1389--1398.

\bibitem{Berkovits:1994vy}
N.~Berkovits and C.~Vafa, {\it {$\mathcal{N}=4$ topological strings}},  {\em
  Nucl. Phys. B} {\bf 433} (1995) 123--180,
  [\href{http://arxiv.org/abs/hep-th/9407190}{{\tt hep-th/9407190}}].

\bibitem{Berkovits:1994ym}
N.~Berkovits, {\it {Vanishing theorems for the selfdual $\mathcal{N}=2$
  string}},  {\em Phys. Lett. B} {\bf 350} (1995) 28--32,
  [\href{http://arxiv.org/abs/hep-th/9412179}{{\tt hep-th/9412179}}].

\bibitem{Ooguri:1995cp}
H.~Ooguri and C.~Vafa, {\it {All loop $\mathcal{N}=2$ string amplitudes}},
  {\em Nucl. Phys. B} {\bf 451} (1995) 121--161,
  [\href{http://arxiv.org/abs/hep-th/9505183}{{\tt hep-th/9505183}}].

\bibitem{Costello:2013sla}
K.~Costello, {\it {Integrable lattice models from four-dimensional field
  theories}},  {\em Proc. Symp. Pure Math.} {\bf 88} (2014) 3--24,
  [\href{http://arxiv.org/abs/1308.0370}{{\tt arXiv:1308.0370}}].

\bibitem{nekrassov1996four}
N.~A. Nekrassov, {\em {Four-dimensional holomorphic theories}}.
\newblock Princeton University, 1996.

\bibitem{Costello:2017dso}
K.~Costello, E.~Witten, and M.~Yamazaki, {\it {Gauge theory and integrability,
  I}},  {\em ICCM Not.} {\bf 06} (2018), no.~1 46--119,
  [\href{http://arxiv.org/abs/1709.09993}{{\tt arXiv:1709.09993}}].

\bibitem{Costello:2018gyb}
K.~Costello, E.~Witten, and M.~Yamazaki, {\it {Gauge theory and integrability,
  II}},  {\em ICCM Not.} {\bf 06} (2018), no.~1 120--146,
  [\href{http://arxiv.org/abs/1802.01579}{{\tt arXiv:1802.01579}}].

\bibitem{Costello:2019tri}
K.~Costello and M.~Yamazaki, {\it {Gauge theory and integrability, III}},
  \href{http://arxiv.org/abs/1908.02289}{{\tt arXiv:1908.02289}}.

\bibitem{Bittleston:2020hfv}
R.~Bittleston and D.~Skinner, {\it {Twistors, the ASD Yang-Mills equations, and
  4d Chern-Simons theory}},  \href{http://arxiv.org/abs/2011.04638}{{\tt
  arXiv:2011.04638}}.

\bibitem{Costello:2020lpi}
K.~Costello and B.~Stefa\'nski, {\it {Chern-Simons origin of superstring
  integrability}},  {\em Phys. Rev. Lett.} {\bf 125} (2020), no.~12 121602,
  [\href{http://arxiv.org/abs/2005.03064}{{\tt arXiv:2005.03064}}].

\bibitem{Ward:1977ta}
R.~S. Ward, {\it {On selfdual gauge fields}},  {\em Phys. Lett. A} {\bf 61}
  (1977) 81--82.

\bibitem{Witten:2003nn}
E.~Witten, {\it {Perturbative gauge theory as a string theory in twistor
  space}},  {\em Commun. Math. Phys.} {\bf 252} (2004) 189--258,
  [\href{http://arxiv.org/abs/hep-th/0312171}{{\tt hep-th/0312171}}].

\bibitem{Mason:2005zm}
L.~J. Mason, {\it {Twistor actions for non-self-dual fields: a derivation of
  twistor-string theory}},  {\em JHEP} {\bf 10} (2005) 009,
  [\href{http://arxiv.org/abs/hep-th/0507269}{{\tt hep-th/0507269}}].

\bibitem{Boels:2006ir}
R.~Boels, L.~J. Mason, and D.~Skinner, {\it {Supersymmetric gauge theories in
  twistor space}},  {\em JHEP} {\bf 02} (2007) 014,
  [\href{http://arxiv.org/abs/hep-th/0604040}{{\tt hep-th/0604040}}].

\bibitem{Penrose:1976jq}
R.~Penrose, {\it {The nonlinear graviton}},  {\em Gen. Rel. Grav.} {\bf 7}
  (1976) 171--176.

\bibitem{Skinner:2013xp}
D.~Skinner, {\it {Twistor strings for $\mathcal{N}$ = 8 supergravity}},  {\em
  JHEP} {\bf 04} (2020) 047, [\href{http://arxiv.org/abs/1301.0868}{{\tt
  arXiv:1301.0868}}].

\bibitem{Sharma:2021pkl}
A.~Sharma, {\it {Twistor action for general relativity}},
  \href{http://arxiv.org/abs/2104.07031}{{\tt arXiv:2104.07031}}.

\bibitem{Capovilla:1991qb}
R.~Capovilla, T.~Jacobson, J.~Dell, and L.~J. Mason, {\it {Selfdual two forms
  and gravity}},  {\em Class. Quant. Grav.} {\bf 8} (1991) 41--57.

\bibitem{Krasnov:2021cva}
K.~Krasnov and E.~Skvortsov, {\it {Flat self-dual gravity}},  {\em JHEP} {\bf
  08} (2021) 082, [\href{http://arxiv.org/abs/2106.01397}{{\tt
  arXiv:2106.01397}}].

\bibitem{Ashtekar:1987qx}
A.~Ashtekar, T.~Jacobson, and L.~Smolin, {\it {A new characterization of half
  flat solutions to Einstein's equation}},  {\em Commun. Math. Phys.} {\bf 115}
  (1988) 631.

\bibitem{Smolin:1992wj}
L.~Smolin, {\it {The $G_\mathrm{Newton}\to0$ limit of Euclidean quantum
  gravity}},  {\em Class. Quant. Grav.} {\bf 9} (1992) 883--894,
  [\href{http://arxiv.org/abs/hep-th/9202076}{{\tt hep-th/9202076}}].

\bibitem{Ray:1973sb}
D.~B. Ray and I.~M. Singer, {\it {Analytic torsion for complex manifolds}},
  {\em Annals Math.} {\bf 98} (1973) 154--177.

\bibitem{Bershadsky:1993cx}
M.~Bershadsky, S.~Cecotti, H.~Ooguri, and C.~Vafa, {\it {Kodaira-Spencer theory
  of gravity and exact results for quantum string amplitudes}},  {\em Commun.
  Math. Phys.} {\bf 165} (1994) 311--428,
  [\href{http://arxiv.org/abs/hep-th/9309140}{{\tt hep-th/9309140}}].

\bibitem{bismut1988analytic1}
J.-M. Bismut, H.~Gillet, and C.~Soul{\'e}, {\it Analytic torsion and
  holomorphic determinant bundles. {I}. {Bott}-{Chern} forms and analytic
  torsion},  {\em Commun.~Math.~Phys.} {\bf 115} (1988), no.~1 49--78.

\bibitem{bismut1988analytic2}
J.-M. Bismut, H.~Gillet, and C.~Soul{\'e}, {\it Analytic torsion and
  holomorphic determinant bundles. {II}. direct images and {Bott}-{Chern}
  forms},  {\em Commun.~Math.~Phys.} {\bf 115} (1988), no.~1 79--126.

\bibitem{bismut1988analytic3}
J.-M. Bismut, H.~Gillet, and C.~Soul{\'e}, {\it Analytic torsion and
  holomorphic determinant bundles. {III}. {Quillen} metrics on holomorphic
  determinants},  {\em Commun.~Math.~Phys.} {\bf 115} (1988), no.~2 301--351.

\bibitem{quillen1985determinants}
D.~Quillen, {\it Determinants of {Cauchy}-{Riemann} operators over a {Riemann}
  surface},  {\em Functional Analysis and Its Applications} {\bf 19} (1985),
  no.~1 31--34.

\bibitem{Costello:2015xsa}
K.~Costello and S.~Li, {\it {Quantization of open-closed {BCOV} theory, {I}}},
  \href{http://arxiv.org/abs/1505.06703}{{\tt arXiv:1505.06703}}.

\bibitem{Gwilliam:2018lpo}
O.~Gwilliam and B.~R. Williams, {\it {Higher Kac-Moody algebras and symmetries
  of holomorphic field theories}},  {\em Adv. Theor. Math. Phys.} {\bf 25}
  (2021), no.~1 129--239, [\href{http://arxiv.org/abs/1810.06534}{{\tt
  arXiv:1810.06534}}].

\bibitem{elliott2020holomorphic}
C.~Elliott and B.~R. Williams, {\it Holomorphic {Poisson} field theories},
  {\em arXiv preprint arXiv:2008.02302} (2020).

\bibitem{Alvarez-Gaume:1983ihn}
L.~Alvarez-Gaume and E.~Witten, {\it {Gravitational anomalies}},  {\em Nucl.
  Phys. B} {\bf 234} (1984) 269.

\bibitem{Frampton:1983ez}
P.~H. Frampton and T.~W. Kephart, {\it {Consistency conditions for Kaluza-Klein
  axial anomalies}},  {\em Phys. Rev. Lett.} {\bf 50} (1983) 1347--1349.

\bibitem{Townsend:1983ana}
P.~K. Townsend and G.~Sierra, {\it {Chiral anomalies and constraints on the
  gauge group in higher dimensional supersymmetric {Yang-Mills} theories}},
  {\em Nucl. Phys. B} {\bf 222} (1983) 493--506.

\bibitem{Zumino:1983rz}
B.~Zumino, Y.-S. Wu, and A.~Zee, {\it {Chiral anomalies, higher dimensions, and
  differential geometry}},  {\em Nucl. Phys. B} {\bf 239} (1984) 477--507.

\bibitem{Wallet:1989wr}
J.~C. Wallet, {\it {Algebraic setup for the gauge fixing of BF and super BF
  systems}},  {\em Phys. Lett. B} {\bf 235} (1990) 71.

\bibitem{Budzik:2022mpd}
K.~Budzik, D.~Gaiotto, J.~Kulp, J.~Wu, and M.~Yu, {\it {Feynman Diagrams in
  Four-Dimensional Holomorphic Theories and the Operatope}},
  \href{http://arxiv.org/abs/2207.14321}{{\tt arXiv:2207.14321}}.

\bibitem{Axelrod:1991vq}
S.~Axelrod and I.~M. Singer, {\it {Chern-Simons perturbation theory}},  in {\em
  {International Conference on Differential Geometric Methods in Theoretical
  Physics}}, pp.~3--45, 1991.
\newblock \href{http://arxiv.org/abs/hep-th/9110056}{{\tt hep-th/9110056}}.

\bibitem{Alexandrov:1995kv}
M.~Alexandrov, A.~Schwarz, O.~Zaboronsky, and M.~Kontsevich, {\it {The geometry
  of the master equation and topological quantum field theory}},  {\em Int. J.
  Mod. Phys. A} {\bf 12} (1997) 1405--1429,
  [\href{http://arxiv.org/abs/hep-th/9502010}{{\tt hep-th/9502010}}].

\bibitem{Tran:2022tft}
T.~Tran, {\it {Toward a twistor action for chiral higher-spin gravity}},
  \href{http://arxiv.org/abs/2209.00925}{{\tt arXiv:2209.00925}}.

\bibitem{Skvortsov:2020gpn}
E.~Skvortsov and T.~Tran, {\it {One-loop finiteness of chiral higher spin
  gravity}},  {\em JHEP} {\bf 07} (2020) 021,
  [\href{http://arxiv.org/abs/2004.10797}{{\tt arXiv:2004.10797}}].

\bibitem{Costello:2022wso}
K.~Costello and N.~M. Paquette, {\it {Celestial holography meets twisted
  holography: 4d amplitudes from chiral correlators}},
  \href{http://arxiv.org/abs/2201.02595}{{\tt arXiv:2201.02595}}.

\bibitem{Okubo:1978qe}
S.~Okubo, {\it {Quartic trace identity for exceptional Lie algebras}},  {\em J.
  Math. Phys.} {\bf 20} (1979) 586.

\bibitem{Okubo:1981td}
S.~Okubo, {\it {Modified fourth order Casimir invariants and indices for simple
  Lie algebras}},  {\em J. Math. Phys.} {\bf 23} (1982) 8.

\bibitem{Costello:2019jsy}
K.~Costello and S.~Li, {\it {Anomaly cancellation in the topological string}},
  {\em Adv. Theor. Math. Phys.} {\bf 24} (2020), no.~7 1723--1771,
  [\href{http://arxiv.org/abs/1905.09269}{{\tt arXiv:1905.09269}}].

\bibitem{McDuff:2012}
D.~McDuff and D.~Salamon, {\em {$J$-holomorphic curves and symplectic
  topology}}.
\newblock American Mathematical Society, 2~ed., 2012.

\bibitem{Adamo:2021bej}
T.~Adamo, L.~Mason, and A.~Sharma, {\it {Twistor sigma models for quaternionic
  geometry and graviton scattering}},
  \href{http://arxiv.org/abs/2103.16984}{{\tt arXiv:2103.16984}}.

\bibitem{Mason:FAv1}
L.~Mason, {\it {Local twistors and the Penrose tranform for homogeneous
  bundles}},  {\em Twistor Newsletter} {\bf 23} (1987) 36--41.

\bibitem{Fradkin:1985am}
E.~S. Fradkin and A.~A. Tseytlin, {\it {Conformal supergravity}},  {\em Phys.
  Rept.} {\bf 119} (1985) 233--362.

\bibitem{Berkovits:2004jj}
N.~Berkovits and E.~Witten, {\it {Conformal supergravity in twistor-string
  theory}},  {\em JHEP} {\bf 08} (2004) 009,
  [\href{http://arxiv.org/abs/hep-th/0406051}{{\tt hep-th/0406051}}].

\bibitem{Chiodaroli:2017ngp}
M.~Chiodaroli, M.~Gunaydin, H.~Johansson, and R.~Roiban, {\it {Explicit
  formulae for {Yang}-{Mills}-{Einstein} amplitudes from the double copy}},
  {\em JHEP} {\bf 07} (2017) 002, [\href{http://arxiv.org/abs/1703.00421}{{\tt
  arXiv:1703.00421}}].

\bibitem{Faller:2018vdz}
J.~Faller and J.~Plefka, {\it {Positive helicity Einstein-Yang-Mills amplitudes
  from the double copy method}},  {\em Phys. Rev. D} {\bf 99} (2019), no.~4
  046008, [\href{http://arxiv.org/abs/1812.04053}{{\tt arXiv:1812.04053}}].

\bibitem{Cangemi:1996rx}
D.~Cangemi, {\it {Selfdual Yang-Mills theory and one loop like - helicity {QCD}
  multi - gluon amplitudes}},  {\em Nucl. Phys. B} {\bf 484} (1997) 521--537,
  [\href{http://arxiv.org/abs/hep-th/9605208}{{\tt hep-th/9605208}}].

\bibitem{Chalmers:1996rq}
G.~Chalmers and W.~Siegel, {\it {The Selfdual sector of {QCD} amplitudes}},
  {\em Phys. Rev. D} {\bf 54} (1996) 7628--7633,
  [\href{http://arxiv.org/abs/hep-th/9606061}{{\tt hep-th/9606061}}].

\bibitem{Nandan:2018ody}
D.~Nandan, J.~Plefka, and G.~Travaglini, {\it {All rational one-loop
  Einstein-Yang-Mills amplitudes at four points}},  {\em JHEP} {\bf 09} (2018)
  011, [\href{http://arxiv.org/abs/1803.08497}{{\tt arXiv:1803.08497}}].

\bibitem{Grisaru:1979re}
M.~T. Grisaru and J.~Zak, {\it {One-loop scalar field contributions to
  graviton-graviton scattering and helicity nonconservation in quantum
  gravity}},  {\em Phys. Lett. B} {\bf 90} (1980) 237--240.

\bibitem{Dunbar:1994bn}
D.~C. Dunbar and P.~S. Norridge, {\it {Calculation of graviton scattering
  amplitudes using string based methods}},  {\em Nucl. Phys. B} {\bf 433}
  (1995) 181--208, [\href{http://arxiv.org/abs/hep-th/9408014}{{\tt
  hep-th/9408014}}].

\bibitem{Siegel:1992wd}
W.~Siegel, {\it {Selfdual $\mathcal{N}$=8 supergravity as closed
  $\mathcal{N}=2$ ($\mathcal{N}=4$) strings}},  {\em Phys. Rev. D} {\bf 47}
  (1993) 2504--2511, [\href{http://arxiv.org/abs/hep-th/9207043}{{\tt
  hep-th/9207043}}].

\bibitem{Dixon:2013uaa}
L.~J. Dixon, {\it {A brief introduction to modern amplitude methods}},  in {\em
  {Theoretical Advanced Study Institute in Elementary Particle Physics}:
  {Particle Physics: The Higgs Boson and Beyond}}, pp.~31--67, 2014.
\newblock \href{http://arxiv.org/abs/1310.5353}{{\tt arXiv:1310.5353}}.

\bibitem{Ball:2021tmb}
A.~Ball, S.~A. Narayanan, J.~Salzer, and A.~Strominger, {\it {Perturbatively
  exact w$_{1+\infty}$ asymptotic symmetry of quantum self-dual gravity}},
  {\em JHEP} {\bf 01} (2022) 114, [\href{http://arxiv.org/abs/2111.10392}{{\tt
  arXiv:2111.10392}}].

\bibitem{Saberi:2020pmw}
I.~Saberi and B.~R. Williams, {\it {Constraints in the BV formalism:
  six-dimensional supersymmetry and its twists}},
  \href{http://arxiv.org/abs/2009.07116}{{\tt arXiv:2009.07116}}.

\bibitem{Bu:2022dis}
W.~Bu and E.~Casali, {\it {The 4d/2d correspondence in twistor space and
  holomorphic Wilson lines}},  \href{http://arxiv.org/abs/2208.06334}{{\tt
  arXiv:2208.06334}}.

\bibitem{Costello:2022upu}
K.~Costello and N.~M. Paquette, {\it {On the associativity of one-loop
  corrections to the celestial OPE}},
  \href{http://arxiv.org/abs/2204.05301}{{\tt arXiv:2204.05301}}.

\bibitem{Guevara:2021abz}
A.~Guevara, E.~Himwich, M.~Pate, and A.~Strominger, {\it {Holographic symmetry
  algebras for gauge theory and gravity}},  {\em JHEP} {\bf 11} (2021) 152,
  [\href{http://arxiv.org/abs/2103.03961}{{\tt arXiv:2103.03961}}].

\bibitem{Strominger:2021lvk}
A.~Strominger, {\it {$w_{1+\infty}$ and the celestial sphere}},
  \href{http://arxiv.org/abs/2105.14346}{{\tt arXiv:2105.14346}}.

\bibitem{Adamo:2021lrv}
T.~Adamo, L.~Mason, and A.~Sharma, {\it {Celestial $w_{1+\infty}$ symmetries
  from twistor space}},  {\em SIGMA} {\bf 18} (2022) 016,
  [\href{http://arxiv.org/abs/2110.06066}{{\tt arXiv:2110.06066}}].

\bibitem{Adamo:2021zpw}
T.~Adamo, W.~Bu, E.~Casali, and A.~Sharma, {\it {Celestial operator products
  from the worldsheet}},  {\em JHEP} {\bf 06} (2022) 052,
  [\href{http://arxiv.org/abs/2111.02279}{{\tt arXiv:2111.02279}}].

\bibitem{Costello:2022jpg}
K.~Costello, N.~M. Paquette, and A.~Sharma, {\it {Top-down holography in an
  asymptotically flat spacetime}},  \href{http://arxiv.org/abs/2208.14233}{{\tt
  arXiv:2208.14233}}.

\end{thebibliography}\endgroup

\end{document}